# SISTEMA DE SENSORIAMENTO SEM FIO APLICÁVEL A DETECÇÃO DE INDÍCIOS DE INCÊNDIOS FLORESTAIS

Lucas Santos Gonçalves, Celso Barbosa Carvalho

# Abstract

In this research work, a hardware and software system is developed that uses wireless sensors to monitor environmental variables such as temperature, gas concentration and luminosity, in order to detect the existence of forest fires. Lora technology was used for wireless sensor networks with communication range that can reach on average up to 5km in urban areas and 10km in rural areas. The developed system also has an integrated web application (dashboard) and that in real time, collects data from wireless sensors, which together form the sensor module, also called device. Then, this data is presented on a map associ- ated with the positioning of each sensor module. The developed system was tested using practical experiments that used flames, gases and lighting, simulating the occurrence of fires. With the tests performed, it was observed the feasibility of the system, hardware/software developed, in detecting the fires in the simulated scenarios. Therefore, it was found that the research is promising, and may advance in the future for the detection of real fires.

**Keywords**: LoRa, LoRaWAN, Server, The Things Network, monitoring.

# 1. Introdução

Os recursos tecnológicos são utilizados para diversos fins, incluindo a prevenção de acidentes e segurança de pessoas ou patrimônio. Decerto, a ocorrência de tragédias ou eventos catastróficos – como o caso de incêndios florestais – é utilizada nas pesquisas, a fim de alertar as autoridades competentes em tempo adequado. A esse respeito, é preciso operar um sistema destinado à descoberta de incêndios na sua fase inicial, quando as chamas ainda não se espalharam.

Para desenvolver um sistema com estas características, faz-se necessário, primeiramente, entender as principais variáveis que caracterizam o início de incêndios, tais como a existência de gases, a elevação da temperatura e, até mesmo, o nível de luminosidade. Ademais, um incêndio pode gerar impactos danosos em termos econômicos, sociais e, principalmente, ambientais, afetando, particularmente, os animais que não têm para onde se abrigar numa situação dessa.

Atualmente, a principal forma de monitorar florestas na detectação de queimadas é por Satélite, com sensores óticos que operam na faixa termal média, com vista à possibilidade do processamento de imagens da superfície terrestre. Vale destacar que cada satélite de órbita produz pelo menos um conjunto de imagens por dia, enquanto os geoestacionários geram algumas imagens por hora. Mesmo porque as regiões, com mais ocorrência de focos de incêndio, são usadas para análise de tendências com o passar dos anos, e o satélite que faz a melhor cobertura dessa área, é tido como referência.

Ancora-se nesse perspectiva, a existência de uma relação entre foco e queimada que dificulta a localização dos focos de incêndio com precisão. Partindo das imagens do satélite de referência, um foco indica a existência de fogo em um elemento de resolução da imagem (pixel), sendo que a área desse pixel pode variar, na escala real, de um a seis quilômetros quadrados. Isto significa – mesmo que o foco seja detectado –, investigar o local exato do foco do incêndio, ou ainda, se existe mais de um foco para ver se toda área já está comprometida com danos que podem ser irreversíveis.

Nesse sentido, encontram-se, ainda, algumas condições que impedem ou prejudicam muito a detecção das queimadas, mesmo com um número grande de satélites. Por exemplo, quando há fogo dentro de uma floresta densa e fechada, sem que as chamas alcancem as copas das árvores, seria em média, equivalente à labaredas de fogo com pelo menos 30 metros de altura. Ou ainda – no caso de haver nuvens cobrindo a região – o satélite não captura uma imagem que comprove com clareza a situação dessa área em questão[1].

Em face disso, Da Maceno (2019) argumenta que, a proposta é apresentar um serviço de monitoramento de indícios de queimadas em florestas, usando tecnologia LoRa (*Long Range*). Trata-se de uma tecnologia de comunicação apropriada para aplicações em áreas externas (*outdoor*), voltadas para áreas remotas sem conexão à internet, geralmente rurais. Mesmo porque não possuem uma boa cobertura de outras tecnologias de comunicação, como as de redes móveis, podendo estabelecer comunicação de uma rede de sensores sem fio, com satélites em órbita.

Portanto, essa tecnologia é classificada como do tipo LPWAN (*Low Power Wide Area*), em que os dispositivos sem fio possuem baixa potência e longo alcance de comunicação. Possui características de baixo consumo para operar em áreas maiores. E por possuir uma plataforma de *Hardware* de fácil configuração com outros microcontroladores, é possível fazer um monitoramento de grandezas que modelam o início de um incêndio, como fumaça e os gases traços, como metano e monóxido de carbono. Assim como a luz e o calor são outras variáveis que ajudam a compor a presença de chamas. Isso é possível implantando um módulo sensor, em locais estratégicos, com sensores que detectam essas grandezas, aumentando a precisão da localização do possível foco de incêndio.

Este trabalho foi construído da seguinte maneira, em: introdução – que apresenta o tema em questão; objetivos – que foram expostos o geral e os específicos do estudo e a estrutura do trabalho; fundamentação teórica – que foram explicados alguns conceitos a respeito dos recursos necessários, dentro da fundamentação teórica; metodologia – que detalha os materiais e os métodos para a realização da pesquisa, compondo a metodologia utilizada e as etapas de desenvolvimento de forma mais prática. E finalmente, têm-se a conclusão e as referências, juntamente com os anexos e os apêncices.

[1] Ibama, 2016.

# 2. Fundamentação Teórica

Nesta sessão, apresentam-se as tecnologias utilizadas neste Trabalho de Conclusão de Curso (TCC), tais como a tecnologia LoRa, os recursos de *hardware* e *software* e os servidores utilizados para a conexão e o acesso ao sistema de sensoriamento desenvolvido.

## 2.1 Princípios Básicos da Tecnologia LoRa

É uma tecnologia de transmissão de dados via rede sem fio, patenteada pela Semtech Corporation (SEMTECH, 2019) e, promovida pela LoRa Alliance (LORA ALLIANCE, 2019). Sua transmissão é realizada por RF (*Radio Frequency*), onde sua faixa de operação não é de uso restrito. Ou seja, não precisa de uma licença para utilizá-la, porém, varia para diferentes regiões do mundo, atuando no Brasil, na Austrália e nos Estados Unidos, na frequência de 915MHz. Possui o nome LoRa por ser a abreviação de *"Long Range"*, sendo, justamente, a sua principal característica. Pois, atua através de uma técnica de modulação de espalhamento espectral, que possibilita uma transmissão de dados a longas distâncias, com baixo consumo de energia. Enquanto o termo LoRaWAN se refere ao protocolo de comunicação, uma rede LPWA (*Low Power Wide Area*) opera na camada de enlace. Com reforço às reflexões, toma-se as inferências de Carvalho et al (2019), Caldas, A. Et al. (2017) e Ferreira, D., Souza, R., Carvalho, C. (2020), a destacar que a maior diferença entre a arquitetura de projetos que utilizam

LoRaWAN e a convencional para projetos IoT (Internet of Things) está na utilização do *gateway* para se conectar a internet.

## 2.2 Gateway LoRaWAN

O *Gateway* LoRaWAN – chamado também de "Concentrador LoRa" –, é um dispositivo, possuindo um chip da Semtech que possibilita a utilização de um conjunto de funções programadas em linguagem C/C++. Estas funções permitem a configuração do *Gateway*, para o envio e recebimento de pacotes, em um conjunto de canais digitais. As modulações utilizadas são a CSS (*Chirp Spread Spectrum*), chamada ainda de "Modulação LoRa" ou FSK (*Frequency-shift keying*).

Nesta linha de raciocínio, apresenta-se os 3 modos de funcionamento do *gateway*: 1) ***'timestamp' –*** o modem é iniciado antes de 1,5 *ms*. É definida uma taxa de transmissão para o pacote recebido e enviado dentro do "tempo de carimbo", que é o tempo para percorrer todo o pacote e enviá-lo para a internet; 2) '**imediato' –** o pacote é emitido o mais rápido possível. Ao transferir o pacote para o *gateway*, praticamente, enviará para internet.

Porém, existe o 'TX-START-DELAY' (pequeno *delay* de transmissão), e então o pacote é emitido; 3) **'acionado' –** o pacote é emitido 1,5 *ms* após confirmação do pacote. Pois, neste modo não há maneira de antecipar o evento acionador e iniciar o circuito analógico para começar transmissão. Antes, é confirmado que o pacote já está de fato pronto para então ser enviado.

O mesmo *gateway* serve como dispositivo de acesso a internet para mais de um *end-point*, em que ele pode operar em diferentes modos e em diferentes canais, dependendo da configuração atribuída ao *end-point* em questão.

## 2.3 Rádio LoRa RHF76-052

O módulo RHF76-052 é de baixo custo e baixa potência, incorporado ao chip SX1276 LoRa da Semtech. Em face disso, a empresa RisingHF[2] que comercializa esse produto direciona o módulo para aplicação em rede de sensores e outros dispositivos IoT. Possui baixo consumo de energia para estender a vida útil da bateria. É um módulo transceptor que possui um modem sem fio RF de longo alcance, ou seja, dentro do módulo se encontra o chip LoRa.

Nessa linha de raciocínio Centenaro (2019) acrescenta que a LoRa oferece comunicação com alta imunidade de interferências e baixo consumo de corrente. Devido a técnica de modulação por espalhamento espectral, faz com que atinja uma sensibilidade maior, o que oferece a vantagem de maior seletividade do sinal, quando comparada às técnicas convencionais de modulação.

Por certo, Bardyn (2016) afirma que essa tecnologia se torna uma grande aliada para redes sem fio de IoT que busca resolver situações relacionadas às mudanças climáticas, ao controle de poluentes, aos desastres naturais, entre outros casos em ambientes externos.

Figura 1 – Módulo LoRa utilizado.

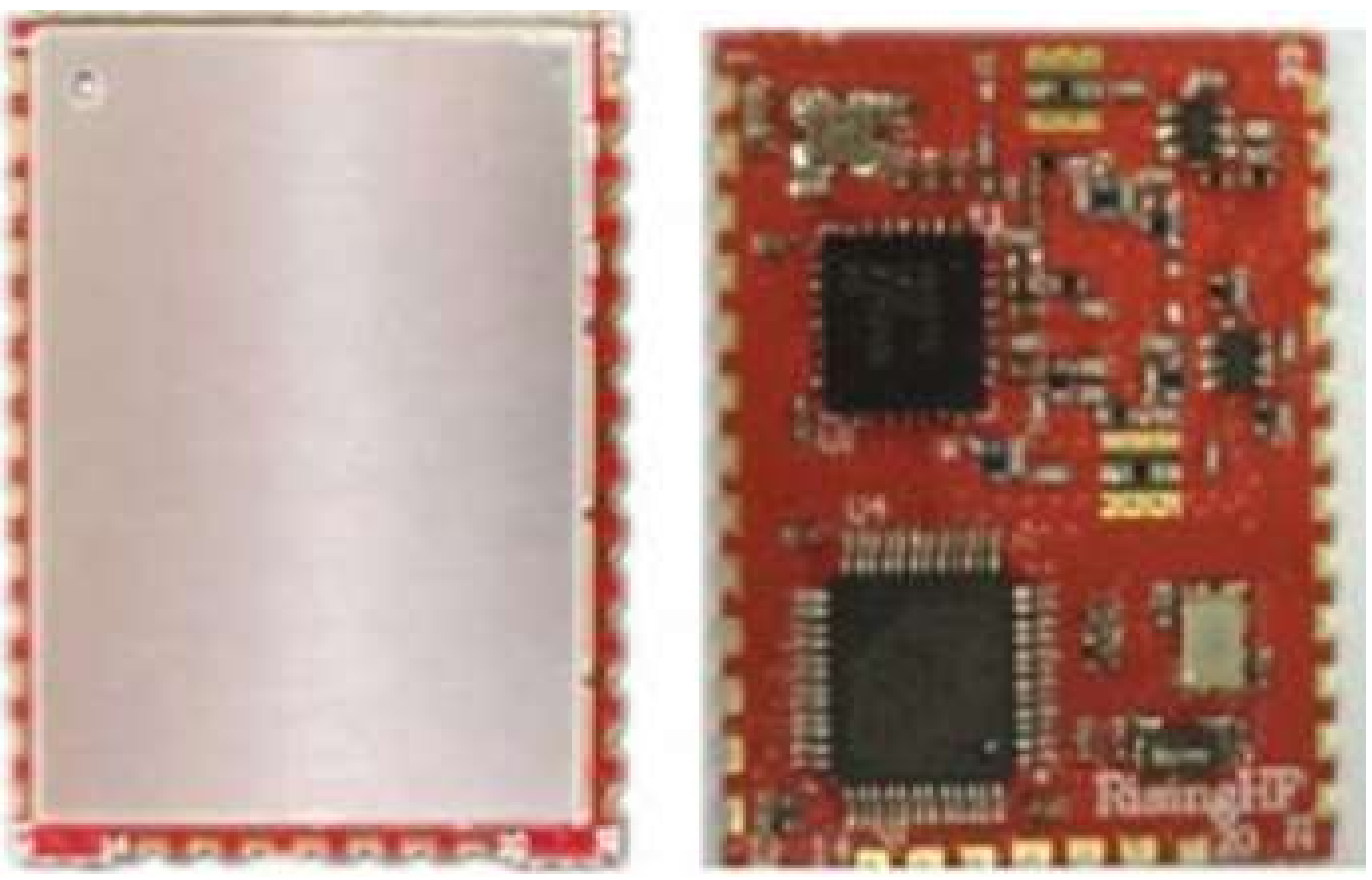

[2] RISINGHF. **How To Use LoRaWAN Modem**. 2015. Disponível em: <http://www.risinghf.com> . Acesso em: 30 novembro.2020.

**Fonte:** RisingHF, 2015

Pode-se usar um MCU (*Micro Control Unit*) *host*, que é um microcontrolador externo, para controlar este modem com comandos AT. A interface de comando simples e avançada oferece uma rápida colocação no mercado, por possibilitar toda a configuração por comunicação serial. O que torna uma plataforma de *hardware* compatível com a maioria dos microcontroladores, possuindo interface UART (*Univer- sal Asynchronous Receiver/Transmitter*) e, contribui também para a inicialização e atualização do firmware via cabo USB.

Tabela 1 – Especificações de Operação.

| **Item** | **Descrição** | **Min** | **Max** | **Unidade** |
|---|---|---|---|---|
| VCC | Alimentação | 1.8 | 3.6 | V |
| T | Temperatura | -40 | +85 | $^{O}C$ |
| P | Potência de entrada RF | - | +10 | dBm |

Figura 2 – Diagrama da pinagem do Módulo.

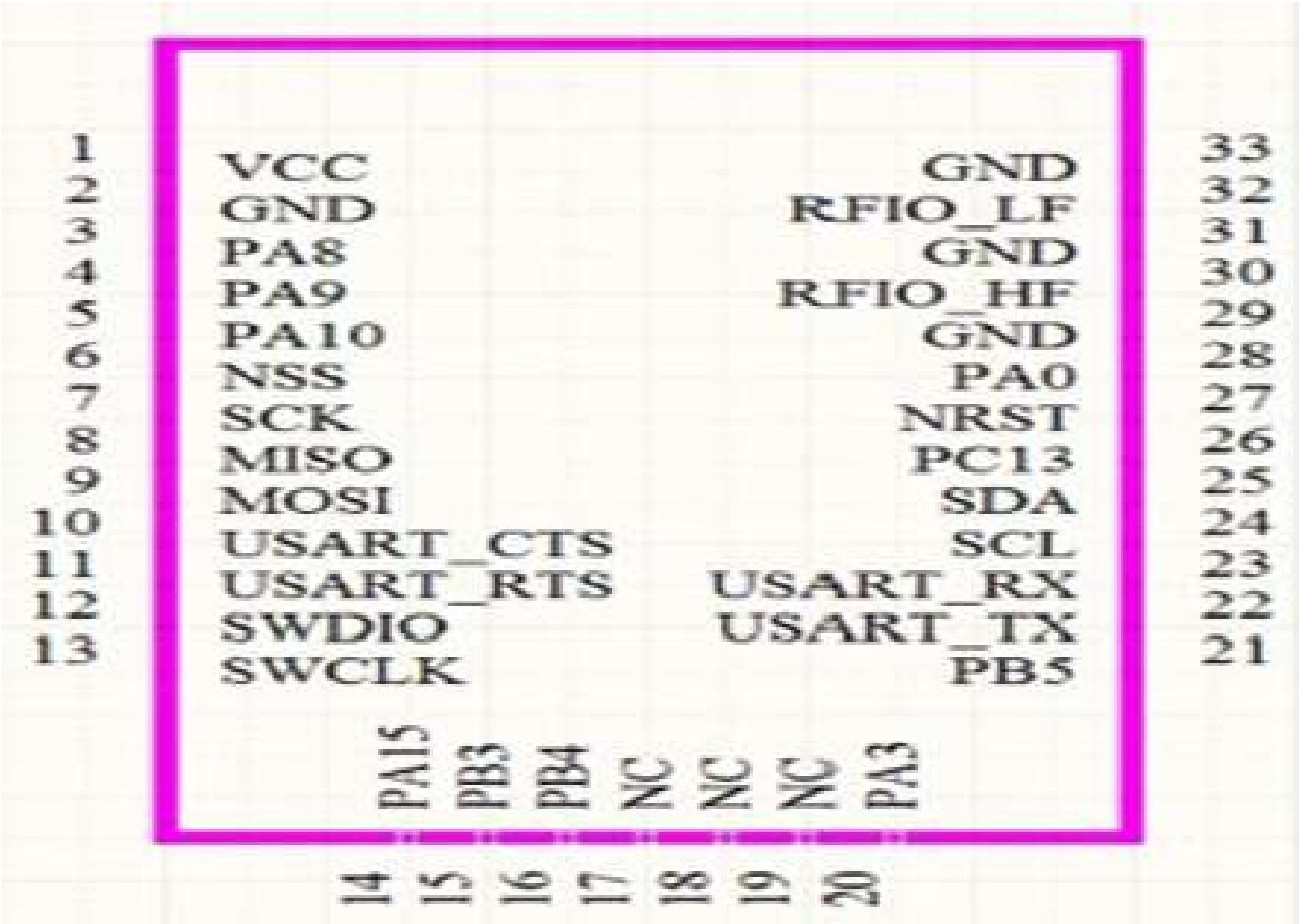

**Fonte:** RisingHF, 2015

É importante remarcar que o RHF76-052 incorpora o modem SX1276 e o MCU STM32L052xx, adequados para *End-point* na rede de IoT. Por sua vez, o módulo possui duas portas de RF (*Radio Frequency*), RFIO-LF e RFIO-HF. A primeira cobre a banda de baixa frequência, ou seja, 434MHz / 470MHz, enquanto a segunda estende a banda de alta frequência, ou seja, 868MHz / 915MHz. Baseado nas poderosas funções e desempenho do SX1276, o RHF76-052 poderia operar em ambas as modulações, FSK e LoRa. Ou seja, no modo LoRa, *bandwidth* (BW) com 62,5kHz, 125kHz, 250kHz e 500kHz podem ser usadas.

Tabela 2 – Descrição das GPIO’s.

| Nº Pino | Item | Descrição |
|---|---|---|
| 1 | VCC | Alimentação |
| 2 | GND | Ground (terra) |
| 3 | PA8 | GPIO-PA8 |
| 4 | PA9 | UART1-TX (Atualização de Firmware OTA) |
| 5 | PA10 | UART1-RX (Atualização de Firmware OTA) |
| 6 | NSS | NSS da interface SPI1 do MCU |
| 7 | SCK | SCK da interface SPI1 do MCU |
| 8 | MISO | MISO da interface SPI1 do MCU |
| 9 | MOSI | MOSI da interface SPI1 do MCU |
| 10 | USART1-CTS | USART1-CTS do MCU |
| 11 | USART1-RTS | USART1-RTS do MCU |
| 12 | SWDIO | SWDIO da interface SWIM para download do programa |
| 13 | SWCLK | SWCLK da interface SWIM para download do programa |
| 14 | PA15 | GPIO do MCU, PA15 |
| 15 | PB3 | GPIO do MCU, PB3 |
| 16 | PB4 | GPIO do MCU, PB4 |
| 17 | NC | GROUND (terra) |
| 18 | NC | GROUND (terra) |
| 19 | NC | GROUND (terra) |
| 20 | PA3/ADC3 | GPIO do MCU, PA3; ou entrada ADC3 |
| 21 | PB5 | GPIO do MCU, PB5 |
| 22 | USART1-TX | GPIO do MCU, PB6 |
| 23 | USART1-RX | GPIO do MCU, PB7 |
| 24 | I2C-SCL | SCL da interface I2C no MCU |
| 25 | I2C-SDA | SDA da interface I2C no MCU |
| 26 | PC13/Wkup2 | pino para acionar  MCU |
| 27 | NRST | Reset trigger entrada para MCU |
| 28 | PA0/AD0 | GPIO do MCU, PA0; ou entrada  ADC0 |
| 29 | GND | GROUND (terra) |
| 30 | RFIO-HF | RF entrada/saída em banda alta (868MHz/915MHz) |
| 31 | GND | GROUND (terra) |
| 32 | RFIO-LF | RF entrada/saída em banda baixa (434MHz/470MHz) |
| 33 | GND | Ground (terra) |

A esse respeito, a interface de dois fios SWIM – *System Wide Information Management)* é sugerida para ser usada para programação, passando a microunidade de controle interna do modem a ser o "microcontrolador escravo" (Slave) e o *Host* MCU é o "microcontrolador mestre" (Master).

Figura 3 – Arquitetura básica.

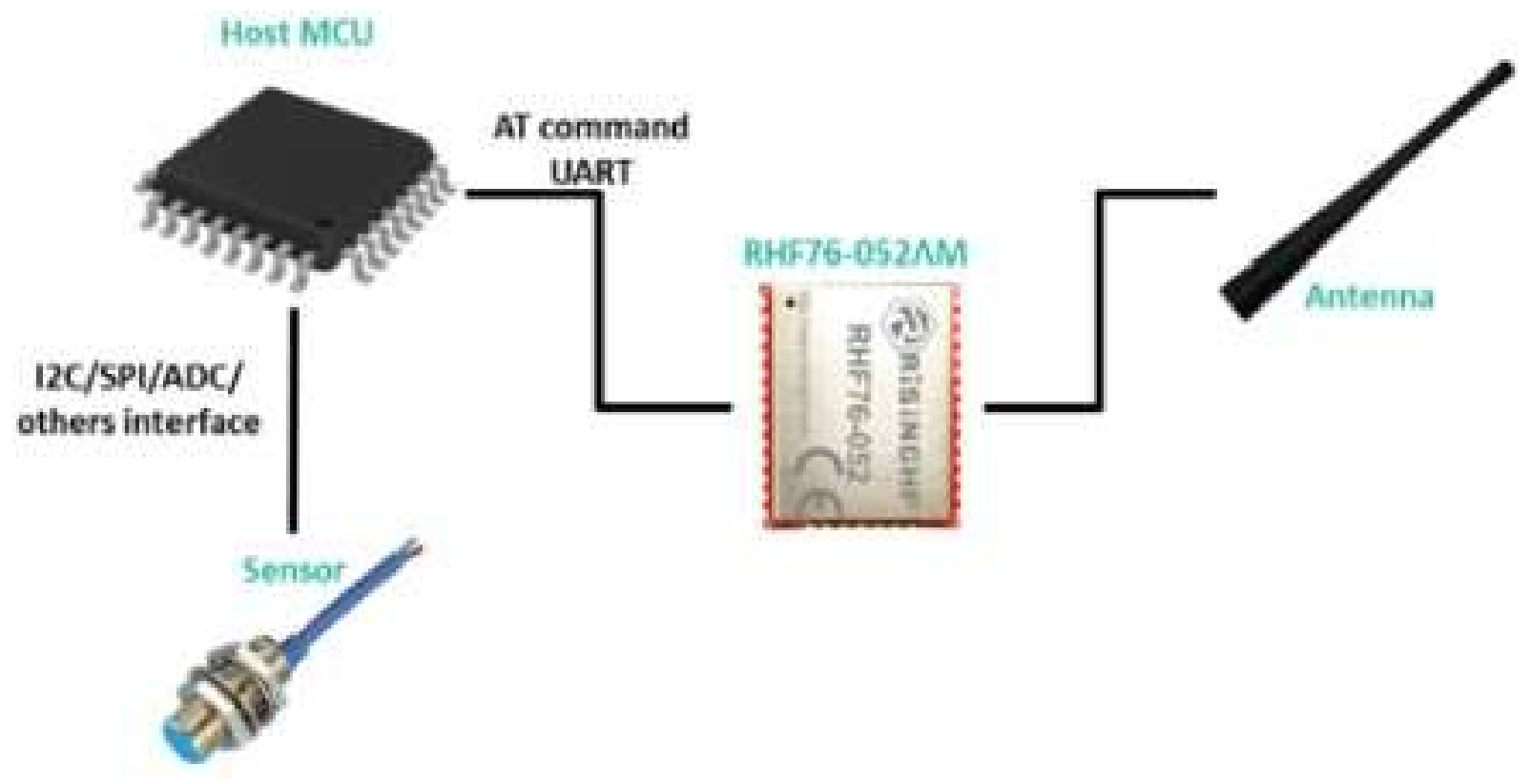

**Fonte:**RisingHF,2015

O módulo é integrado ao sensor por meio das interfaces, acessadas ao GPIOs (General Purpose Input/Output) do *host* microcontrolador. Com isso é possível configurar o módulo, usando os comandos AT.

## 2.4 Modulação da tecnologia LoRa

A tecnologia LoRa utiliza a modulação CSS (*Chirp Spread Spectrum*), desenvolvida, originalmente, para aplicações de radar e, utilizada em aplicações militares. Por isso, foram, inicialmente, chamados de sinais Chirp (*Compressed*

*High Intensity Radar Pulse*), mas depois se consolidou com a nomenclatura CSS para aplicações que não envolvem radar. Portanto, essa característica somada a técnica de espalhamento espectral (*Spread Spectrum*) deu origem à modulação LoRa.

Nessa linha de raciocínio, destaca-se que essa modulação foi aplicada no meio militar pelo fato de que um sinal SS (*Spread Spectrum*) poder ser facilmente ocultado, devido sua DEP (densidade espectral de potência) ser muito baixa. Essa característica faz com que o sinal seja capaz de ficar abaixo do ruído de fundo no equipamento.

Figura 4 – Exemplo Prático de uma aplicação de um sinal SS.

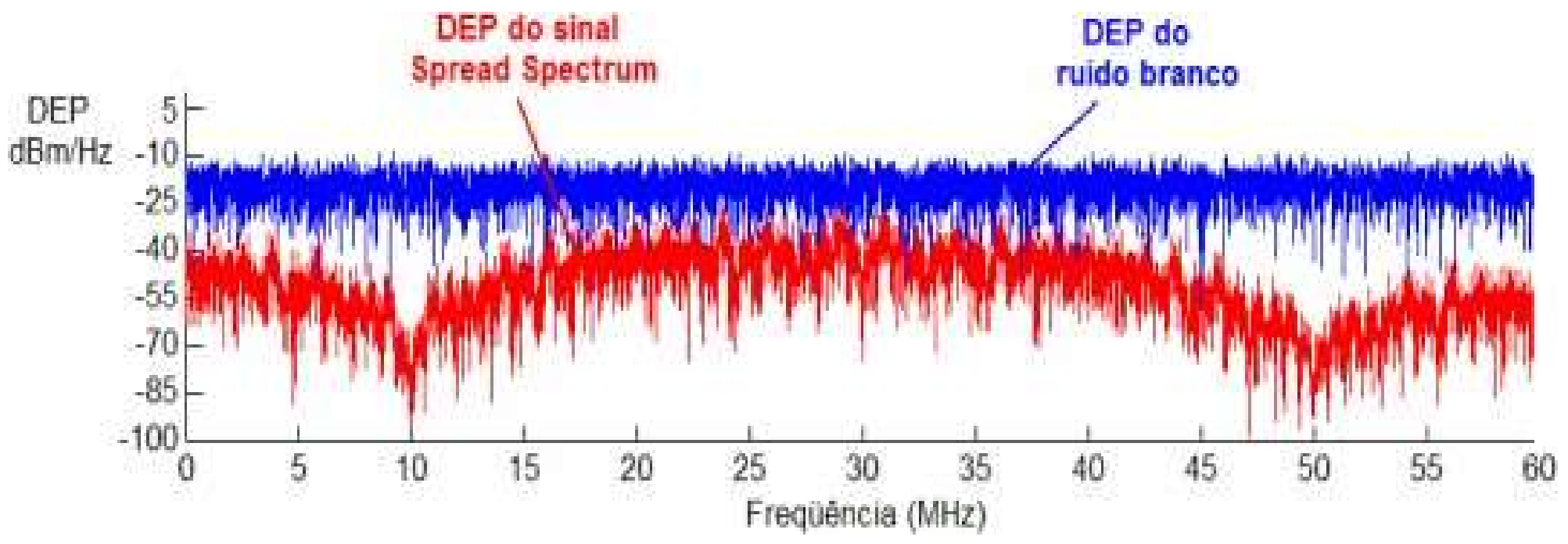

**Fonte:** Guimaraes, 2012.

Um sinal com espalhamento espectral é aquele que ocupa uma largura de faixa muito maior que a necessária. A largura de faixa ocupada é, até certo ponto, independente da taxa de informação. O que determina – se um sinal é ou não um sinal SS –, é a forma de geração do sinal modulado.

Considerando que há um sinal com uma taxa de bit ($T_b$) pequena, o mesmo sinal terá que passar por um conversor para polarizá-lo. Em seguida, Guimarães (2012) assevera que será multiplicado pela sequência de espalhamento a taxa de *bit* dessa sequência, chamada de taxa de chip ($T_c$), como ilustrado na figura 5. Portanto, diz-se que a duração de um *bit* do sinal da informação é equivalente a duração de um *chip* da sequência de espalhamento.

Figura 5 – Esquema da modulação CSS.

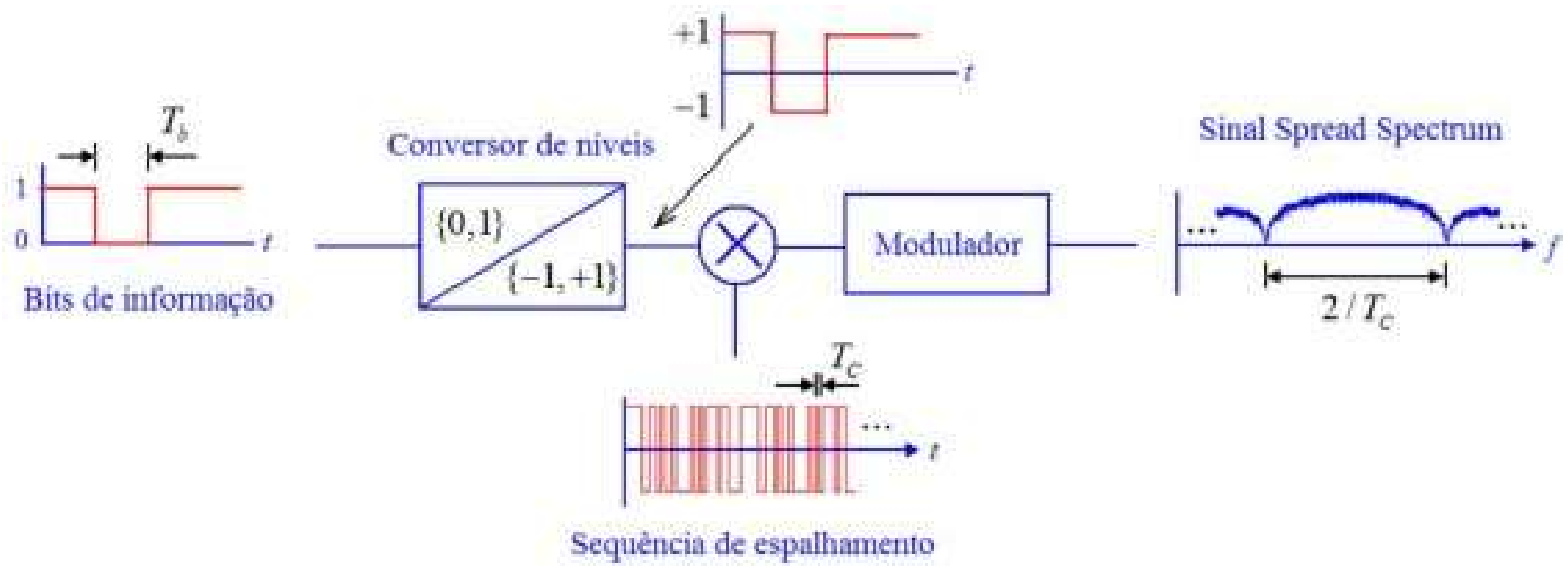

**Fonte:** Guimaraes, 2012

No diagrama mostrado, percebe-se que a banda ocupada pelo sinal de saída é função de $Tc$ e se trata de uma banda maior, nessária à transmissão dos dados. Pois, o período é menor no domínio da frequência, sendo possível ver a largura de banda maior. Portanto, segundo a definição dada, trata-se realmente de um sistema com espalhamento espectral.

Mesmo porque se alterar $T_b$, a largura de faixa não se altera, pois quem a governa é $T_c$, desde que $T_c > T_b$. É por esta razão que, na definição de um sinal *Spread Spectrum* a banda do sinal espalhado é independente da taxa de *bits* de informação até certo ponto. Isto porque se a taxa de *bits* que começa a ficar com valor comparável à taxa da sequência de espalhamento, influênciando na banda do sinal original. Assim, não se tem mais um sinal *SS.*

## 2.4.1 Principais atributos de um sinal *Spread Spectrum*

### 2.4.1.1 Baixa densidade espectral de potência

Nessa acepção, sabe-se que, quanto menor a densidade espectral de potência (DEP) de um sinal, menor será a concentração de potência por faixa de frequência. Por baixa DEP, entende-se uma distribuição de determinada potência em uma grande faixa de frequências. Toma-se como exemplo, na parte “a” da figura 6, um sinal de potência “P” que é transmitido numa pequena banda. Dessa forma, atribui-se este sinal a um transmissor de faixa estreita com alta DEP. Por outro lado, na parte “b” da figura, tem-se a potência “P” distribuída em uma faixa bastante elevada, característica típica de um sinal SS com baixa DEP.

Figura 6 – Esquema para comparar High DEP e Low DEP.

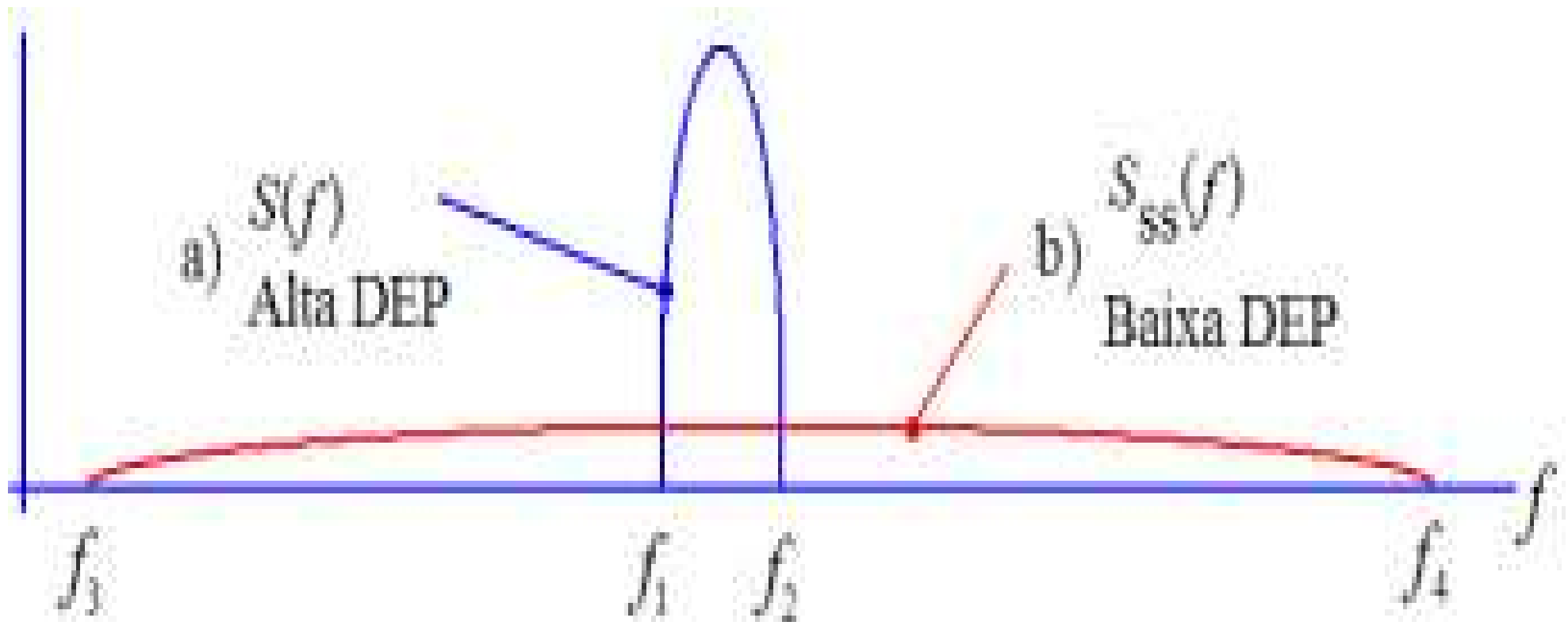

**Fonte:** Guimaraes, 2012

2.4.1.2 Baixa probabilidade de interceptação

A baixa probabilidade de interceptação (*Low Probability of Interception, LPI*) pode ocorrer devido a duas características de um sinal SS:

A baixa densidade espectral de potência pode tornar um sinal SS “invisível” a um receptor não intencional, tornando-se difícil a interceptação de um sinal não detectado.

Quanto maior o comprimento da sequência de espalhamento (quanto maior o número de *chips* em um período desta sequência), maior a dificuldade de geração de uma réplica pelo interceptador, o que também dificulta a interceptação.

## 2.4.2 Parâmetros da modulação LoRa

A relação entre a taxa de *bits* de dados ($T_b$) desejada para a modulação LoRa, é expressa por meio da equação (3.1):

$$T_b = SF \cdot \frac{\frac{1}{2^{SF}}}{BW} \quad (3.1)$$

Onde :

SF = Fator de Espalhamento

BW = Largura de Banda (Hz)

Período dos símbolos (Ps) pode ser definido pela expressão (3.2):

$$P_s + 2^{SF}/BW \text{ (seg)} \quad (3.2)$$

A partir da equação (3.2) é possível definir a taxa de símbolos ($T_s$) como:

$$T_s = BW/2^{SF} \text{ (símbolo/seg)} \quad (3.3)$$

A partir dessas equações define-se a taxa de *chips* ($T_c$) como:

$$T_c = T_s \, . \, 2^{SF} \text{ (Símbolo/seg)} \quad (3.4)$$

Ao substituir a equação (3.3) na (3.4), observa-se que em LoRa a taxa de *chip* depende apenas da largura de banda: a taxa de *chip* é igual à largura de banda (um *chip* por segundo por *Hertz* de largura de banda).

A modulação do LoRa inclui também um código correção de erros de tamanho variável, melhorando a robustez do sinal transmitido em detrimento da redundância e, fornecendo recursos de recuperação contra a corrupção de *bits*. Isto significa dizer que é implementado por meio de diferentes taxas de codificação (*Code Rate, CR*). Desta forma a taxa nominal de bits $T_b$ como:

$$Tb = SF . \frac{\frac{4}{4+CR}}{\frac{2^{SF}}{BW}} \quad (3.5)$$

Onde:

SF = Fator de Espalhamento

BW = Largura de Banda (Hz)

CR = Taxa de codificação

## 2.5 Classes do Protocolo de Comunicação

Toma-se as inferências de Oliveira (2017), em que afirma ser o protocolo de comunicação LoRaWAN, provê três classes diferentes para os *end-points*, usadas de acordo com a finalidade da aplicação. As características variantes entre as classes são a latência de transmissão e o consumo de energia, associado também à duração da bateria do dispositivo LoRa. As classes são A, B e C que são mais bem compreendidas quando comparadas, como é o caso da figura 7.

Figura 7 – Características das classes.

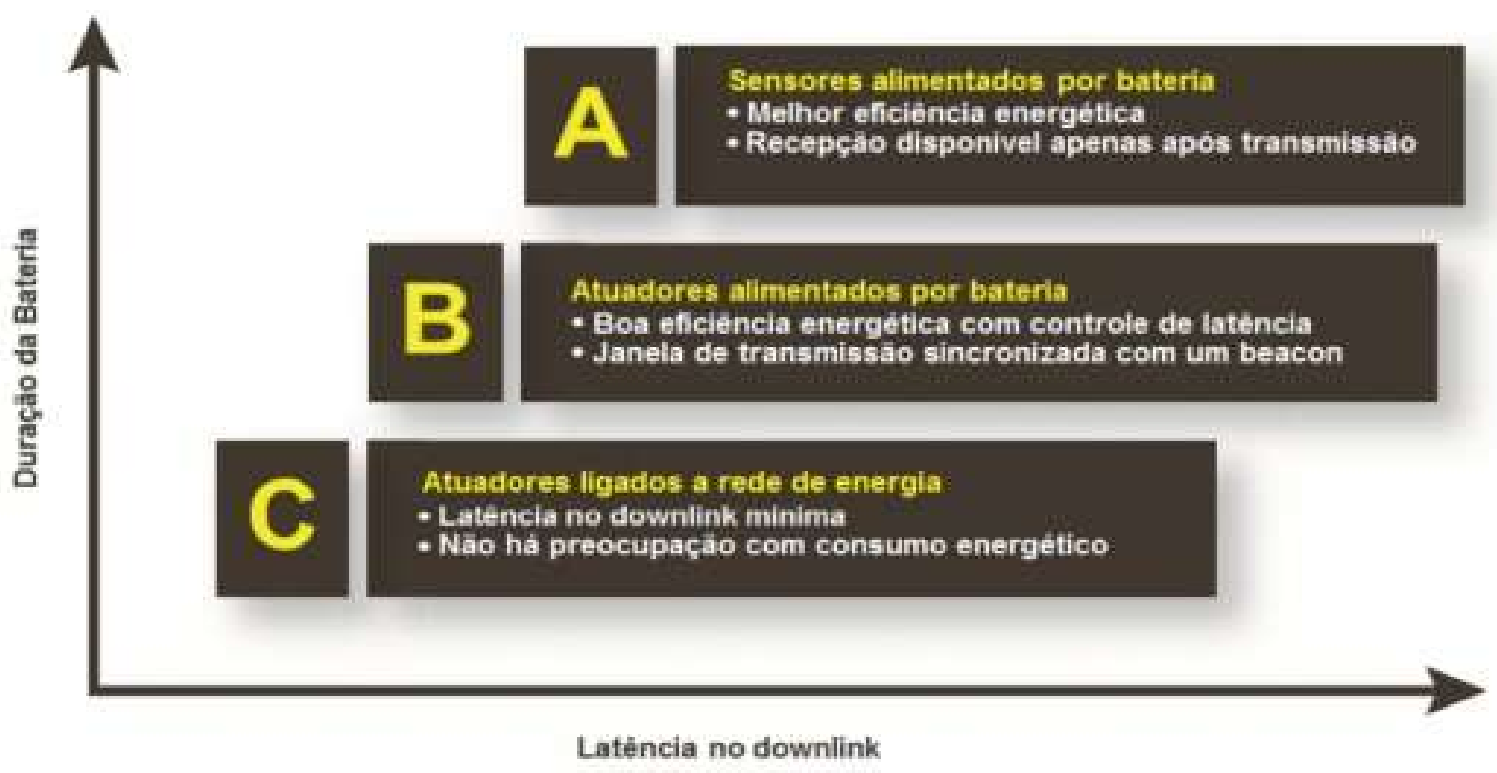

**Fonte:** Lora Allinance, 2020

Nesse ponto, a classe A permite que o dispositivo consiga enviar o pacote de dados para o *gateway* (*uplink*) e receber uma confirmação (*downlink*) dentro da mesma "janela de recebimento", o que caracteriza uma comunicação bidirecional. Graças a isso, esta é a classe que apresenta a melhor eficiência energética, em que o sinal de *downlink* é enviado apenas se houver o recebimento do sinal de *uplink*.

Por sua vez, a classe B possui o dispositivo de recebimento de confirmação de envio que depende da latência de transmissão (atraso para confirmar recebimento, após o envio) do sinal de *uplink.* Pois, nessa classe, são abertas mais janelas de recebimento, independente do sinal de *uplink*, o que gera um aumento no nível de atividade do dispositivo.

Ao que concerne à classe C, a maior diferença está no fato de que o dispositivo passa a trabalhar praticamente de forma contínua com a janela de recepção, exceto quando está sendo feita a transmissão, visto que, neste caso, a janela é aberta com precisão. Dessa maneira, a Lora Alliance[3] mostra que a latência é mínima pois há mais chances de captura do sinal em um tempo mais curto, porém a eficiência energética é reduzida drasticamente.

## 2.6 Tipos de incêndios

Para sensoriamento de incêndios florestais, é importante saber sobre os tipos de incêncidios existentes e suas principais características. Basicamente existem três tipos que podem também se categorizar por classe. De acordo com as instruções técnicas do Corpo de Bombeiro Militar (IT-CBM), ele são classificados como classe A, porque a queima acontece por causa de algum tipo de combustível sólido. Em outras palavras, é o caso de resíduos em decomposição ou galhos e folhas secas no solo, entre outros materiais mais reagentes ao fogo, em que:

1) O primeiro é chamado de Fogo de Solo ou Subsolo – que se propaga pela matéria orgânica em decomposição e pelas raízes. Quase sempre se queimam lentamente e em combustão incandescente (pouca chama ou ausência dela) por não dispor de muito oxigênio. Pode também se enquadrar na classe A;

2) O segundo é chamado de Fogo de Superfície – que se propaga pelo combustível encontrado no solo, incluindo o lixo, as ervas, os arbustos, os galhos secos e as folhas, mas não imersos no solo. Estes são ainda considerados de classe A, por seus ativos serem combustíveis sólidos.

3) O terceiro é chamado de Fogo de Copas – que é a fase mais avançada do incêndio, iniciado com o fogo de superfície, chegando às copas das árvores e espalhando-se mais rapidamente devido ao vento. Atualmente, esse é o tipo de incêndio que é possível detectar com imagem de satélite, contido também na categoria de incêndio classe A.

Para fazer o sensoriamento de indícios de incêndio, optou-se que o projeto atenda a cenários mais condizentes com o segundo tipo – Fogo de Superfície –, que está mais presente a identificação da presença da fumaça, da luz das chamas e do calor, mas ainda em uma dimensão menor. Com isso foram introduzidos alguns sensores para detectar tais grandezas.

[3] LORA Alliance Technical Marketing Workgroup. **LoRaWAN. What is it?** A technical overview of LoRa and LoRaWAN. San Ramon, CA, 2015. Disponível em: <https://www.loraalliance.org/portals/0/documents/whitepapers/LoRaWAN101.pdf>. Acesso em: 15 abril. 2019.

### 2.6.1 Sensor de Temperatura

O sensor DHT22[4] detecta valores analógicos de temperatura e umidade. Nesse sentido, cabe lembrar que, na biblioteca "DHT.h", é possível fazer a solicitação da leitura do valor referente a cada uma dessas grandezas separadamente. No caso, para auxiliar na detecção de chamas, a grandeza mais relevante é a temperatura, ou seja, a energia térmica que é dissipada quando há uma fonte de calor. Este sensor opera na detectação de valores analógicos de temperatura dentro da faixa de -40 a 80 graus *Celsius* com margem de erro de 0,5 graus, medindo a umidade do ar nas faixas de 0 a 100 porcento, com margem de erro de 2 porcento.

Figura 8 – Sensor de temperatura.

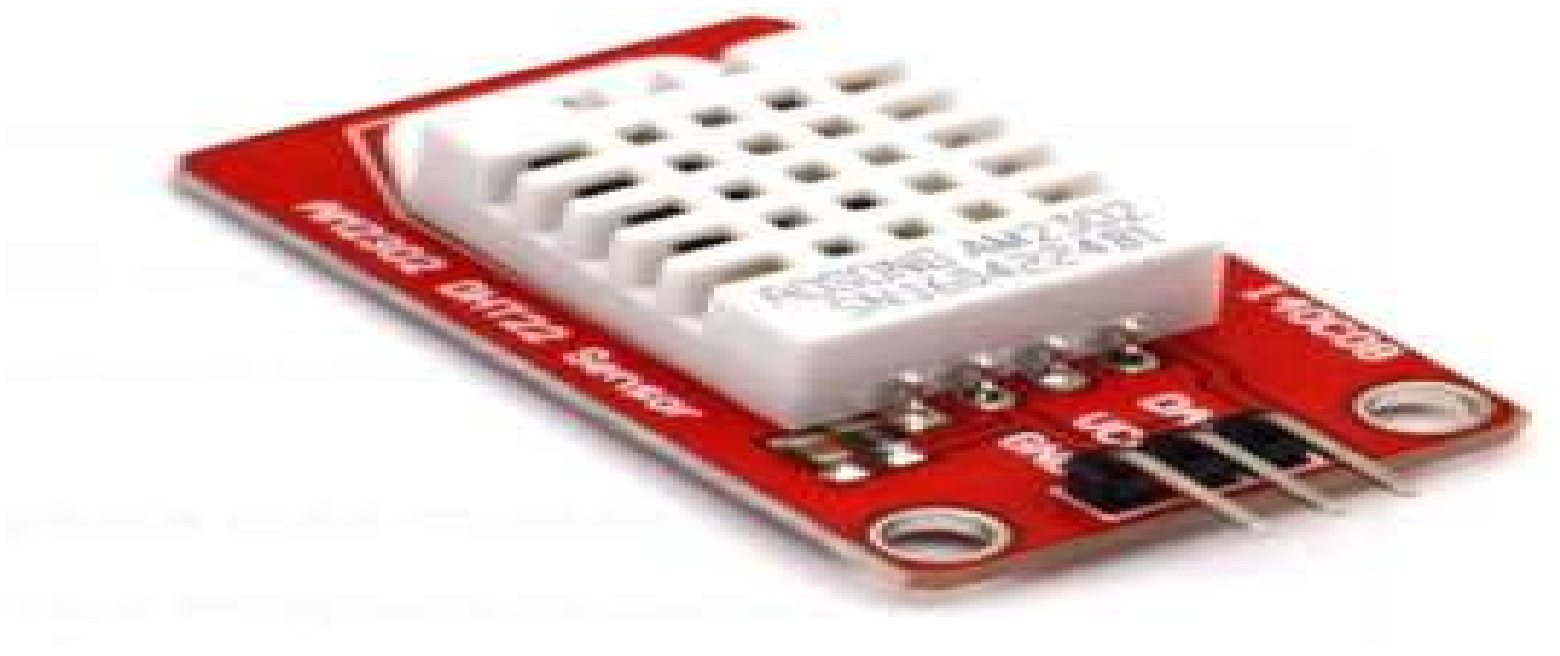

**Fonte:** Filipeflop,2019.

Tabela 3 – Especificações técnicas do sensor de temperatura.

| Características | Valores |
|---|---|
| Faixa de medição(temperatura) | $-40^{o}$ a $80^{o}$ celsius |
| Precisão (temperatura) | $0{,}5^{o}$ celsius |
| Tempo de resposta | 2 seg |

[4] FILIPEFLOP. Sensor de Umidade e Temperatura DHT22. Disponível em: <https://www.filipeflop.com/produto/sensor-de-umidade-e-temperatura-am2302-dht22/#tab-blogrelacionados> Acesso em: 19 março. 2019

| | |
|---|---|
| Tensão de funcionamento | 3 a 5V (DC) |
| Consumo de corrente | 2,5mA a 5V |

### 2.6.2 Sensor de Chamas

Este sensor é bastante usado na detectação de fontes de chamas que dissipam energia em forma de calor e luz. No entanto, a forma de detectar esse sensor é por meio do comprimento de onda da luz, pelos espectros entre 760 e 1100 nm (nanometros) da fonte luminosa de chama. Seu ângulo de detecção é de 60 graus, possuindo um potenciômetro de ajuste de sensibilidade.

Por essa via de raciocínio, esse sensor possui saídas analógica e digital, sendo a saída digital, em estado baixo (saída digital 0), quando há presença de alguma fonte de chama e; em estado alto (saída digital 1), quando não há.

À vista disso, é assumido que o sensor funciona com uma "lógica invertida", refletindo na sua saída analógica também. Se porventura algo o danifique em campo, ele irá detectar valores altos, mesmo porque a resistência dentro dele passa a tender ao infinito, acarretando em uma leitura muito alta.

Por isso, é indicado que este sensor seja mantido com "lógica invertida", com vista ao que corresponderá a detecção deste evento, sendo de valor menor, o que impede de haver a possibilidade de o sensor fazer uma leitura errada.[5]

Figura 9 – Sensor de chamas.

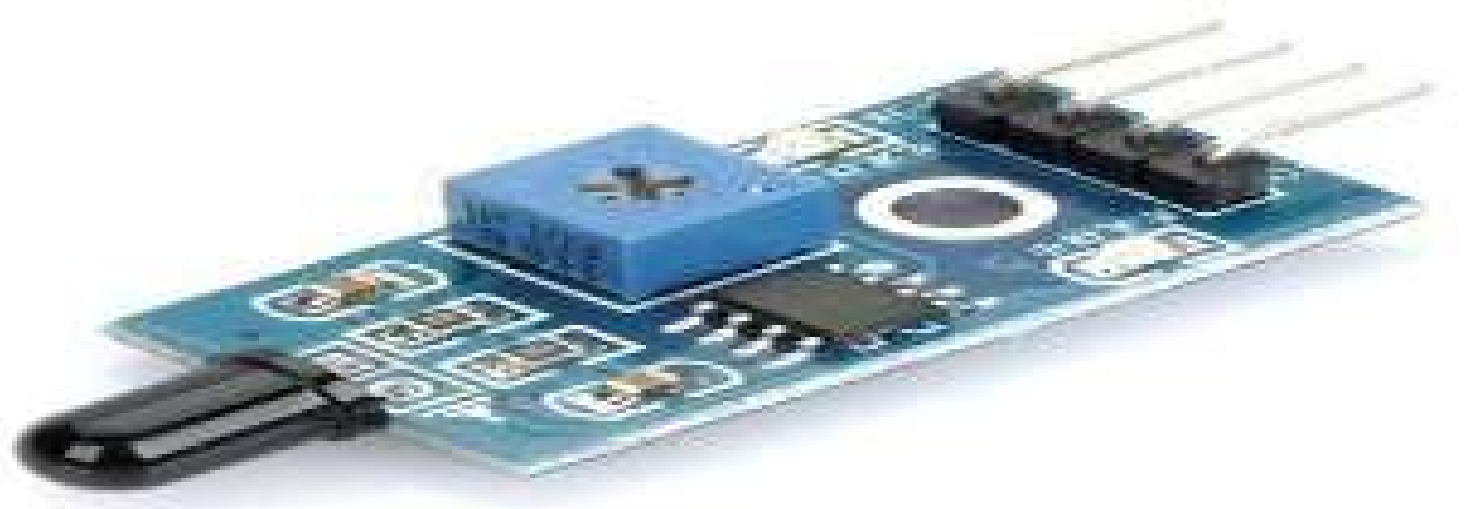

**Fonte:** Filipeflop, 2019.

Tabela 4 – Especificações técnicas do sensor de chamas.

| **Características** | **Valores** |
|---|---|

[5] Idem.

| | |
|---|---|
| Dimensões do sensor | 3,2 x 1,4 cm |
| Tensão de funcionamento | 3,3 a 5V (DC) |
| Consumo de corrente | 15mA a 5V |

### 2.6.3 Sensor de Gás

O sensor MQ-135 é bastante usado em projetos de detecção de gases, pois contempla os principais gases inflamáveis e/ou tóxicos, como amônia, dióxido de carbono, benzeno, óxido nítrico e ainda, fumaça ou álcool. No caso se trata da fase inicial de um incêndio, em que o gás principal é, nesse momento, o dióxido de carbono ou a fumaça que sobe com o efeito da combustão.

Em face no que já foi colocado, Fearnside (1997) reforça que isso pode carregar os chamados gases traços, como metano (CH4), monóxido de carbono (CO) e nitroso de oxigênio (N2O), dependendo do tipo de vegetação. Esse sensor é capaz de detectar concentrações na faixa de 10 a 1000ppm (partículas por milhão) e possui um potenciômetro para regular a sua sensibilidade, que fica integrado a ele no módulo, como mostra a figura 10.

### 2.6.4 Princípio de funcionamento do Sensor

Em tal perspectiva, Candido (2019) demonstra que, na sua grande maioria, os sensores de gás da família MQ são sensores catalíticos. Eles funcionam por intermédio do fenômeno de oxidação catalítica que ocorre em uma bobina de fio de platina tratada, a qual fica encapsulada dentro do sensor.

Tal perspectiva, sofre oxidação, quando o gás invade os espaços da superfície do invólucro, alterando a sua resistência à medida que a proporção do gás aumenta, devido a dissipação de energia em calor que contribui para a degradação da sua resistividade. Nesse sentido, percebe-se na figura 11 que é possível visualizar melhor a estrutura do sensor

Figura 10 – Sensor de gás.

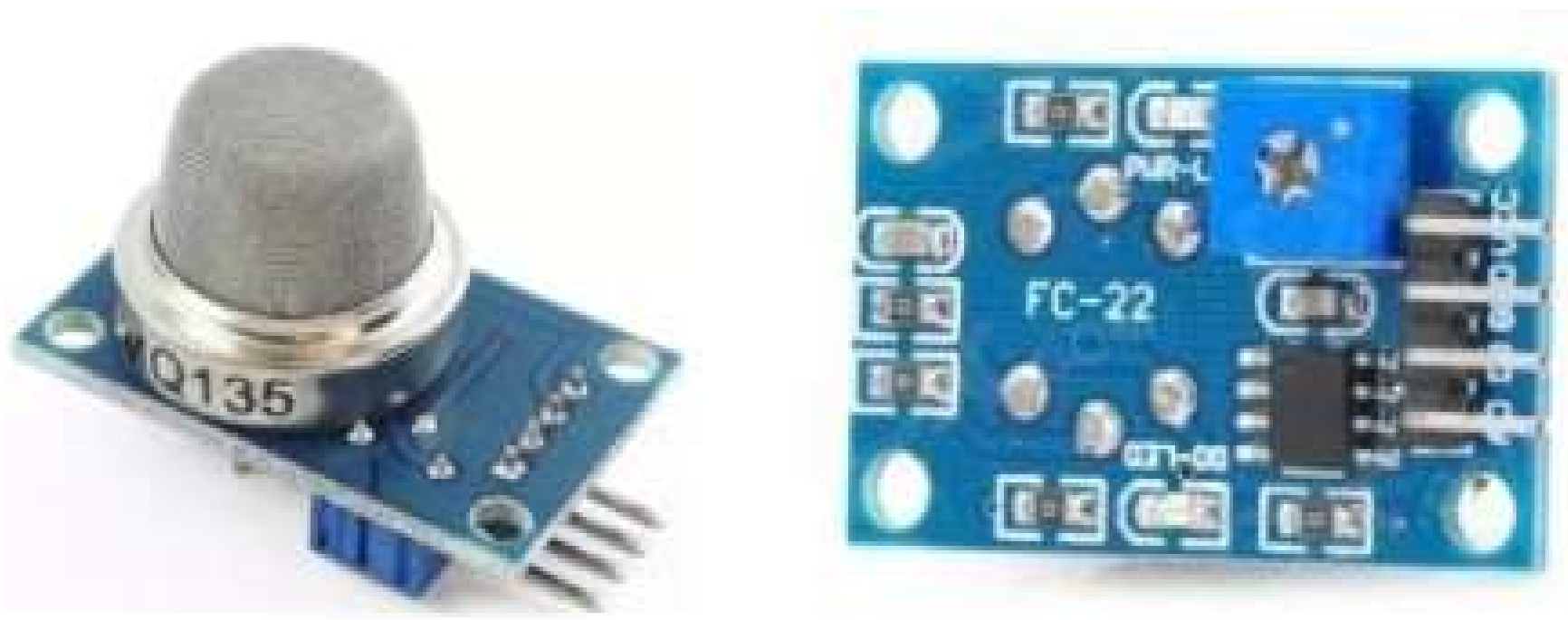

**Fonte:** Filipeflop, 2019.

Tabela 5 – Especificações técnicas do sensor de gás.

| Características | Valores |
|---|---|
| Dimensões do modulo sensor | 32mmm x 22mm x 27mm |
| Tensão de funcionamento | 0 a 5V (DC) |
| Temperatura de operação | -10 a 70 °C |
| Consumo de corrente | 150mA a 5V |

Figura 11 – Esquema por dentro do sensor.

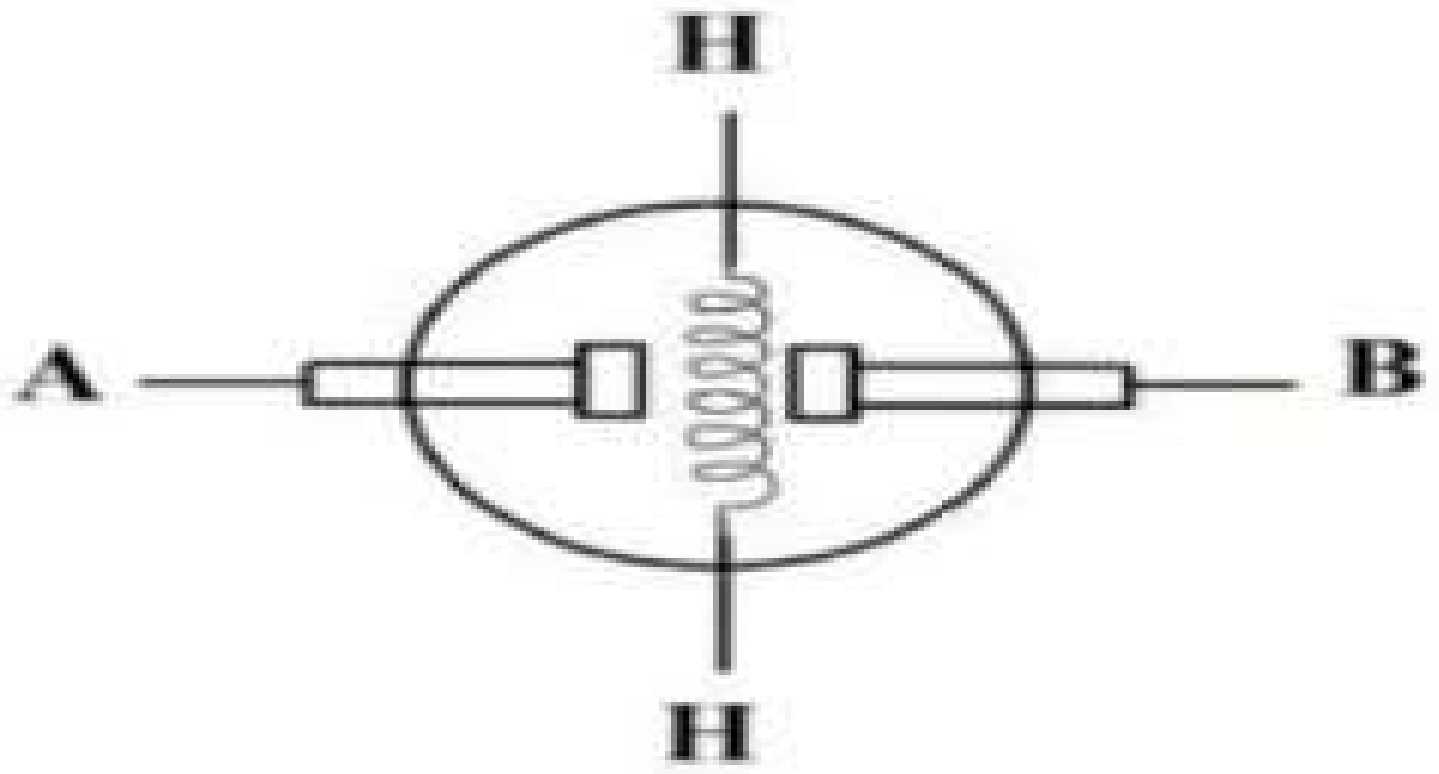

**Fonte:** Filipeflop, 2019.

Na vertical, tem-se, entre as duas extremidades, a bobina que gera a resistência entre os contatos A e B quando energizada, por isso a alimentação de 5V fica conectada em um H, e no outro o GND (0V), corforme Tabela 5. Quanto maior for a concentração de gás, menor é a resistência entre os contatos A e B, portanto a tensão no ponto B, passa a tender a 5V. Com isso é possível esquematizar um circuito que deixe essa saída analógica mais evidente, como na figura a seguir.

Figura 12 – Esquema interno de um sensor da família MQ.

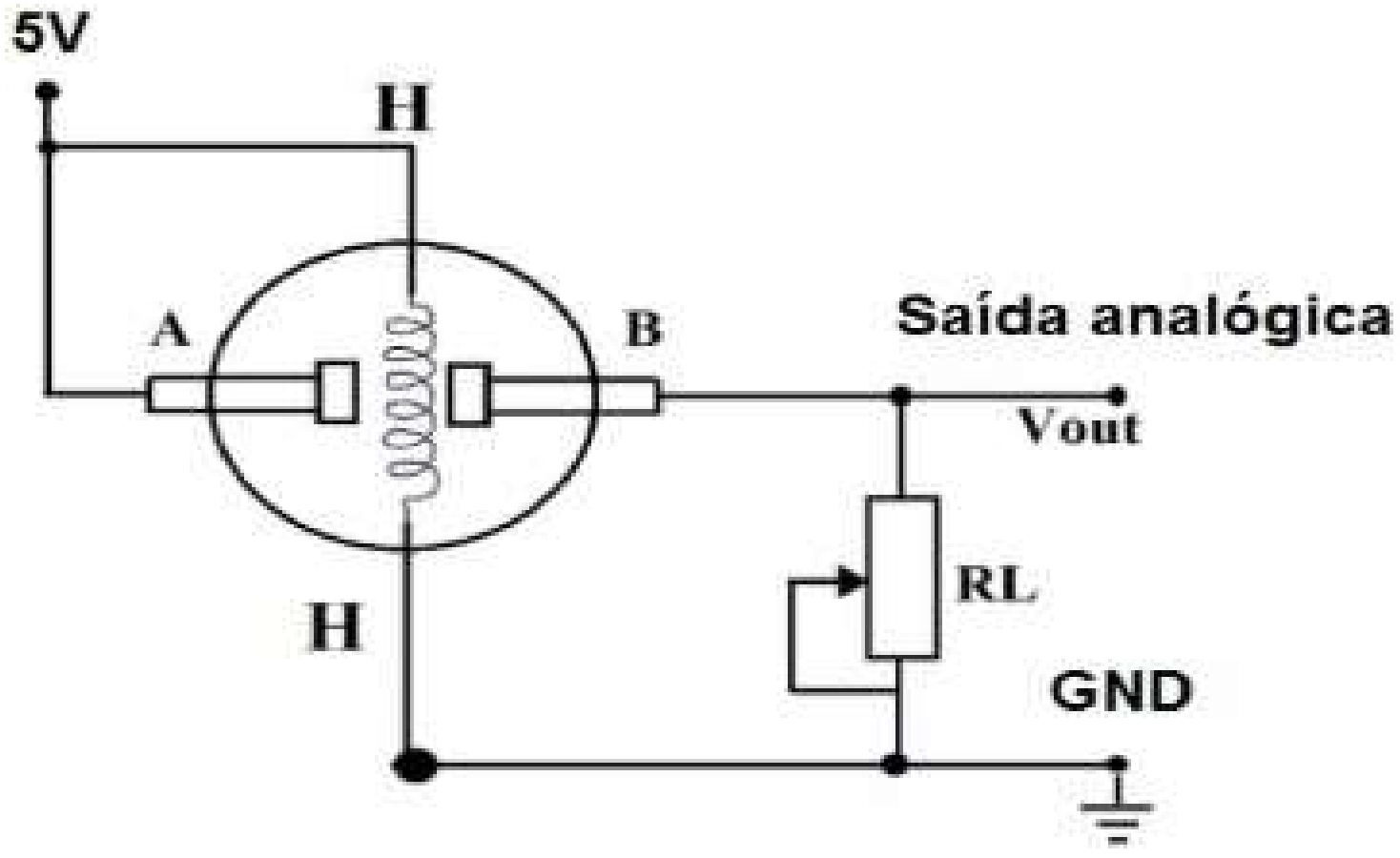

**Fonte:** Filipeflop, 2019.

Associa-se o aumento da tensão em cima do resistor RL como sendo equivalente ao aumento da concentração de gás no ambiente. Na figura a seguir, é possível observar a estrutura física do circuito explicado.

Figura 13 – Interior do encapsulamento do sensor de gás.

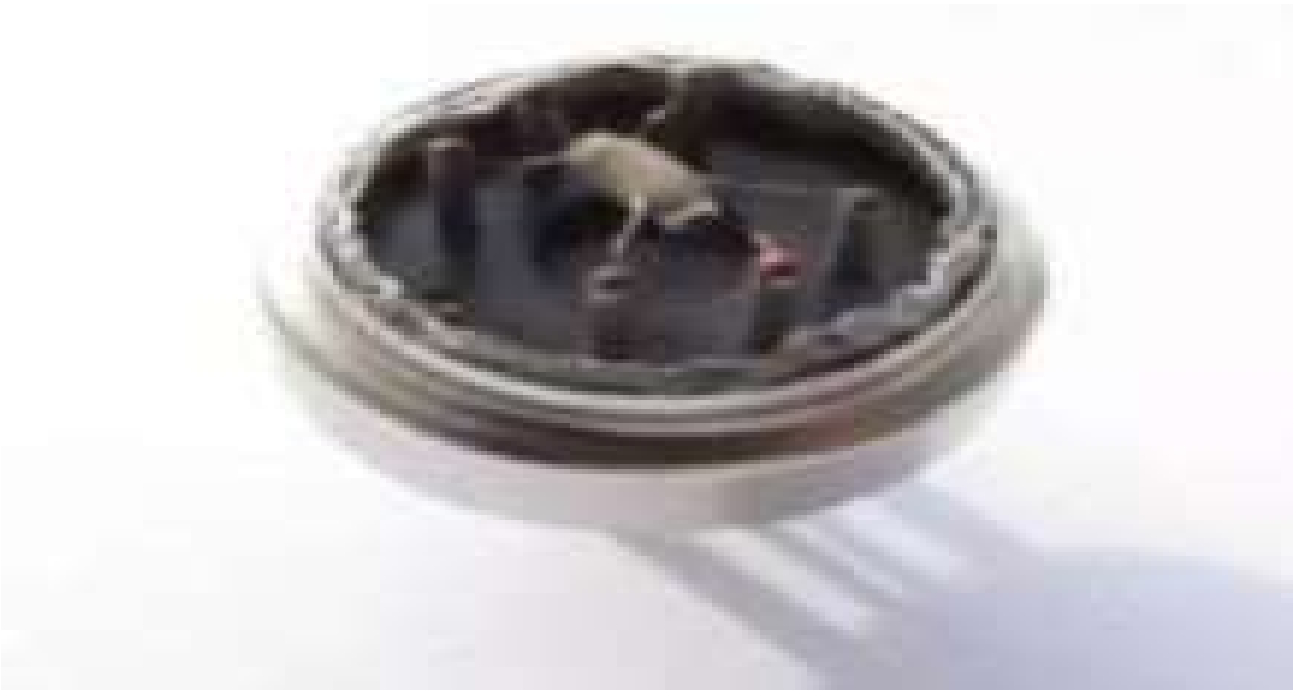

**Fonte:** Filipeflop, 2019.

## 2.7 Arduino Mega 2560

Foi usado o Arduino Mega, por ter maior capacidade de processamento e maior quantidade de memória que o Arduino Uno. Ele também possui grande acessibilidade para estudantes e projetistas amadores, com o objetivo de realizar interações dentro de um ambiente físico por meio da integração de *hardware* e *software*, uma vez que é disponibilizada uma plataforma de desenvolvimento gratuita e compatível com os principais sistemas operacionais como Windows e Linux.

Com base nessa definição, Souza (2014) menciona que a placa é composta basicamente por um microcontrolador ATmega2560, circuitos de entradas e saídas digitais totalizando 54 pinos, dentre eles, 15 como saídas PWM e 16 entradas analógicas, com 4 pares de portas Tx e Rx para comunicação serial. Possui fácil conexão a um computador sem a necessidade de equipamentos ou acessórios extras além de um cabo USB e do IDE (Integrated Deve- lopment Environment) que utiliza uma linguagem de programação baseada em C/C++, disponibilizando uma vasta gama de bibliotecas e funções úteis para aplicações diversas.

Figura 14 – Placa do Arduino Mega.

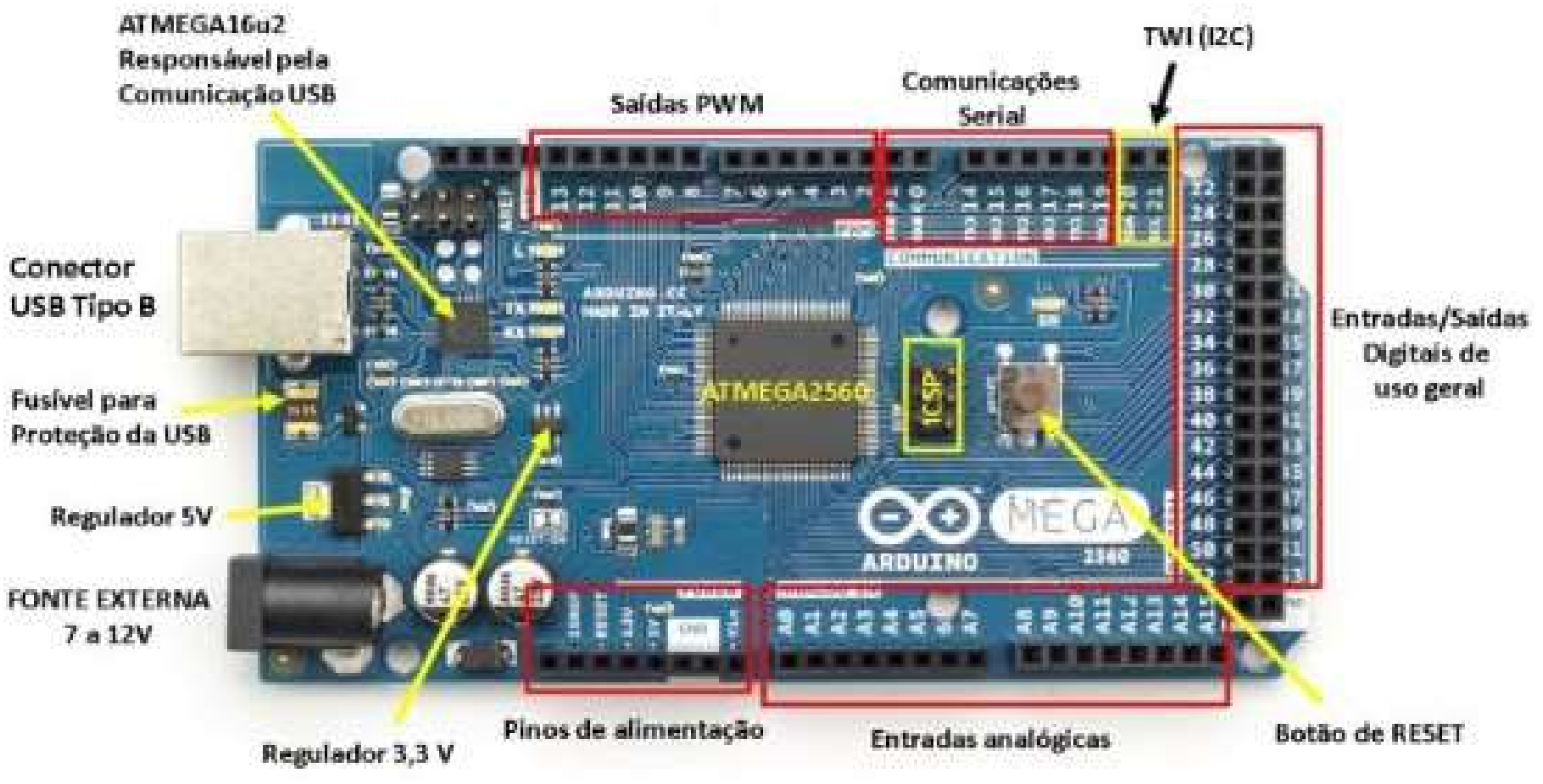

**Fonte:** Filipeflop, 2019.

**Conector USB –** esta porta é utilizada para conectar ao computador, utilizando o cabo USB que acompanha o Arduino (idêntico aos de impressora, USB-A de um lado e USB-B do outro). Funciona para transmissão de dados (embarcar os programas do computador para o chip assim como alimentação DC).

**Fonte Externa** – porta para alimentação DC do circuito. Pode ser utilizada uma fonte ou uma bateria para ligar o circuito, a tensão é 9V. O *plug* que conecta é de 2,1mm de diâmetro, em que a parte de fora do conector é o negativo e o positivo no meio.

Tabela 6 – Especificações Técnicas do Arduino Mega 2560.

| Características | Arduino Mega |
|---|---|
| Microcontrolador | ATmega2560 |
| Tensão Operacional | 5 V |
| Tensão de Alimentação (recomendado) | 7 a 12 V |
| Tensão de Alimentação (limites) | 6 a 20 V |
| Pinos I/O digitais | 54 |
| Pinos de entrada analógica | 16 |
| Corrente contínua por pino I/O | 40mA |
| Corrente contínua para o pino 3.3 V | 50mA |
| Memória flash | 256KB |
| SRAM | 8KB |
| EEPROM | 4KB |
| Frequência de clock | 16MHz |
| Porta USB | Não |

**Pinos de Alimentação –** alimentação de 3,3V e de 5V, sendo dois pinos com 5V. Podem ser uteis para acionar entradas digitais, acender LEDs, ligar sensores, atuadores, etc.

**Pinos de entrada analógica (A0-A15) –** pinos de entrada analógica (pinos de A0 a A15). São 16 pinos muito uteis para ligação com dispositivos analógicos como sen- sores, por exemplo. Resumindo, fazem leituras de tensão de um sensor e os converte em número entre 0 e 1023.

**Pinos digitais IN/OUT –** as entradas e saídas digitais são os pinos 22 a 53. Os outros pinos são destinados para comunicação serial e saída PWM.

## 2.8 The Things Network

A *The Things Network* (TTN) é uma rede baseada em *cloud servers*, ou seja, é um servidor hospedado em nuvem que visa permitir que os dispositivos de baixo consumo de energia usem *gateways* de longo alcance para conectar-se a uma rede descentralizada de código aberto para trocar dados com as aplicações. O TTN possibilita trabalhar com dados, registrar as aplicações, criando os endereços dos dispositivos e *gateways* para poder fazer o gerenciamento das aplicações.

Figura 15 – Dispositivos para registrar no TTN.

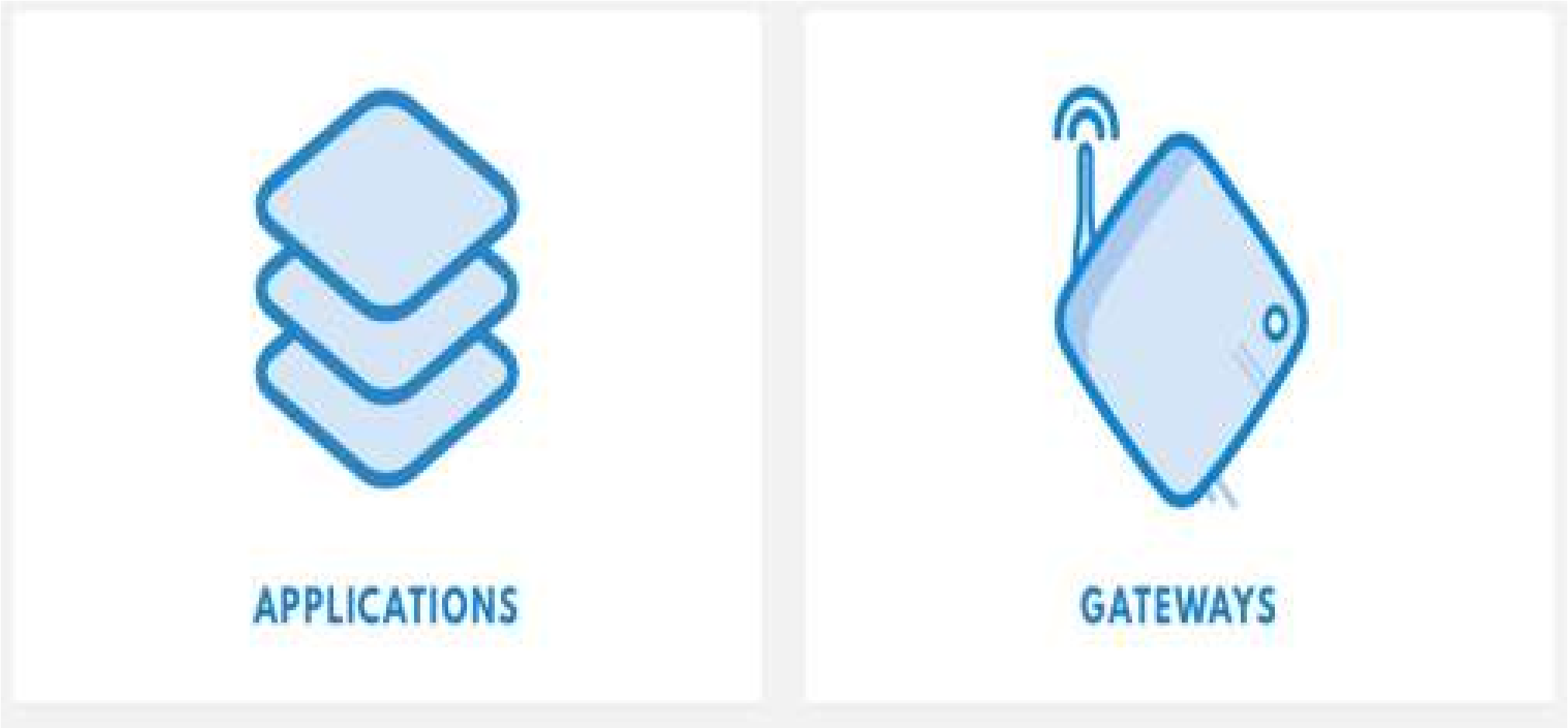

**Fonte:** Lucas Gonçalves,2020

Na aplicação, os *End-points* registrados são chamados de "*Devices*" que irão se comunicar com os *gateways*.

Figura 16 – Aplicações Registradas.

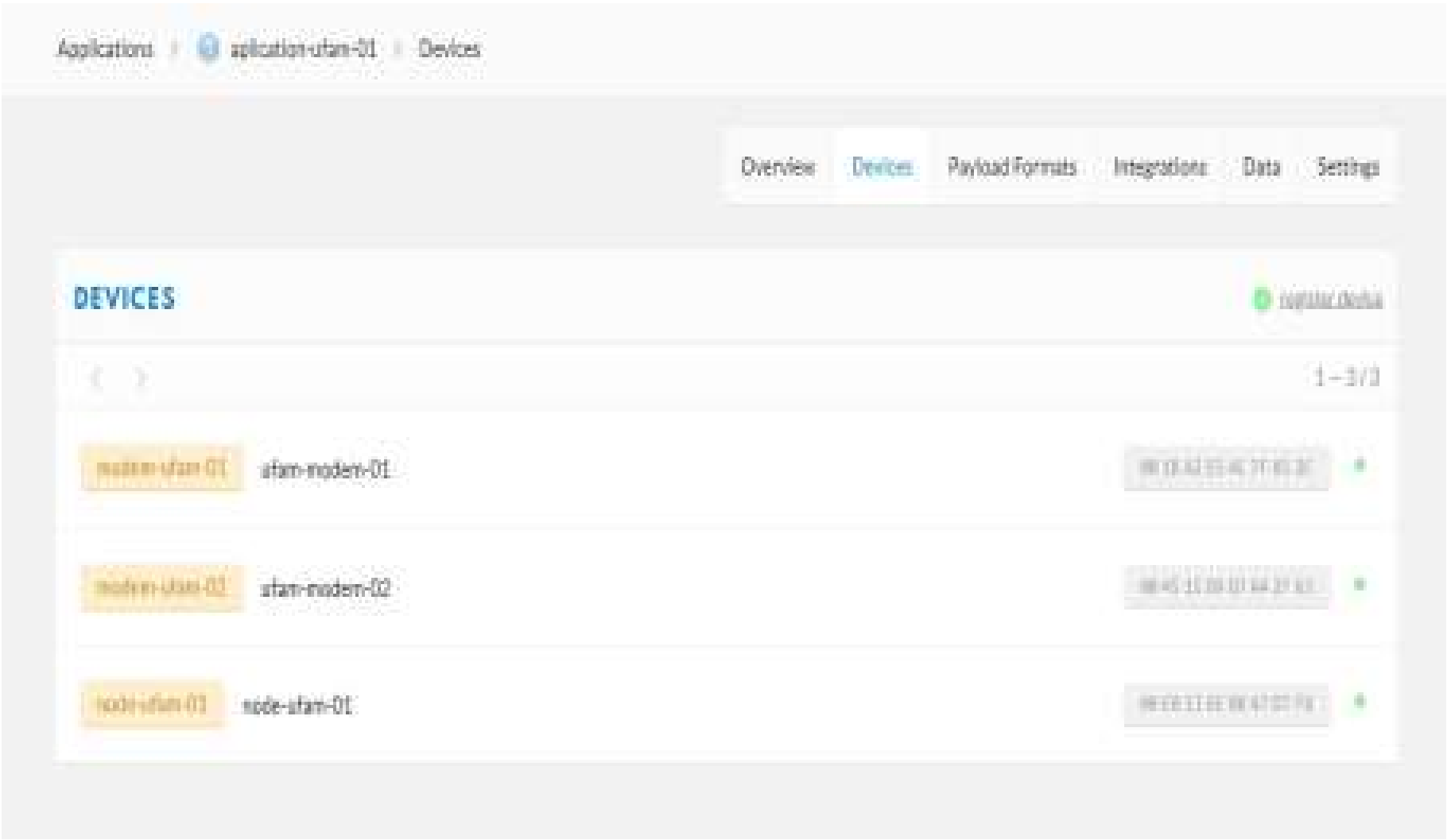

**Fonte:** Lucas Gonçalves,2020

A própria interface do servidor TTN possibilita fazer a decodificação dos dados, chamados de *payload*, que na computação, refere-se à carga de uma transmissão de dados. É a parte principal dos dados transmitidos, da qual se omitem algumas informações utilizadas para facilitar a entrega, como cabeçalhos e metadados. Ou seja, é apenas a carga útil, que carrega os valores detectados pelo *End-point.*[6]

## 2.9 Aplicação *Web* para monitoramento visual

Foi desenvolvido um servidor local do tipo Node.JS que se trata de um interpretador de código JavaScript, funcionando do lado do servidor. Isso significa que, com o Node.JS, é possível criar aplicações *Javascript* para rodar como uma aplicação independente (*Standalone Application*) em uma máquina. Com isso é possível importar dados da internet o que é bastante útil para desenvolvimento *web.*

### 2.9.1 Javascript para Backend e Frontend

O *Backend* é a parte do *software* responsável por todas as funcionalidades da aplicação. Ou seja, tudo que precisa estar integrado para gerar um sistema funcional, como manipulações de dados, cálculos, processamento, banco de dados e etc. Já o *Frontend* está associado com a interação do usuário, ou seja, a forma de aplicação será acessada, exibindo informações processadas pelo *Backend*, seja mostrando gráficos, tabelas, animações ou outros tipos de elementos que compoem uma interface de usuário, dependendo da finalidade do sistema.

[6] THE THINGS NETWORK. Device Registration. Disponível em: https://www.thethingsnetwork.org/docs/devices/registration.html. Acesso em: 20 outubro. 2020.

### 2.9.2 Socket.io para comunicação em tempo real

Como mencionado anteriormente, as informações do *Backend* devem ser exibidas no *Frontend*. Para que isso seja possível para uma aplicação de monitoramento, é im- portante que essas informações sejam as mais atuais e precisas. Por isso, foi utilizado o recurso fornecido pela biblioteca de desenvolvimento de aplicações em tempo real, Socket.io.

Essa biblioteca utiliza comunicação bidirecional, o que significa dizer que quando há alguma atualização dos dados no *Backend*, no mesmo instante essa mesma atualização é requisitada no *Frontend* e renderizada na interface.

### 2.9.3 Node.JS para servidor

Apesar de recente, o Node.js já é utilizado por grandes empresas no mercado de tecnologia, como Netflix, Uber e LinkedIn.

O principal motivo de sua adição a este trabalho é a sua alta capacidade de escala. Além disso, sua arquitetura e flexibilidade, o tornam uma boa escolha para implementação de Microsserviços. Por isso, é bastante utilizado, com fácil compatibilidade com outras plataformas, inclusive com os principais fornecedores de produtos e serviços em nuvem. Uma vez que já têm suporte para desenvolvimento de soluções escaláveis, utilizando o Node.js, como é o caso do servidor TTN.

## 2.10 Dashboard

*Dashboard* é como um painel, é um tipo de interface gráfica para o usuário que geralmente oferece visualizações rápidas dos principais indicadores de desempenho relevantes para um objetivo ou processo específico.Em outro contexto, "painel" pode ser associado a um "relatório de progresso", e para esta aplicação, a localização do dispositivo e os valores detectados pelos sensores, são as informações mais importantes, e precisam ser exibidas e acessadas de forma simples e rápida.

### 2.10.1 API da Plataforma Google Maps

Foi usada uma API da plaforma Google para exportar as funcionalidades do Google Maps, onde são consumidos os dados requisitados do *Backend*.

### 2.10.2 Frameworks utilizados para Frontend

Para o *Frontend* foi necessário utilizar as tecnologias principais e mais comuns para desenvolvimento *web*. Que é o caso do uso do HTML5 (*Hypertext Markup Language*), para criar a estrutura da página onde ficaria a interface, *javascript* para fazer as interações com os elementos exibidos, o CSS3 (*framework Bootstrap*) para montar os componentes com mais detalhes, e por ultimo, o (*Cascading Style Sheets*) que possui um conjunto de bibliotecas e funções para padronizar os elementos de UI (*User Interface*) da aplicação.

# 3 Metodologia (Materiais e Métodos)

## 3.1 Arquitetura Geral do sistema

A arquitetura do sistema é um modelo sistema cliente/servidor, em que a diferença é um concentrador LoRa (*gateway*) que faz o direcionamento dos dados obtidos pelos dispositivos LoRa, para o servidor TTN. Por sua vez, é integrado a uma aplicação *web* executada em um servidor Node.JS, a qual é acessada pelo cliente possibilitando um monitoramento em tempo real.

Figura 17 – Arquitetura do Sistema

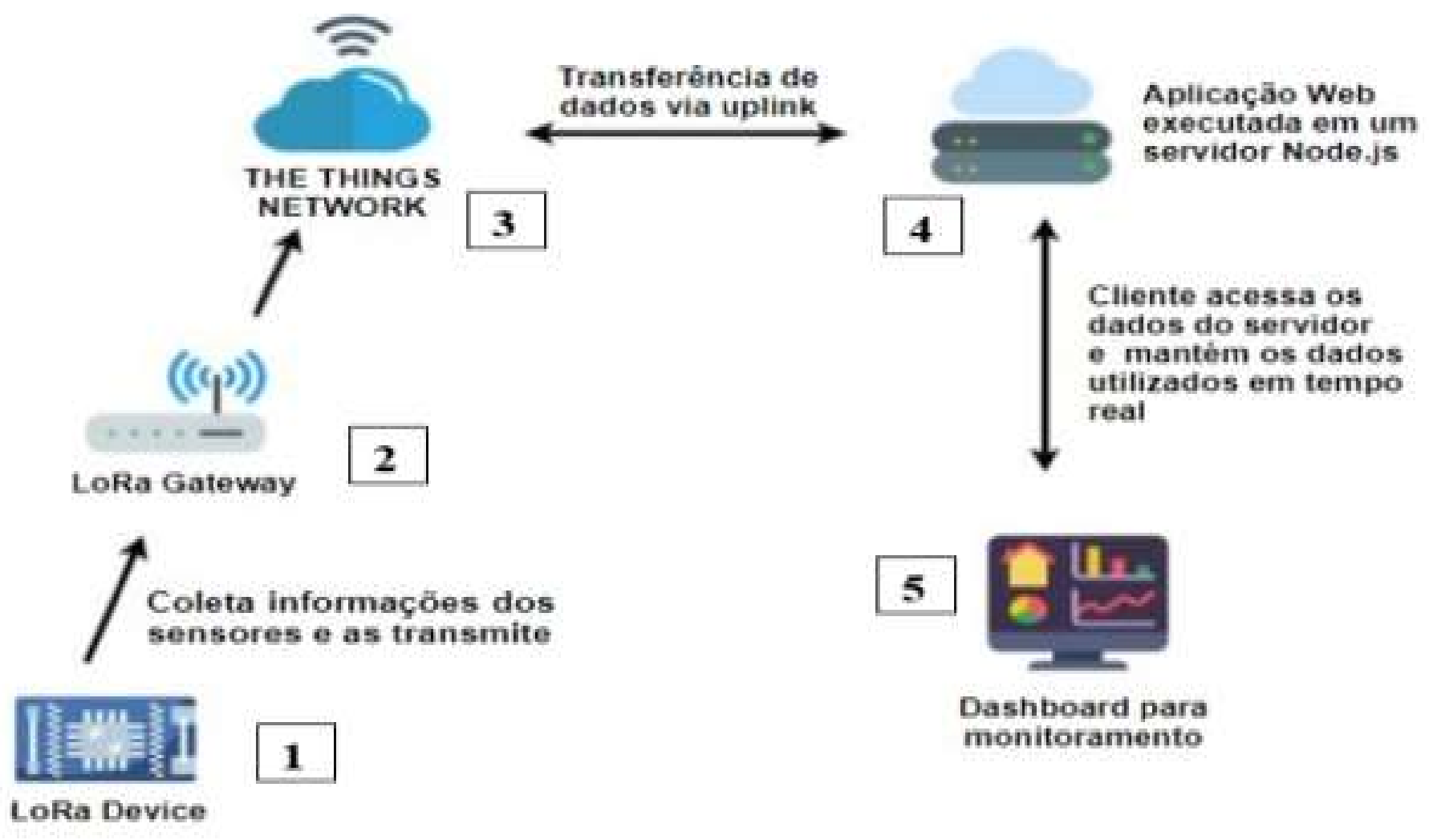

**Fonte:** Lucas Gonçalves,2020

**(1) LoRa Device –** é um *device* adaptado à tecnologia LoRa, composto pelo Arduino MEGA, sensores e o módulo transmissor LoRa. O evento é detectado, com base no valor analógico obtido pelos sensores e, interpretado e instruído pelo microcontrolador do Arduino, por meio de comandos AT, a enviar o valor codificado, através do transmissor LoRa, para o *gateway*.

**(2) LoRa Gateway –** é o concentrador de dados cuja responsabilidade é servir de intermediário de comunicação entre um *LoRa device* e a internet, de forma que o sistema se mantém conectado. Também chamado de *LoRaWAN Base Station*, tem uma função análoga a uma e-NodeB ou BTS (*Base Transceiver Station*), atuando com procolo de rede LoRaWAN.

**(3) The Things Network (TTN) –** é o servidor hospedado em nuvem, exclusivo para aplicações LoRaWAN. Basta criar uma conta, cadastrar os *devices* e o *gateway*. Com isso, serão gerados endereços/códigos em hexadecimal, que são usados para identificar cada *device*, e assim poder monitorar o tráfego de dados no servidor.

**(4) Aplicação Web –** é uma aplicação utilizando uma API (*Application Programming Interface*) da plataforma do *Google Maps*. Foi utilizada para integração entre servidor TTN e o servidor Node.JS onde a aplicação é executada, possibilitando a transferência dos dados e exibi-los em tempo real pela interface desenvolvida.

**(5) Dashboard –** é a interface gráfica que possibilita a visualização dos dados sensoriados de uma forma mais intuitiva para interpretação do usuário, evidenciando a localização do *device* e os valores detectados pelos sensores.

# 3.2 LoRa Device

## 3.2.1 Conexões internas do módulo

Para todo projeto de IoT, deve haver um microcontrolador e um módulo que possibilite que a comunicação com os sensores seja realizada. Em casos de projetos *indoor* é muito comum que tenha um módulo *Wi-Fi*, porém nesse caso em específico, a tecnologia LoRa já se encarrega.

Até porque ela é a responsável pela comunicação, que ocorre graças ao módulo RF SX1276, possuindo duas portas RFIO-HF (*High Frequency*) e RFIO-LF (*Low Frequency*), integradas ao MCU (*Micro Controler Unit*) STM32L052. Elas podem ser programas para transmitir pacotes, por meio das interfaces de comunicação serial, permitindo um dispositivo para auxiliar como Arduino. Este envia comandos seriais e recebe mensagens de resposta a respeito do status da transmissão.

A Microunidade de Controle (Modem - STM32L052) possui todas essas interfaces com protocolos de comunicação serial, por meio do protocolo SPI (*Serial Peripheral interface*), semelhante ao I2C (*Inter-integrated-circuit*).

Figura 18 – Esquema das conexões internas do Módulo usado.

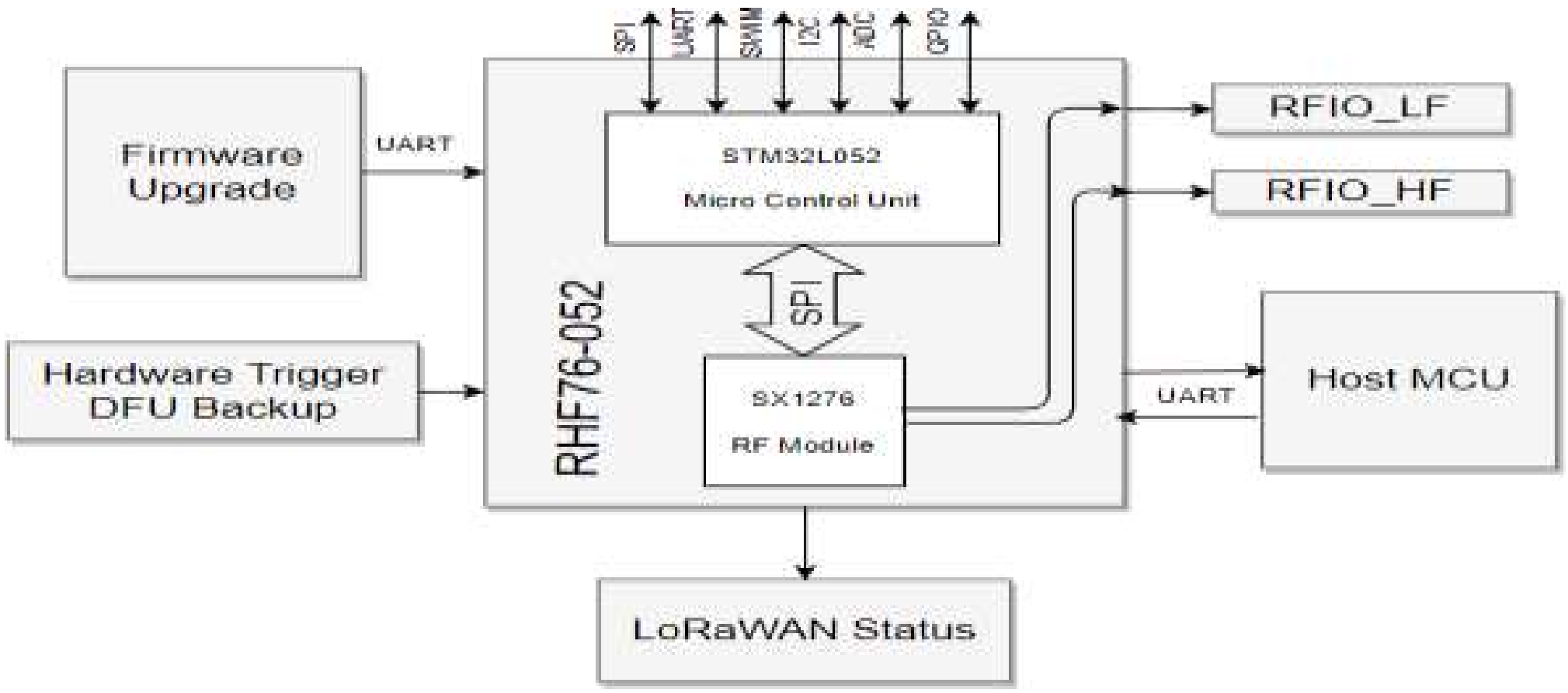

**Fonte:** Lucas Gonçalves, 2020

Com efeito, é uma forma diferente de protocolo de comunicação serial, especialmente, projetado para a conexão entre microcontroladores (um *master* e um *slave*), como neste caso, em que o microcontrolador do Arduino (Host MCU) passa as instruções para o MCU do módulo que, por sua vez, comanda o módulo RF interno.

3.2.1.1 Acionamento por DFU

O Pin14 (GPIO-PA15) é usado para ativar o modo DFU (*Device Firmware Update*) para atualização do *Firmware* (FW) do dispositivo que pode ser ativado por meio de comando AT para recuperação de configurações anteriores ou retornar ao padrão de fábrica.

3.2.1.2 Atualização de Firmware por UART

Protocolo de comunicação serial simples permite que o *host* se comunique com o dispositivo auxiliar. O UART (*Universal Asynchronous Reception and Transmission*) suporta transmissão de dados bidirecional, assíncrona e serial, possuindo duas linhas de dados, uma para transmitir (TX) e outra para receber (RX). Importa dizer que pode operar de três modos: 1) *simplex* – transmissão de dados em uma direção; 2) *half duplex* – transmissão de dados em qualquer direção, mas não simultaneamente; 3) *full duplex* – transmissão de dados.

#### 3.2.1.3 LoRaWAN status

O Pin16 (GPIO-PB4) é usado para mostrar o *status* de processamento do *End-Point LoRa*. É conectado um LED a este pino, responsável pela sinalização, quando há uma transmissão de uma mensagem.

#### 3.2.1.4 Antena para transmissão

Antena Monopolo Espiral Omnidirecional possui uma estrutura confiável para redes de comunicações de longo alcance, específica para o range de frequência de 900 a 930 MHz.

#### 3.2.1.5 Parâmetros de configuração do módulo

O módulo LoRa é muito adequado para aplicações de longa duração com autonomia de bateria. Pode ser usado no ingresso em uma rede LoRaWAN, com o modo ABP (*Activation By Personalization*) ou OTAA (*Over The Air*) que serão explicados a seguir.

Para participar de uma rede LoRaWAN, cada dispositivo final deve ser personalizado e ativado. Até porque a ativação de um dispositivo final pode ser alcançada de duas maneiras: 1) OTAA – Através da Ativação pelo ar quando um dispositivo final é implantado ou redefinido; 2) ABP – Ativação por personalização, em que as duas etapas de personalização e ativação do dispositivo final são executadas como uma única etapa.

No modo ABP, são utilizadas duas chaves – NwkSKey e o AppSKey – que precisam ser conhecidos com antecedência, as quais são as chaves de endereçamento de rede e da aplicação, respectivamente. Ou seja, elas têm a função de alterar a identificação da rede e do dispositivo.

Com o modo OTAA, as chaves são AppKey e AppEui, as quais são apenas as chaves correspondentes à identificação da aplicação. Ou seja, neste modo não é possível alterar o endereço da rede, mas precisa de duas chaves para identificar a aplicação.

### 3.2.1.6 Comandos AT

Atualmente, existem 3 tipos de modem LoRaWAN na família de produtos da RisingHF, todos baseados no módulo RHF76-052 do RisingHF e suporta o mesmo conjunto de comandos AT. Todos são dispositivos (*"Plug and Play"*), com o uso de qualquer um deles, pode-se fazer testes e análises rápidas para qualquer rede LoRaWAN. Além do módulo, é necessário um PC ou MCU *host* para enviar o comando AT para controlar o módulo. Os comandos que formam necessários para configurar a microunidade de controle no módulo são mostrados na tabela 7, em seguida.

Tabela 7 – Comandos AT utilizados.

| **Comandos AT** | **Descrição** |
|---|---|
| AT | Comando para iniciar configuração |
| HELP | lista os comandos e funções |
| FDEFAULT | Restaura a configuração de taxa de dados |
| RESET | Reinicia o software |
| DFU | Modo para atualização do software |
| LOWPOWER | Habilita modo de baixo consumo |
| MSGHEX | Envio de mensagem em Hexadecimal sem confirmação de recebimento |
| CMSGHEX | Envio de mensagem em Hexadecimal com confirmação de recebimento |
| CH | Habilita o canal de frequência |
| ADR(ON/OFF) | Habilita/desabilita funcionalidade taxa de dados adaptável |
| DR | Habilita taxa de dados |
| RXWIN1 | Habilita frequência para janela de recebimento 1 |
| RXWIN2 | Habilita frequência para janela de recebimento 2 |
| MODE | Habilita o modo de operação (LWABP/LWOTAA) |
| ID | Endereços de identificação (DevAddr/DevEui/AppEui) |
| KEY | Chaves de acesso (NWKSKEY/APPSKEY/APPKEY) |
| CLASS | Definir classe (A/B/C) |
| DELAY | Intervalo de tempo entre envio e recebimento |

# 3.3 LoRa Gateway

RHF2S008 é um Gateway Industrial LoRa/LoRaWAN feito para aplicações IoT/M2M (*Machine to Machine*). Em tal circurstância, Biral (2015) infere que as interfaces suportam Ethernet e 3G/4G/LTE (*Long Term Evolution*). Isto significa o Gateway externo LoRaWAN de alto desempenho e 8 canais RHF2S008 P4G – é um (Gateway) inteligente desenvolvido completamente, com base no protocolo LoRaWAN. Transceptor multicanal SX1301 de alto desempenho.

O principal *chip* de controle é a plataforma ARM Cortex-A53 e a frequência principal é de 1.2GHz. O Gateway pode ser conectado à rede através de redes Ethernet ou 4G/3G/2G.

A fonte de alimentação é fácil de instalar no local por PoE (*Power over Ethernet*), equipado com o módulo GPS (*Global Positioning System*) interno, podendo fornecer sinal PPS (*Pulse Per Second*) de alta precisão para o sistema.

A aparência do *Gateway* adota o tipo de caixa de metal. O RHF2S008P4G é um *Gateway* externo de nível industrial, adequado para ambientes industriais complexos, a faixa de temperatura de trabalho atinge de -40 a 75 °C.

O RHF2S008B4G é um *Gateway* IoT implantado para redes LPWAN baseadas em LoRaWAN. Ele integra uma bateria como fonte de energia de backup. Quando a energia externa é cortada, ela também pode ser continuamente alimentada por uma fonte de energia de backup.

Esse *Gateway* industrial possui um núcleo ARM Cortex-A53 de 1,2 GHz de alto desempenho e integra um chip de processamento de banda base SX1301 LoRa. Ele suporta 8 canais multi-SF (*multi-Spread Factor*), 1 canal SF único e uma alta velocidade. A potência de saída pode ser de até 27dBm e a sensibilidade é de -142dBm a 300bps. Sob um ciclo de transmissão específico e duração dos dados, um único *Gateway* pode acessar dezenas de milhares de nós.

O *mainframe* RHF2S008B4G possui um tamanho pequeno e uma confiabilidade muito alta, que pode ser facilmente implantada em ambientes externos ou em um ambiente complexo.

O diagrama de blocos funcionais do *Gateway* RHF2S008B4G LoRaWAN é mostrado abaixo na Figura 19 e inclui principalmente duas partes: o *host* RHF2S008B4G, chamado de "Gateway Encapsulado" na ilustração e o módulo de fonte de alimentação (PoE).

Figura 19 – Esquema das conexões internas do Gateway usado.

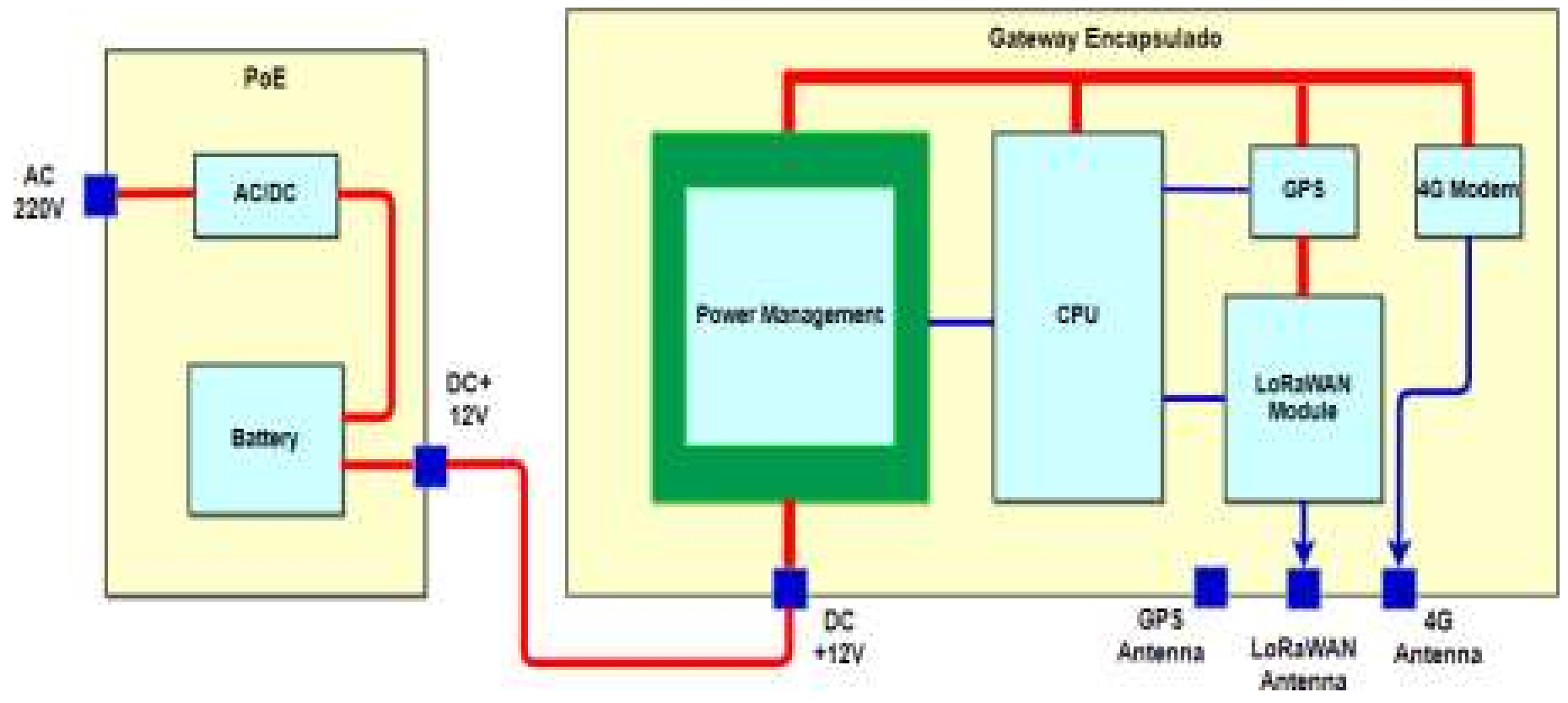

**Fonte:** Lucas Gonçalves, 2020

# 3.4 Servidor em nuvem TTN

É uma plataforma que oferece uma rede aberta e colaborativa, onde qualquer usuário pode contribuir, desde que possua os dispositivos compatíveis com a tecnologia LoRa e que faça o cadastro na estrutra em nuvem *The Things Network*. O restante é gerenciado automaticamente, assim que são registradas as aplicações, tornando acessível todos os dados enviados dos *end-points*.

## 3.4.1 Levantamento de dados pelo módulo sensor

Basicamente o *firmware* foi desenvolvido, com a estipulação de condições para fazer a transmissão de alguma leitura dos sensores. Ou seja, os sensores fazem as leituras dos valores analógicos de temperatura (calor), concentração de gás (fumaça) e luminosidade (chamas). A medida em que esses valores começam a mudar bruscamente, são enviadas atualizações das detecções dos sensores.

De forma genérica, a variável definida para cada sensor, armazena o valor lido/detectado

sensor = ValorLido

Tendo o valor real, são feitas as condições. Em face disso, os valores de comparação para estipular como valores de risco ou não, são definidos dependendo da aplicação.

se (valorAlerta <= sensor < valorRisco)

ou

se (sensor >= valorRisco)

Dentro de cada condição dessa, é adicionada a ação/instrução, com uma mensagem e em seguida é enviado o valor analógico detectado pelo sensor.

serial.imprime("Ficar em alerta");

serial.imprime ("Possível área em perigo");

Como o valor é analógico e se trata do valor real, é necessária fazer a conversão desse valor para a base hexadecimal, para que seja enviado.

serial.imprime (sensor,HEX)

O valor detectado, codificado para a base hexadecimal é armazenado em uma *string* e assim é possível enviar pela serial do arduino.

string = sensor,HEX;

O comando AT para instruir o módulo rádio LoRa, a enviar o pacote é o seguinte:

serial.imprime (AT + MSGHEX = "string")

Para estipular os valores limitantes das condições, foram feitos testes em cenários reais para verificar quais indicavam alguma alteração significativa na medição das grandezas de cada sensor. Tais testes serão expostos nos resultados.

### 3.4.2 Comunicação entre módulo sensor e servidor TTN

Os dados enviados pelo *end-points* chegam no servidor TTN, por meio do Gateway. Internamente, há todas as opções de canais e frequências para serem habilitadas. Uma vez habilitado, o gerenciamento das aplicações é realizado automaticamente, para não haver direcionamento de dados para outra aplicação que não seja a correta. Para configurar é necessário acessar:

✓ o Gateway por meio do roteador para obter o seu enderço de IP (*Internet Protocol*), buscando pelo *hostname* “rhf2s008”;

✓ o Gateway remotamente por SSH (*Secure Shell*), usando uma das ferramentas *“PuTTY” ou “TeraTerm”* , no sistema operacional Windows;

Nesse caso, usou-se a ferramenta “PuTTY”, colocando-se o *username* padrão, no campo de endereço IP, junto com o endereço IP da seguinte forma “rxhf@IP_Address”. Importa dizer que a porta 22 é definida por padrão, na Figura 20.

Figura 20: Configuração da ferramenta PuTTY

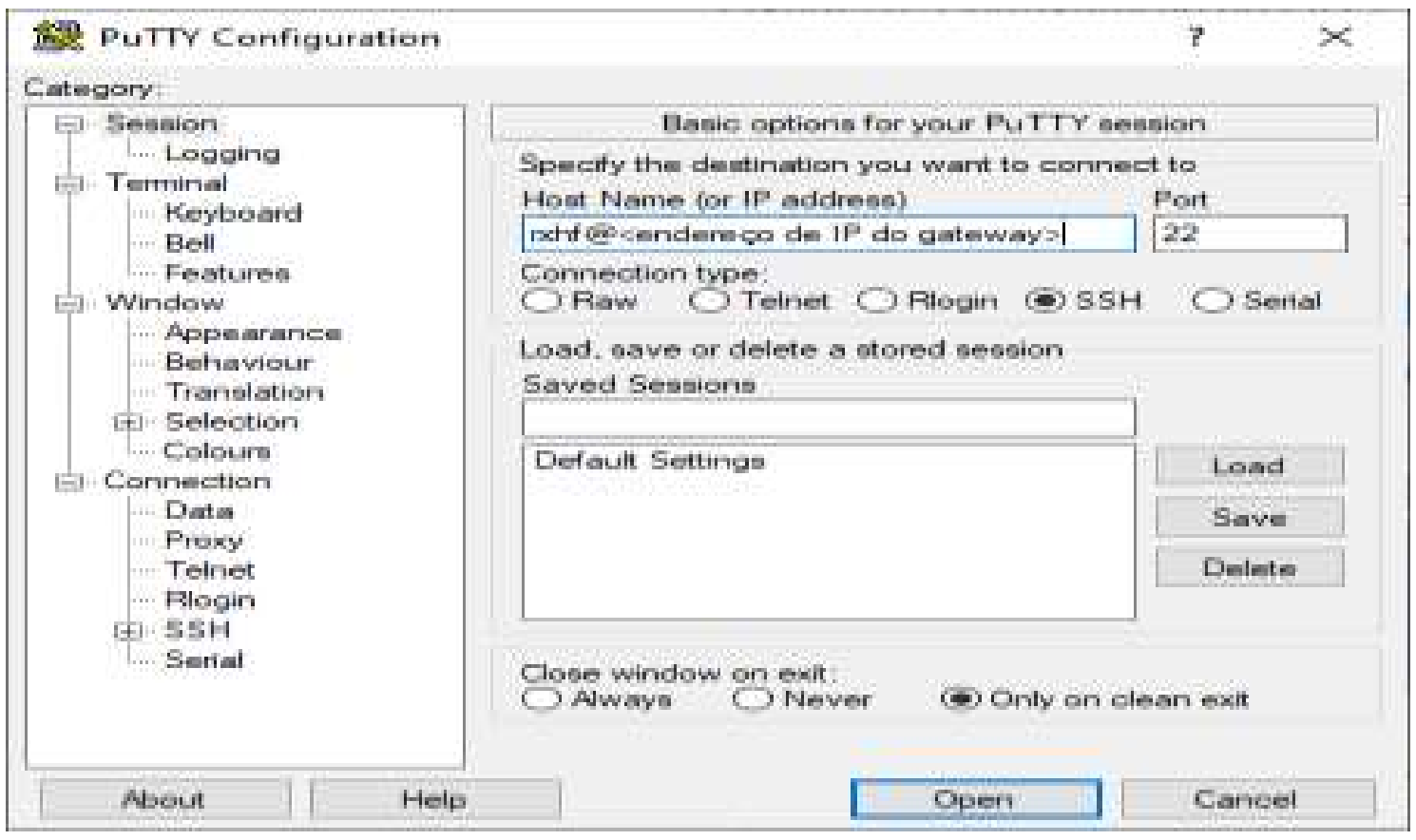

**Fonte:** Lucas Gonçalves, 2020

Ao abrir, solicita informar a senha “risinghf” de acesso aos arquivos de configuração do *Gateway.* Após a sua efetivação, acompanha-se o seguinte *path*, para acessar o arquivo de configuração de canais de frequência na faixa dos 900MHz: “/home/rxhf/risinghf/test/cfg/freq_conf_920.json”. No caso de acessar/habilitar qualquer outra faixa de frequência, segue a hierarquia de arquivos da Figura 21.

Figura 21: Hierarquia de arquivos para configuração de frequência

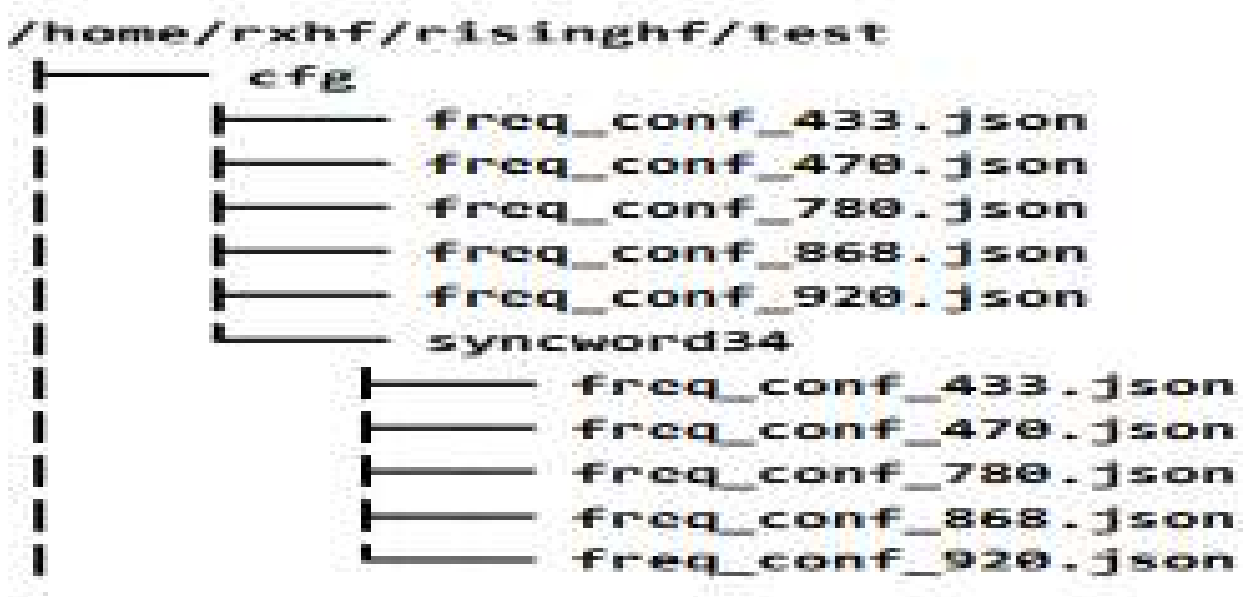

**Fonte:** risinghf, 2015

Com isso é possível habilitar canais de frequência e desabilitar outros. Neste caso, utilizará a faixa de frequência de 915MHz, liberada no Brasil.

## 3.5 Integração cliente/servidor

O servidor TTN disponibiliza várias formas de integração com outras plataformas, para descentralizar os dados obtidos pelas suas aplicações, o que possibilita o redireciona mento desses dados para outras aplicações externas. Porém, utilizar as formas sugeridas que já vêm pré configuradas para facilitar o acesso, podem gerar custos e ainda fazem o projeto depender de outras API’s (*Application Programming Interface*) com o estabelecimento da conexão necessária entre servidor e a aplicação. Isto sugere a existência de possíveis problemas, como alguma instabilidade na API ou no servidor, podendo prejudicar o funcionamento do sistema.

Pensando nisso, foi feito um servidor local que faz a intermediação entre o TTN (servidor *web*) e a aplicação *web* (cliente). Este servidor local é basicamente responsável por fazer o espelhamento do servidor *web*, tornando tudo visível para o desenvolvedor do projeto. Pois, dessa maneira, torna-se possível requisitar os dados no formato desejado para o acesso em tempo real.

### 3.5.1 Estrutura de implementação do servidor local

O projeto conta com uma aplicação na camada de servidor, em que as requisições são concentradas e processadas. Desenvolvida a utilização da linguagem de programação *JavaScript* – que compõe a base da pilha de tecnologias usadas –, possibilita a aplicação quanto ao recebimento de requisições de sensores e usuário final.

O *framework* NodeJS, que utiliza o *JavaScript* como sua linguagem oficial, permite a criação de diversos tipos de aplicações, dentre elas, a aplicação do servidor deste projeto. Ela tem a responsabilidade de receber as informações contidas na nuvem TTN (*The Things Network*), processá-las e disponibilizá-las, em seguida, na interface de usuário.

O *framework Express.js* – parte integrante do desenvolvimento da aplicação do servidor – , é responsável por criar rotas de endereçamento. Essas rotas definem quais páginas mostrar ou até quais dados devem ser retornados para o requisitante, sendo uma para *web*, smartphones, smartwatches, aplicativo para TV digital dentre outros.

A biblioteca Socket.io compõe a camada mais acima da pilha de tecnologias, pois basicamente auxilia no direcionanamento dos dados entre *backend* e *frontend* através das rotas criadas. É por isso que essa biblioteca e o *framework Express.js* são bastante usados de forma integrada.

### 3.5.2 Requisição dos dados do servidor web em tempo real

Como mencionado anteriormente, uma das responsabilidades da aplicação *web* é exibir ao usuário final informações coletadas pelos sensores. Esses dados são obtidos de um provedor chamado TTN (*The Things Network*).

Figura 22 – Pilha de tecnologias utilizadas para o desenvolvimento backend.

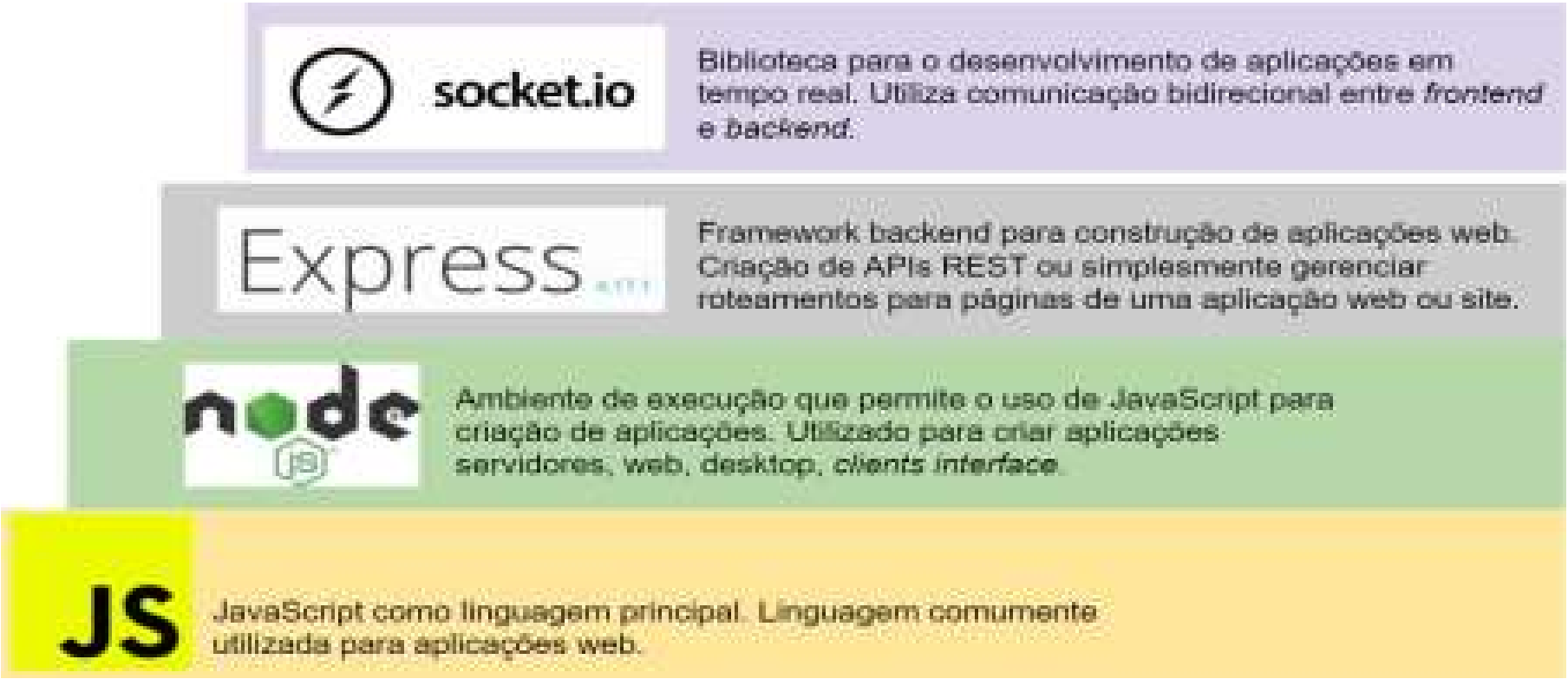

**Fonte:** Lucas Gonçalves, 2020

E mais, os sensores captam as informações do ambiente instaladas e as enviam para o provedor TTN, a aplicação do servidor "percebe" que existem dados novos no provedor. Em seguida, faz as devidas tratativas, mostrando ao usuário em um dashboard onde está sendo exibido um mapa com a posição de cada sensor e seus respectivos valores.

Isso é possível através do código criado em javascript, pelos desenvolvedores do The Things Network, onde já é feita a obtenção dos dados, ao informar os endereços referentes à aplicação registrada no TTN.

Figura 23 – Código javascript para obter dados da aplicação do TTN.

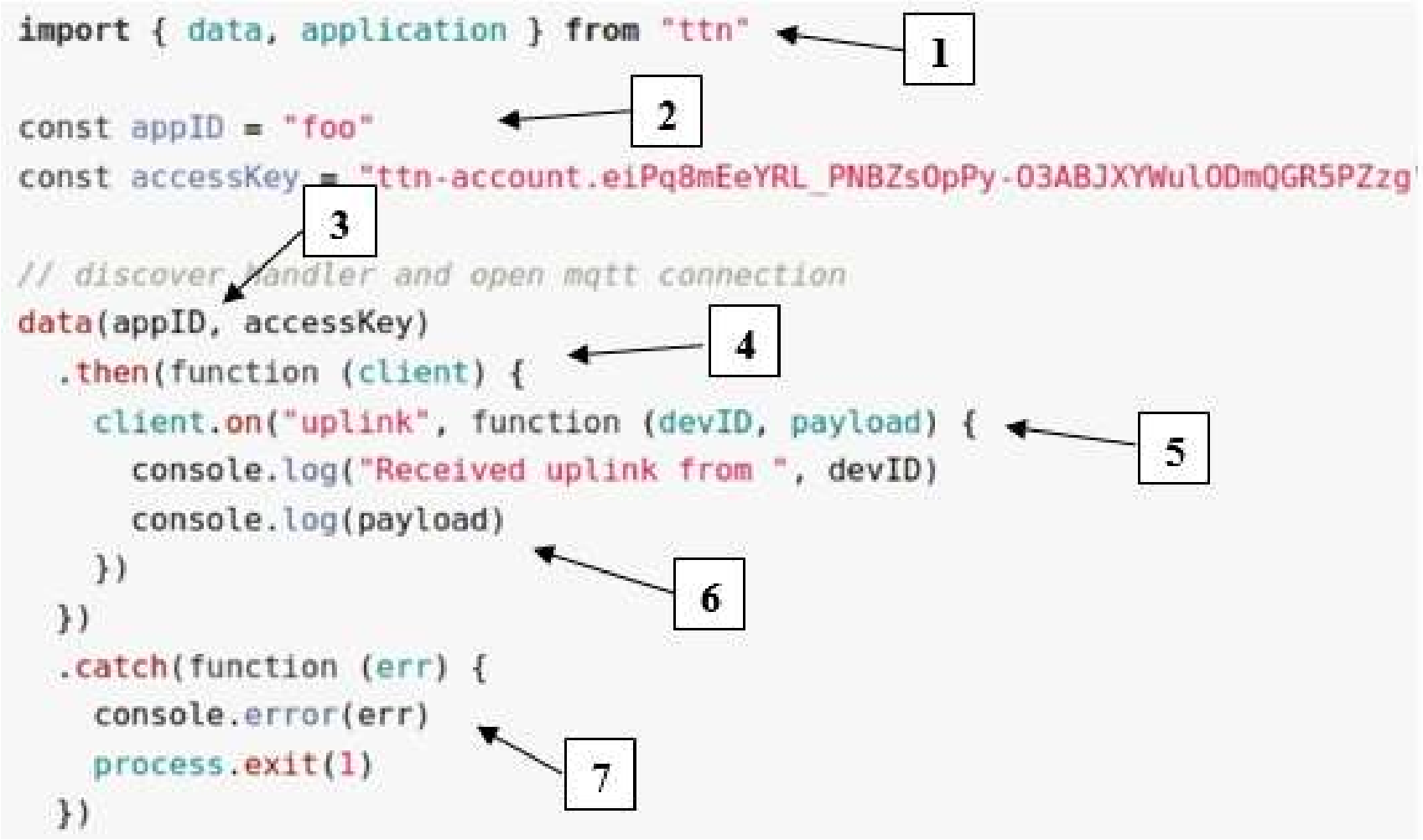

**Fonte:** Lucas Gonçalves, 2020

Com referência a figura 24, a forma que o servidor local verifica a existência de dados novos no servidor web TTN, é explicada nos 7 passos a seguir: **(1)** Importação das variáveis necessárias para fazer a requisição de dados ; **(2)** Declaração das constantes de indentificação e acesso; **(3)** Colocação da variável "data" em função das constantes, conectando-se ao TTN ; **(4)** Em javascript, quando se tem o método *".then"* significa que é o caminho verdadeiro, ou seja, se houver algum dado novo, é como se o cliente ficasse online; **(5)** Aplicação fica em alerta, neste ponto. Pois, a variável *"client"* já indica que há dados, armazenando os dados de *uplink* na variável "payload", referente ao dispositivo *"devID";* **(6)** visulização da informação no console; **(7)** O método *"catch"* é o contrário do *"then",* pois serve para capturar os erros e ajudar no tratamento dos mesmos.

## 3.6 Acessibilidade à aplicação web

Essa aplicação representa o funcionamento de todo um sistema, existindo por trás toda essa integração de recursos de *software*. E para poder acessar a aplicação, é necessário primeiro ligar o servidor com o comando *"npm start".*

Figura 24 – Servidor NodeJS ligado

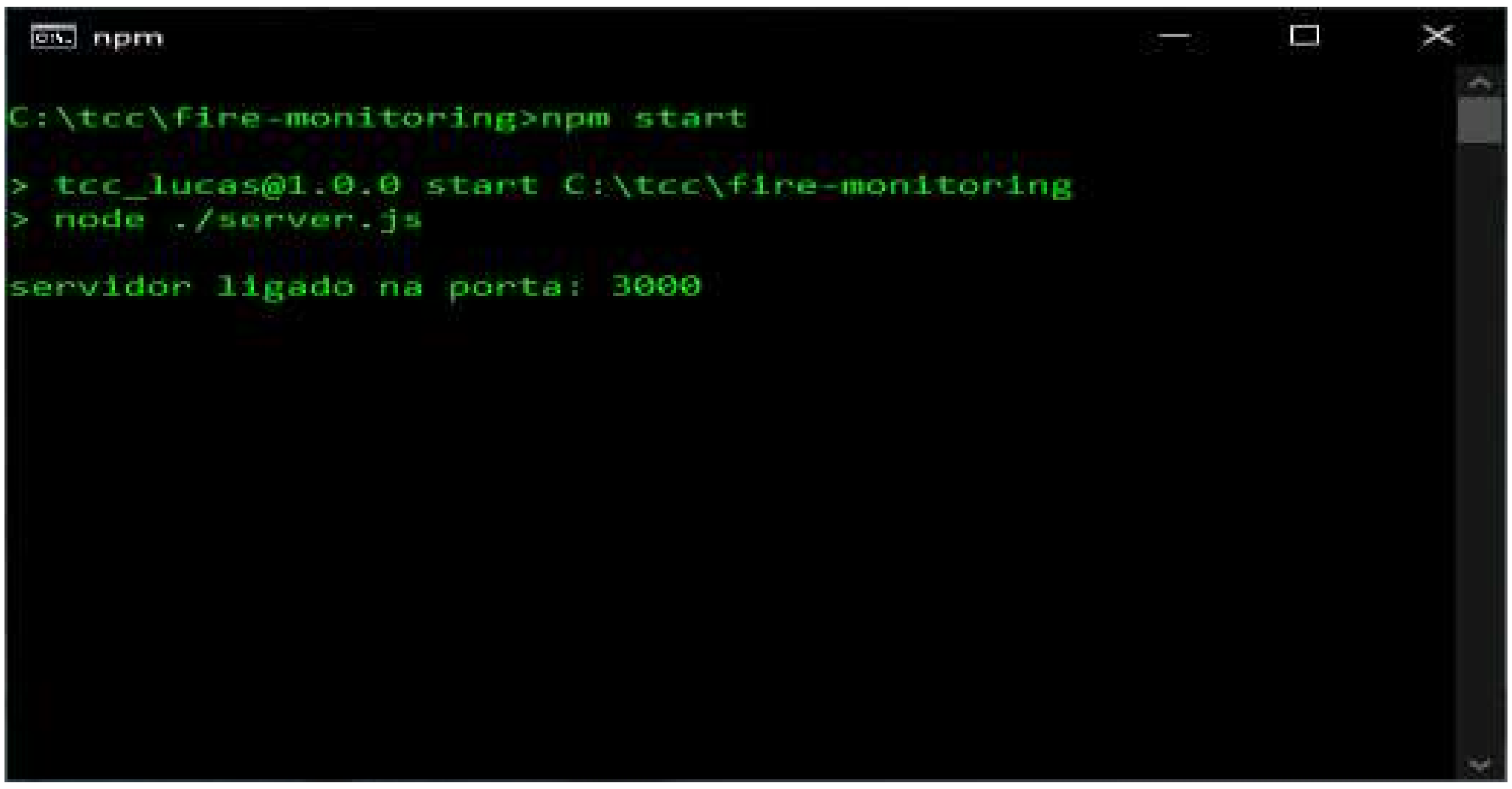

**Fonte:** Lucas Gonçalves, 2020

Com isso, basta acessar o local *host*, apontando para a porta "3000" com rota definida como *"/dashboard",* de modo que o servidor se encontra (figura 25), na forma de navegador: “localhost:3000/dashboard”

Figura 25: Acesso à aplicação web no navegador Chrome

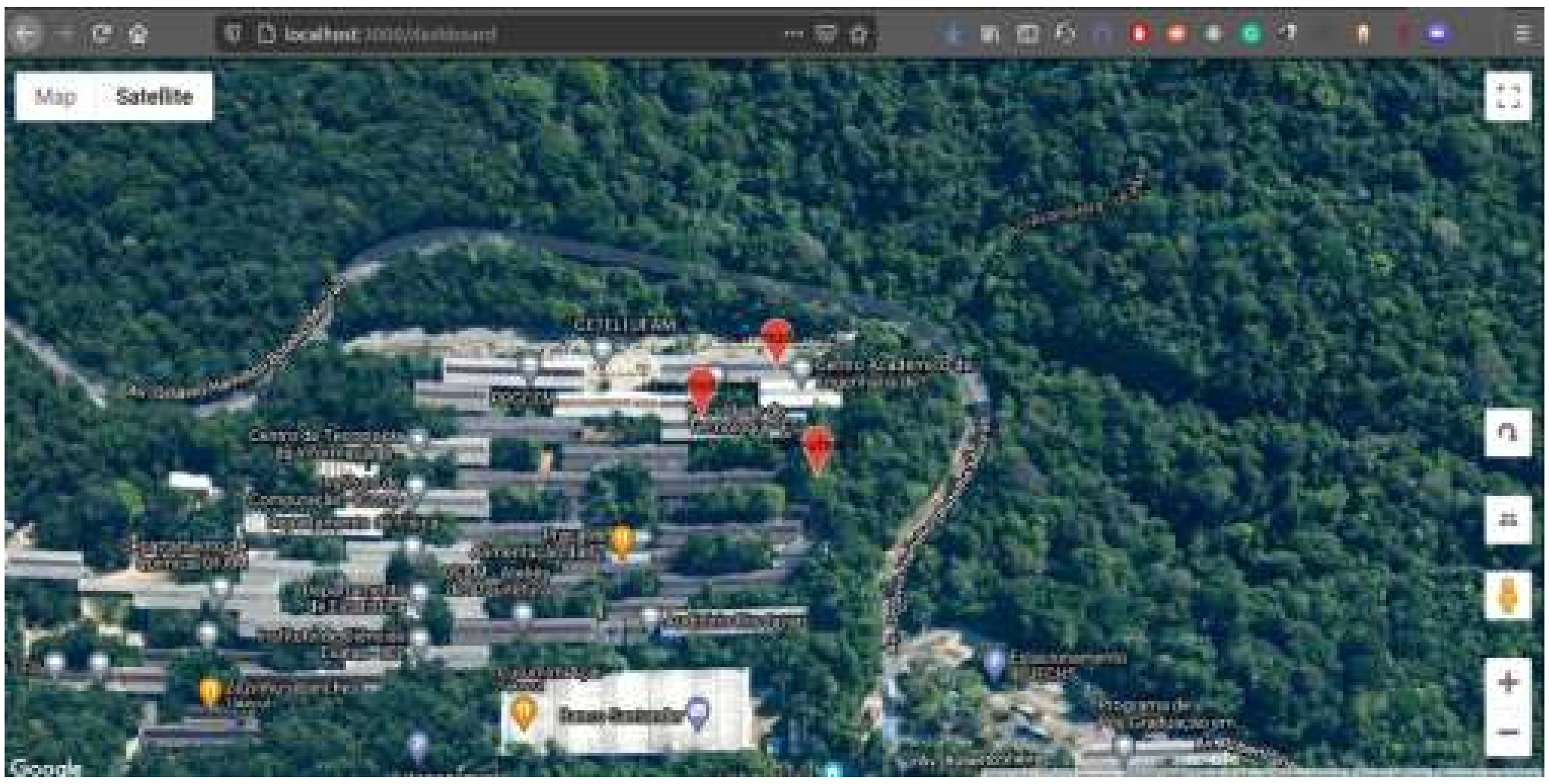

**Fonte:** Lucas Gonçalves, 2020

Percebe-se que há algumas funcionalidades do Google Maps, com a excessão dessa janela chamada de *"Modal"*, abrindo por cima do mapa, ao clicar em cima de um *"Marker"* do mapa, que representam os *end-points* LoRa nos pontos de teste.

### 3.6.1 Estrutura do Dashboard para o Frontend

A etapa de apresentação de informações para o usuário final foi desenvolvida com a utilização das tradicionais tecnologias de desenvolvimento *web*, sendo elas: HTML5, CSS3, *JavaScript*. Para utilização da API do Google Maps, responsável por exibir o mapa expandido na tela, criou-se uma conta no Google Cloud Platform para obter uma chave de autorização para uso da API.

### 3.6.2 Estrutura do Dashboard para o Frontend

A etapa de apresentação de informações para o usuário final foi desenvolvida com a utilização das tradicionais tecnologias de desenvolvimento *web*, sendo elas: HTML5, CSS3, *JavaScript*. Para utilização da API do Google Maps, responsável por exibir o mapa expandido na tela, criou-se uma conta no *Google Cloud Platform* para obter uma chave de autorização para uso da API. Para visualizar o *dashboard* da figura 26, basta clicar no “*Marker*” que são os pontos vermelhos da figura 25, os quais representam os *End-points,* que inicialmente se encontravam posicionados na UFAM próximo ao Ceteli.

Figura 26 – Acesso à aplicação Web no browser

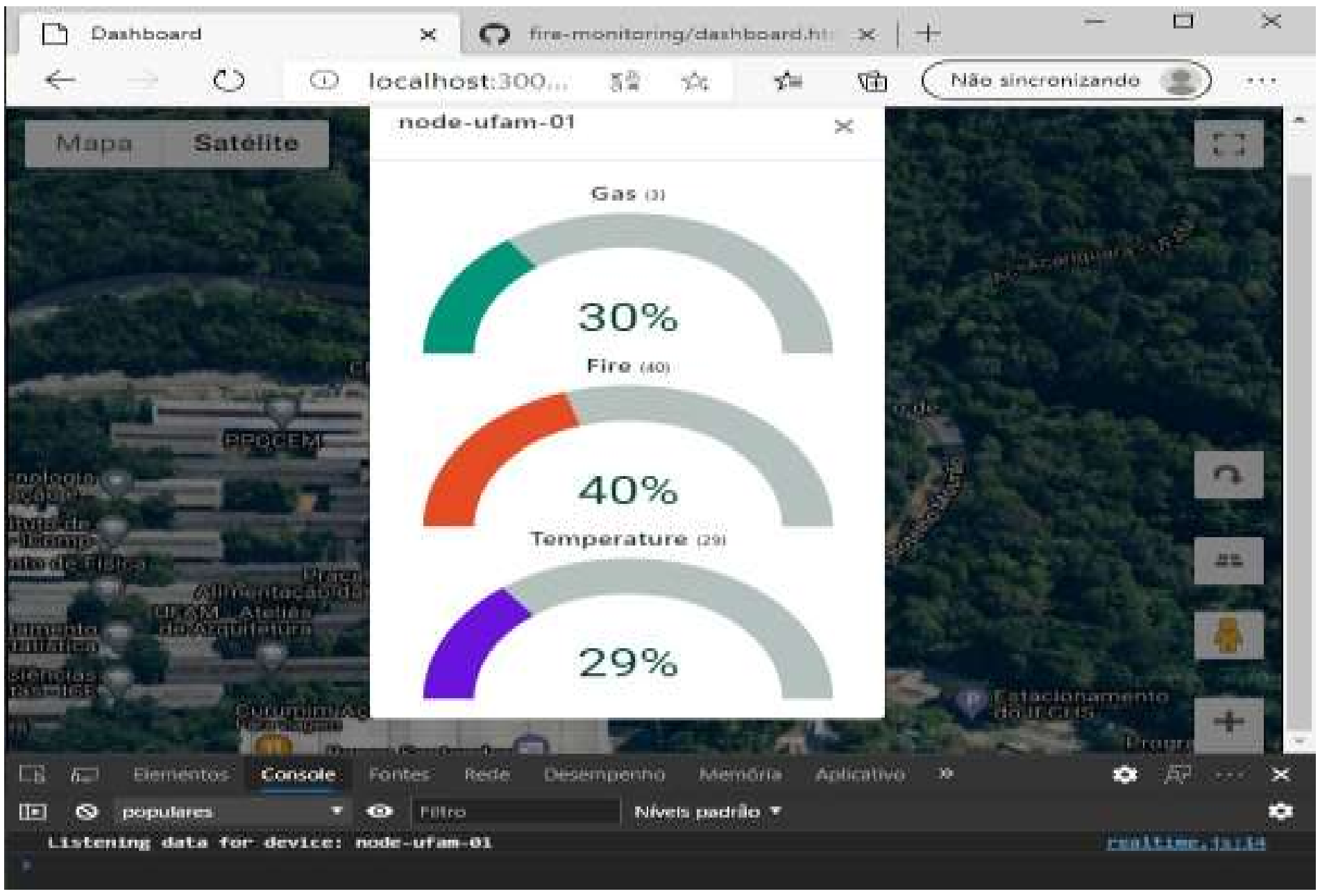

**Fonte:** Lucas Gonçalves, 2020

# 3.7 Cenários de teste e resultados

## 3.7.1 Cenário 1 – Testes com Software CoolTerm

A configurar do módulo LoRa usa os comandos AT de um *software open source* de comunicação serial, possíbilitando a aplicação de comandos interrogativos para extrair informações de configurações internas da unidade de controle do modem LoRa, assim como atribuir valores diferentes aos valores pré setados. Os comandos principais para funcionamento geral são os de inicialização, identificação, modo de operação e canais de frequência. Basta conectá-lo via porta USB e configurá-lo através do CoolTerm.

Figura 27 – Modem LoRa.

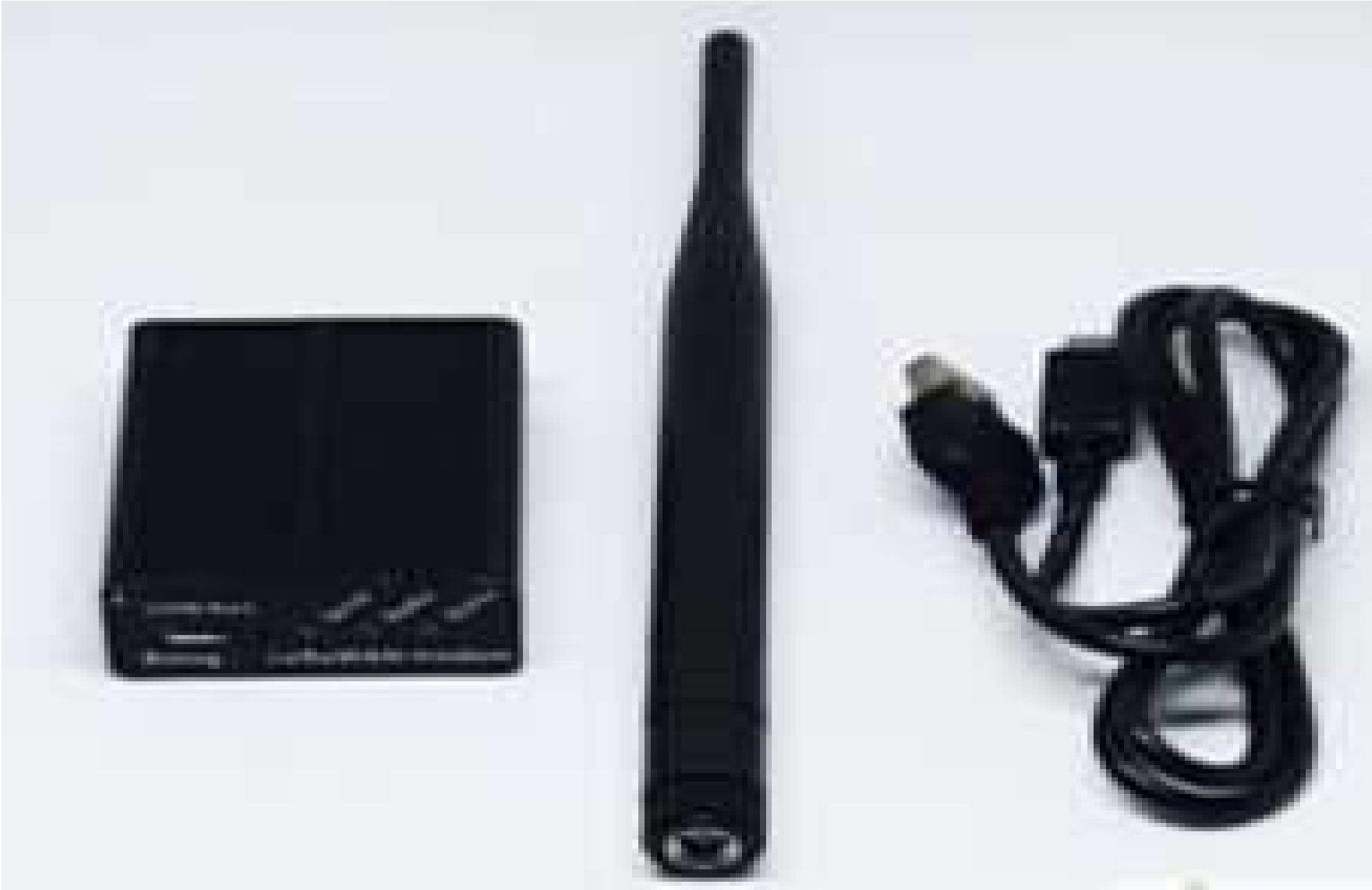

**Fonte:** RisingHF,

O *software* possibilita uma comunicação transparente com o modem, pois a cada comando enviado é retornado imediatamente a resposta de confirmação. Outro recurso interessante e bem útil é a possibilidade de configurar o modem com todos os comandos de uma vez só. Dessa forma é salvo como um *script*, e não será mais necessário mandar cada comando por vez, apenas se quiser alterar algo. Isso foi muito útil para agilizar os testes.

Os parametros principais em aspecto de rede são os referentes à taxa de transmissão (DR), banda de frequência passante (BW) e fator de espalhamento espectral (SF), que nada mais são que os parâmetros compostod da modulação CSS, podendo alterar para modulção FSK.

Neste primeiro momento não houve participação dos sensores, mas foram enviadas mensagens aleatórias em Hexadecimal, e visualizadas na aplicação criada no servidor TTN. Logo abaixo na figura 28, é possível notar um exemplo em que é enviado um comando de verificação, sendo dado imediatamente o retorno.

Figura 28 – Comando de verificação de parâmetros.

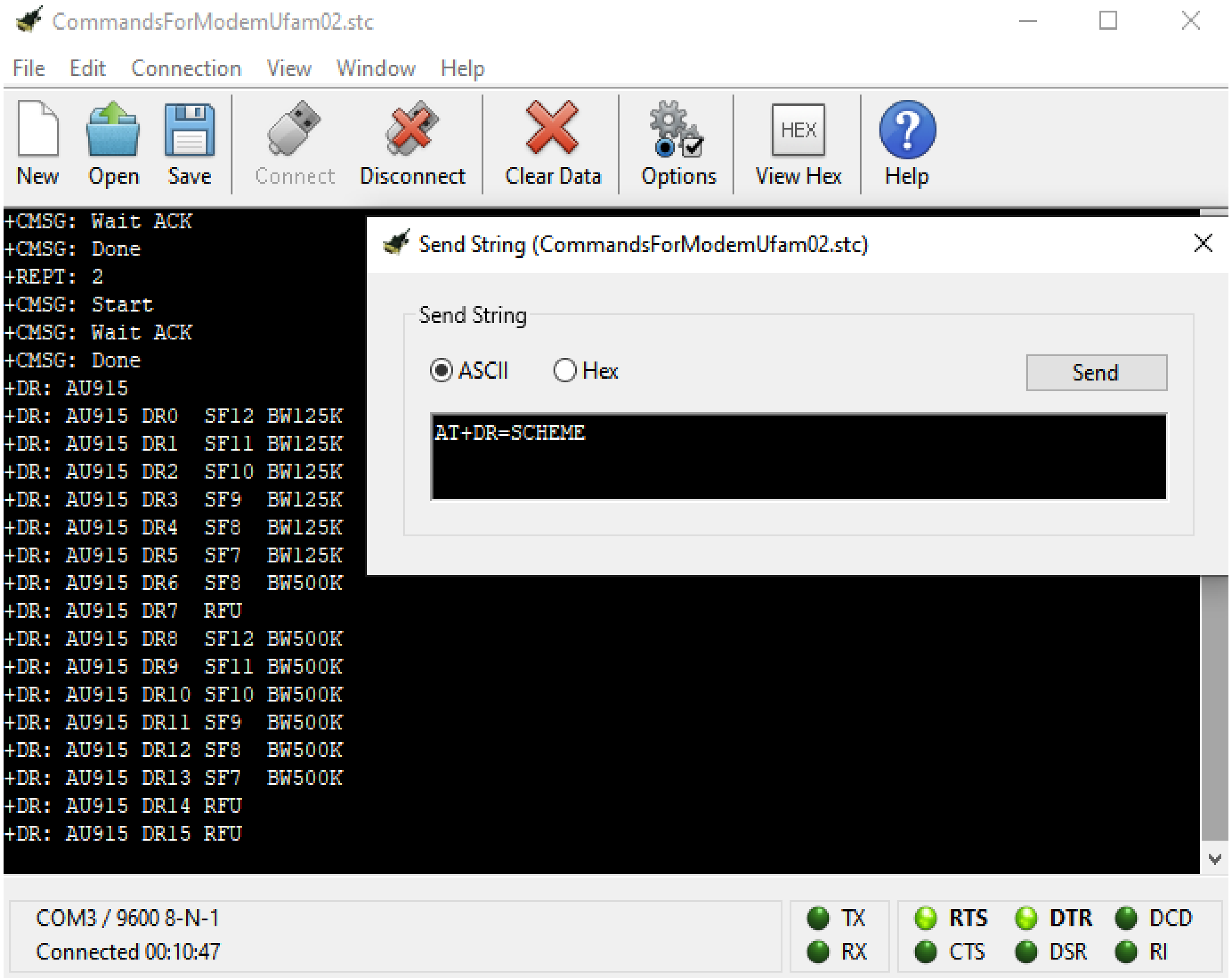

**Fonte:** Lucas Gonçalves, 2020

Percebeu-se que o gateway não permite a visualização do payload enviado pelo device. Mas confirma apenas o recebimento entregando os metadados automáticos do próprio servidor e mostra um payload criptografado. Decerto, apenas o mantenedor da aplicação e os colabores podem dar permissão ao acessar do payload enviado de fato.

Na configuração de teste, para a rede foi montada a seguinte estrutura, como é mostrado na figura 29, onde se encontra a sala de projetos disponibilizada para pesquisa e desenvolvimento. Nela há pontos de rede cabeada, onde foi conectado um cabo de rede até a entrada de rede do roteador e uma das portas do roteador foi conectada ao PoE o qual energiza o gateway e faz a conexão com ele e a internet

Figura 29 – Arquitetura do cenário de teste

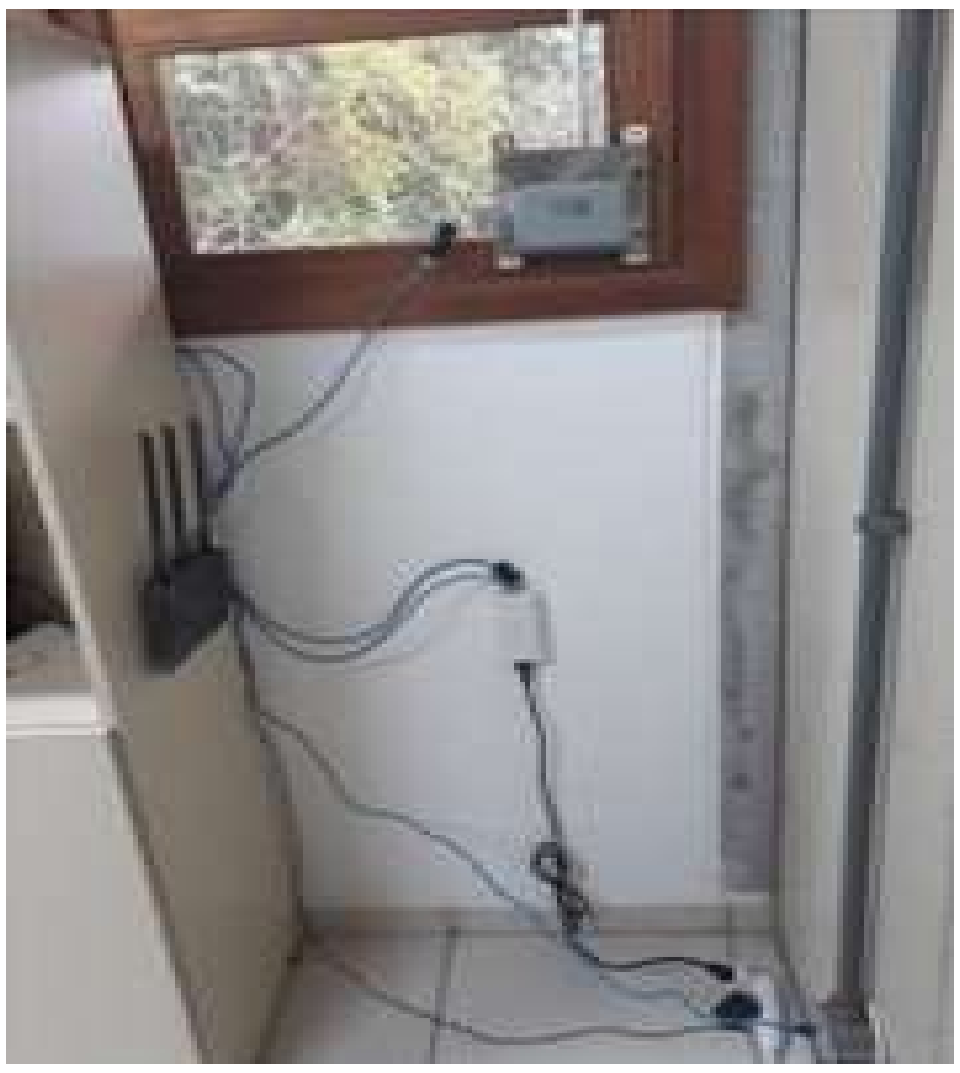

**Fonte:** Lucas Gonçalves, 2020

### 3.7.2 Cenário 2 – Teste com Arduino

No segundo momento, foi desenvolvido um protótipo do *end-point* LoRa, adaptado com um arduino. A conexão para ligar o módulo rádio LoRa foi baseada no seguinte circuito.

Figura 30 – Circuito do Rádio LoRa.

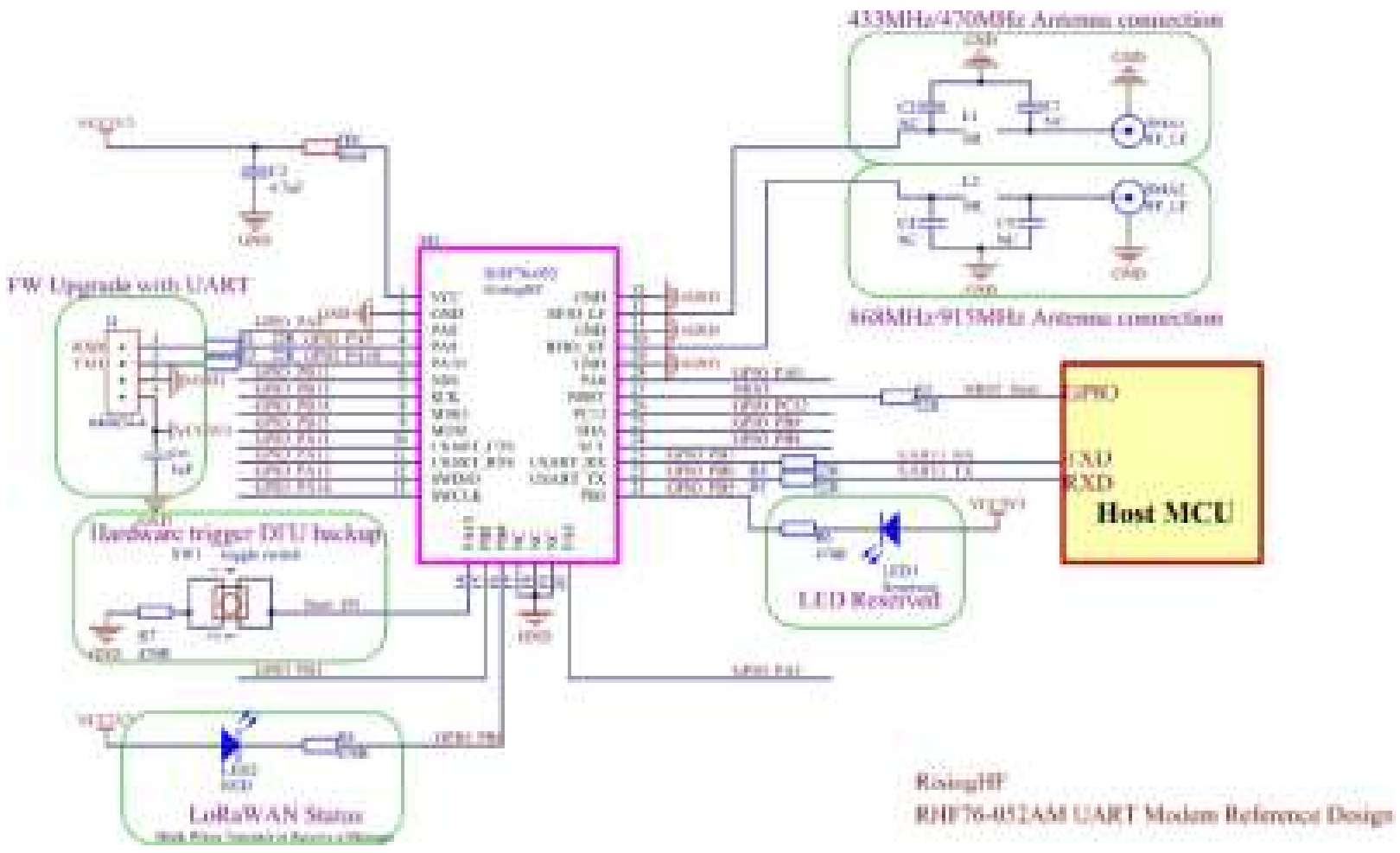

**Fonte:** RisingHF,2015

De acordo com o esquema anterior do módulo, foram seguidas apenas as partes essenciais do circuito para funcionamento.

Figura 31 – Primeiro protótipo do Circuito do Rádio LoRa.

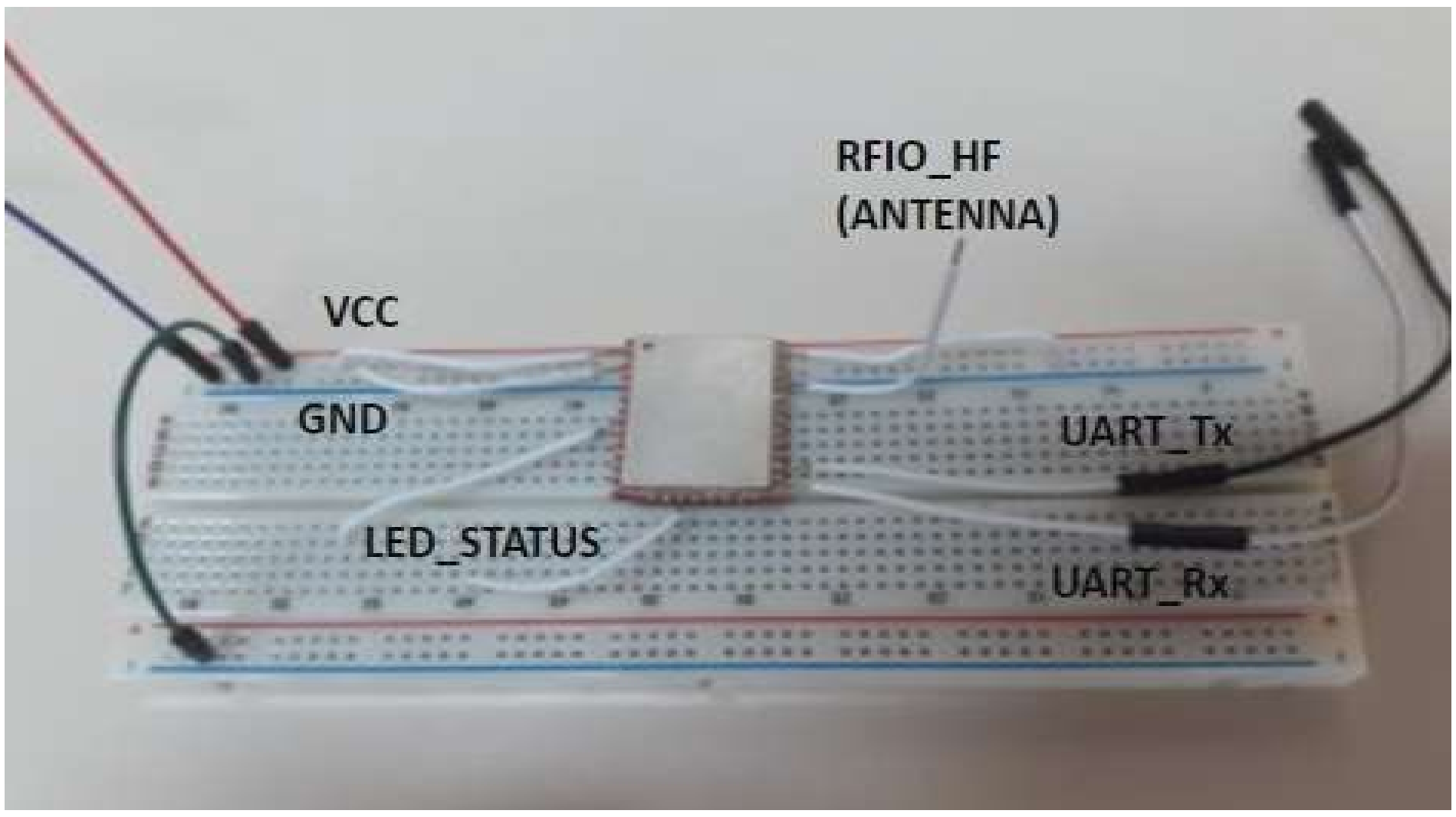

**Fonte:** Lucas Gonçalves, 2020

Com isso, a conexão serial com o arduino foi possível e o primeiro protótipo do *end-point* completo ficou assim:

Figura 32 – Primeiro teste com o primeiro protótipo do *End-Point*.

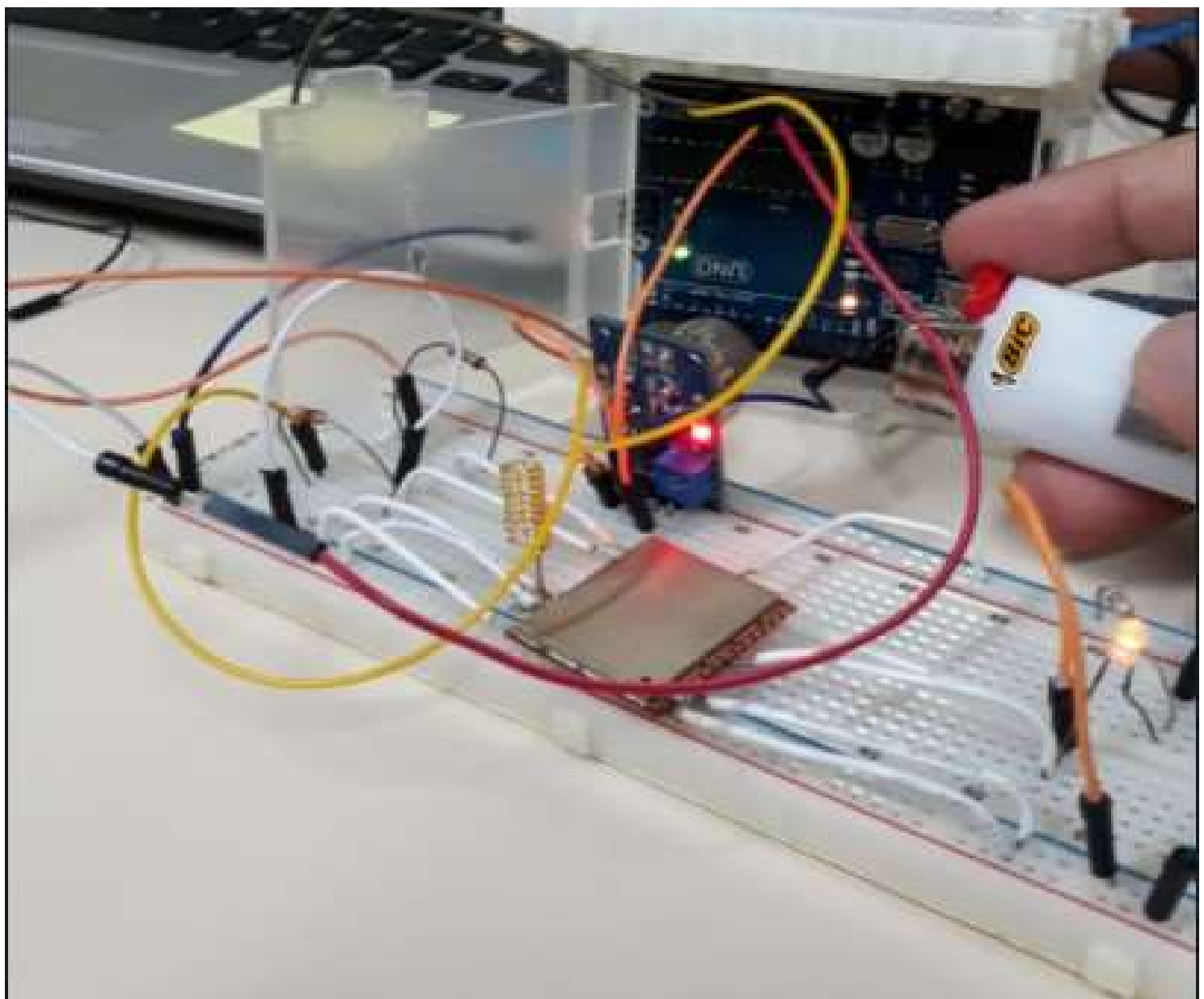

**Fonte:** Lucas Gonçalves, 2020

Figura 33 – Monitoramento pelo monitor serial.

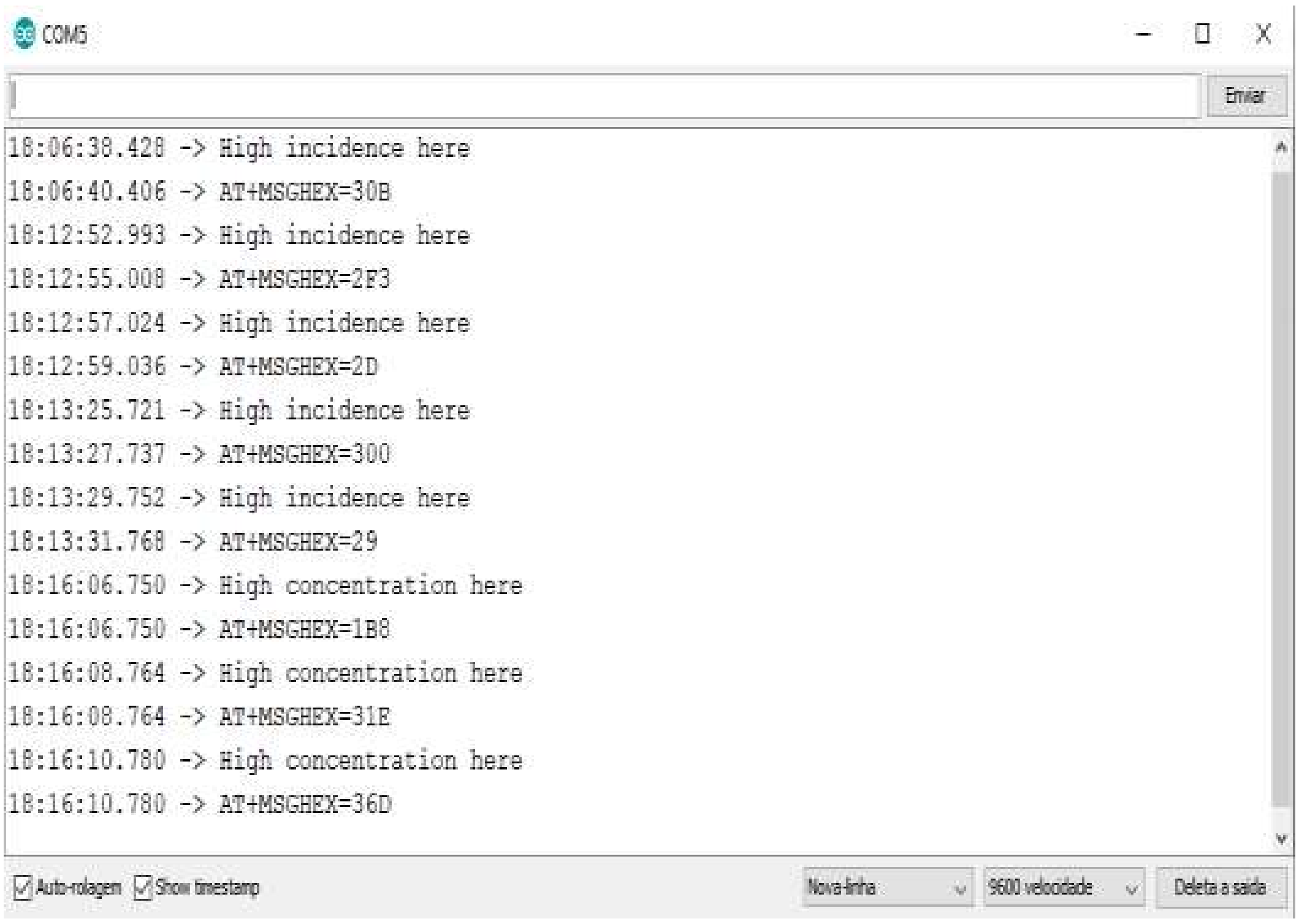

**Fonte:** Lucas Gonçalves, 2020

Nesse contexto, a estrutura de rede se manteve, mas o *end-point* foi adaptado e integrado ao sensor de gás para realizar os testes. Nesse momento, foi feito o primeiro teste em uma simulação controlada e monitorada pelo monitor serial da IDE do Arduino.

No *firmware* de teste embarcado, constava todos comandos de configuração, anteriormente testados no modem, acrescentando as condições com base no *status* das portas analógica e digital do sensor MQ-135. Com as instruções embarcadas no Arduino, já foi possível fazer o envio dos dados detectados, para o servidor *web*, como mostra na figura do primeiro protótipo, onde o *led* de *status* está aceso indicando que houve transmissão. Havendo indicação de transmissão, o próximo passo é confirmar o recebimento do pacote no servidor web TTN, o qual deve obter o pacote de *uplink* transmitido, como aparece na Figura 34.

Figura 34 – Visualização do upload dos dados para o TTN.

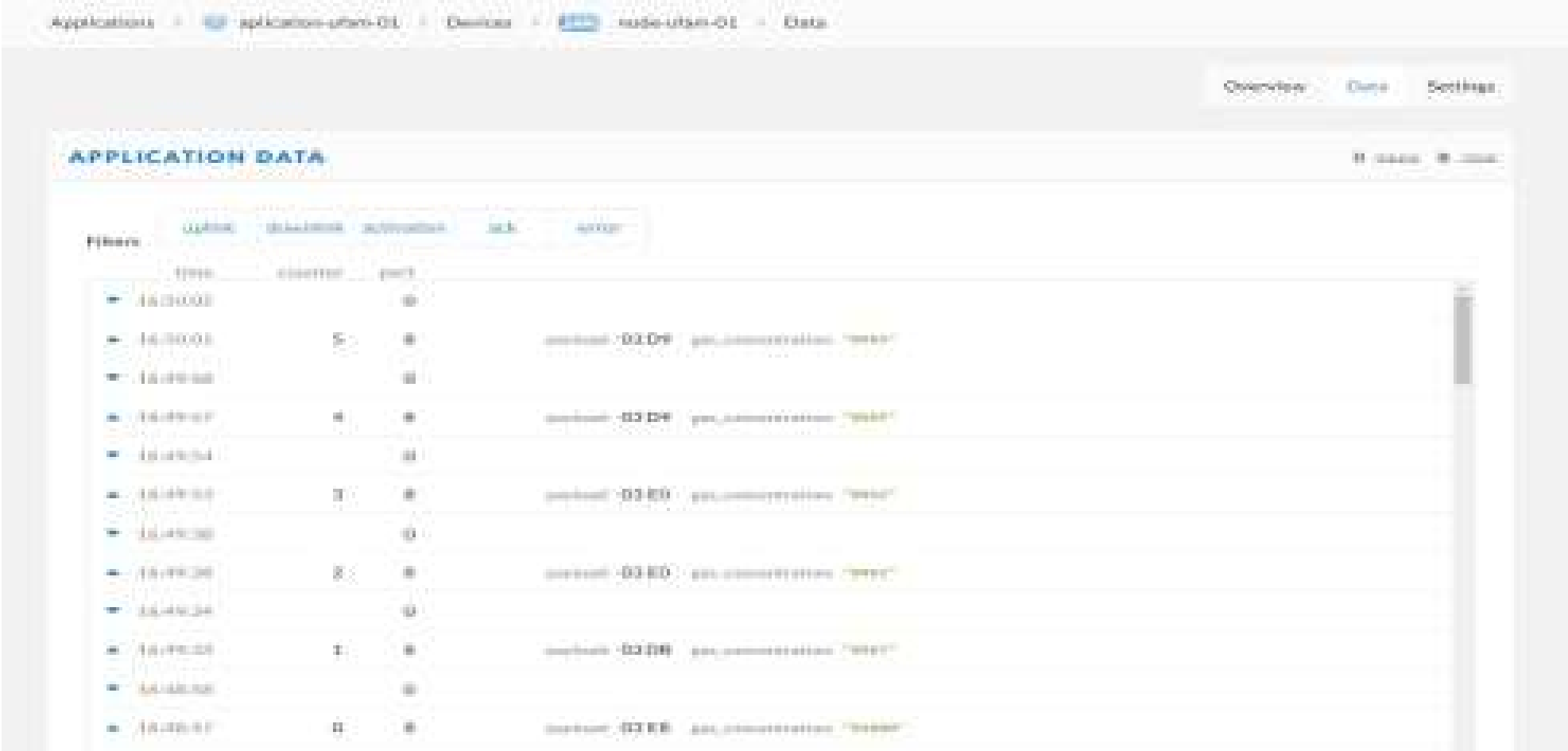

**Fonte:** Lucas Gonçalves, 2020

Para os dados chegarem no formato adequado no servidor TTN, é necessário editar o formato do *payload*, ou seja, é enviada uma *string* em hexadecimal. Mas para fazer a leitura dela, é preciso decodificá-la a fim de mostrar o valor real novamente. Feito isso, o *payload* é organizado de forma que facilite a conversão do pacote, já que é recebido em hexadecimal, a função recebe em *bytes.* Isto significa que é dividido a cada 8 *bits,* fazendo duplas de dígitos hexadecimais, considerando o *bit* mais significativo sendo o da esquerda.

Figura 35 – Visualização do decoder na plataforma TTN.

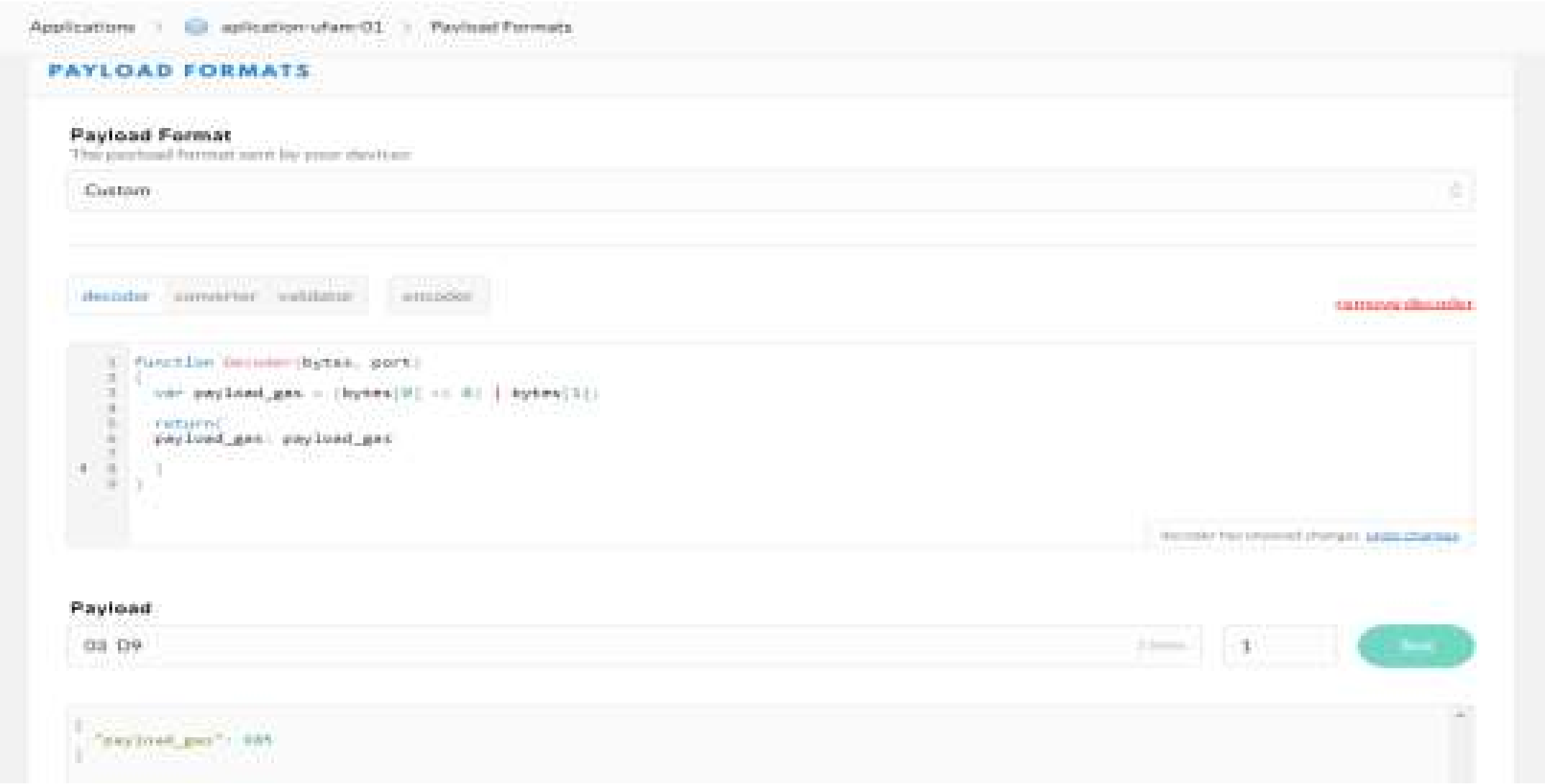

**Fonte:** Lucas Gonçalves, 2020

### 3.7.3 Cenário 3 – Teste de Alcance

Este teste foi voltado para descobrir a qual distância do *gateway*, na qual o módulo sensor é capaz de enviar os dados. Foi realizado deixando o *gateway* conectado em um ponto fixo, mudando a posição do módulo sensor para pontos mais distantes e verificando se os pacotes enviados chegavam no TTN. Foram feitas paradas por volta de 100, 200, 300 e 700 metros de distância em relação ao *gateway,* segundo o software Google Earth Pro.

Os pontos foram escolhidos com o intuito de fazer uma triangulação que pudesse testar um trecho urbano e outro trecho mais natural, com uma extensão florestal maior. Os pontos escolhidos foram:

Figura 36– Teste de alcance a 100 metros.

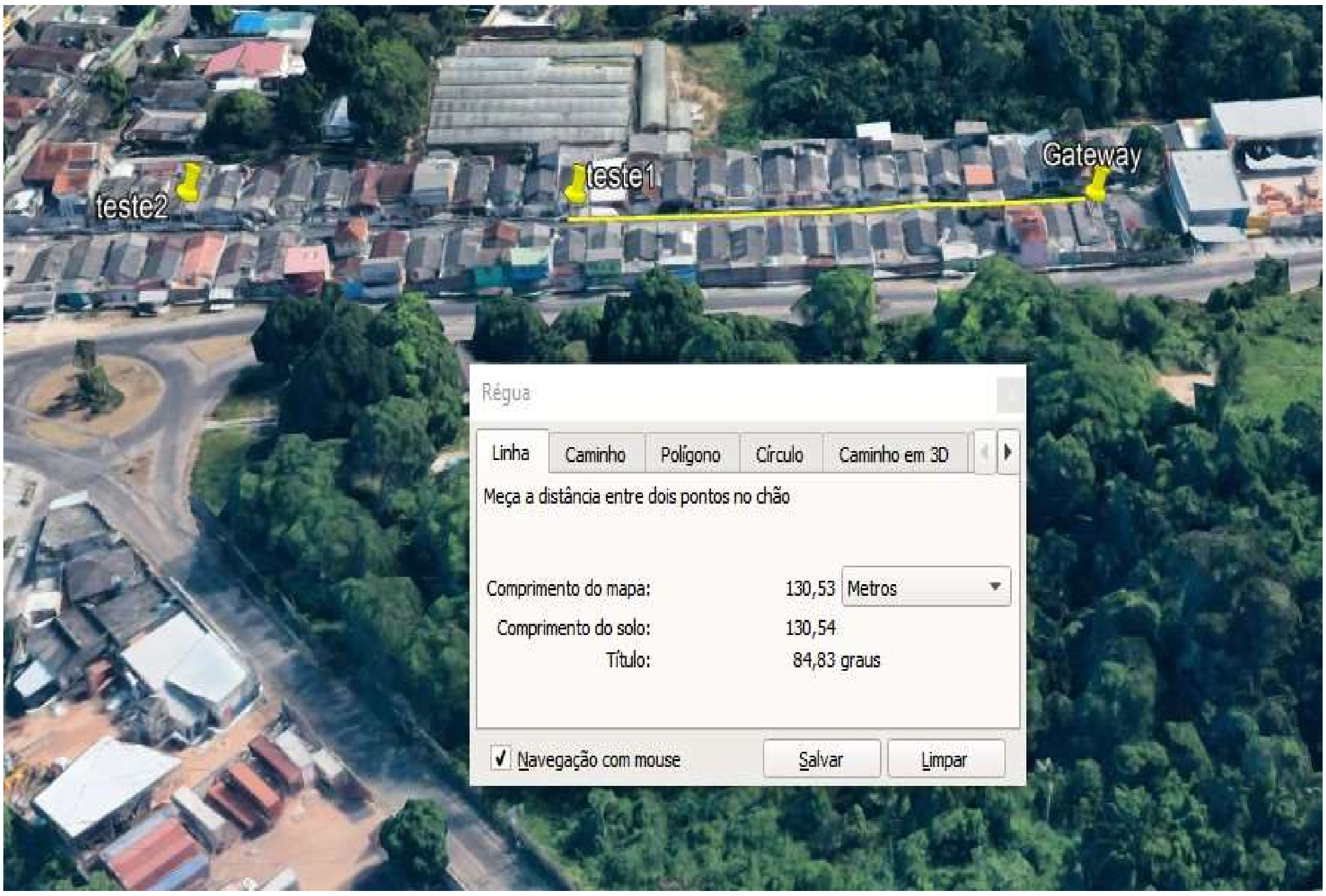

**Fonte:** Lucas Gonçalves, 2020

Figura 37 – Teste de alcance a 200 metros.

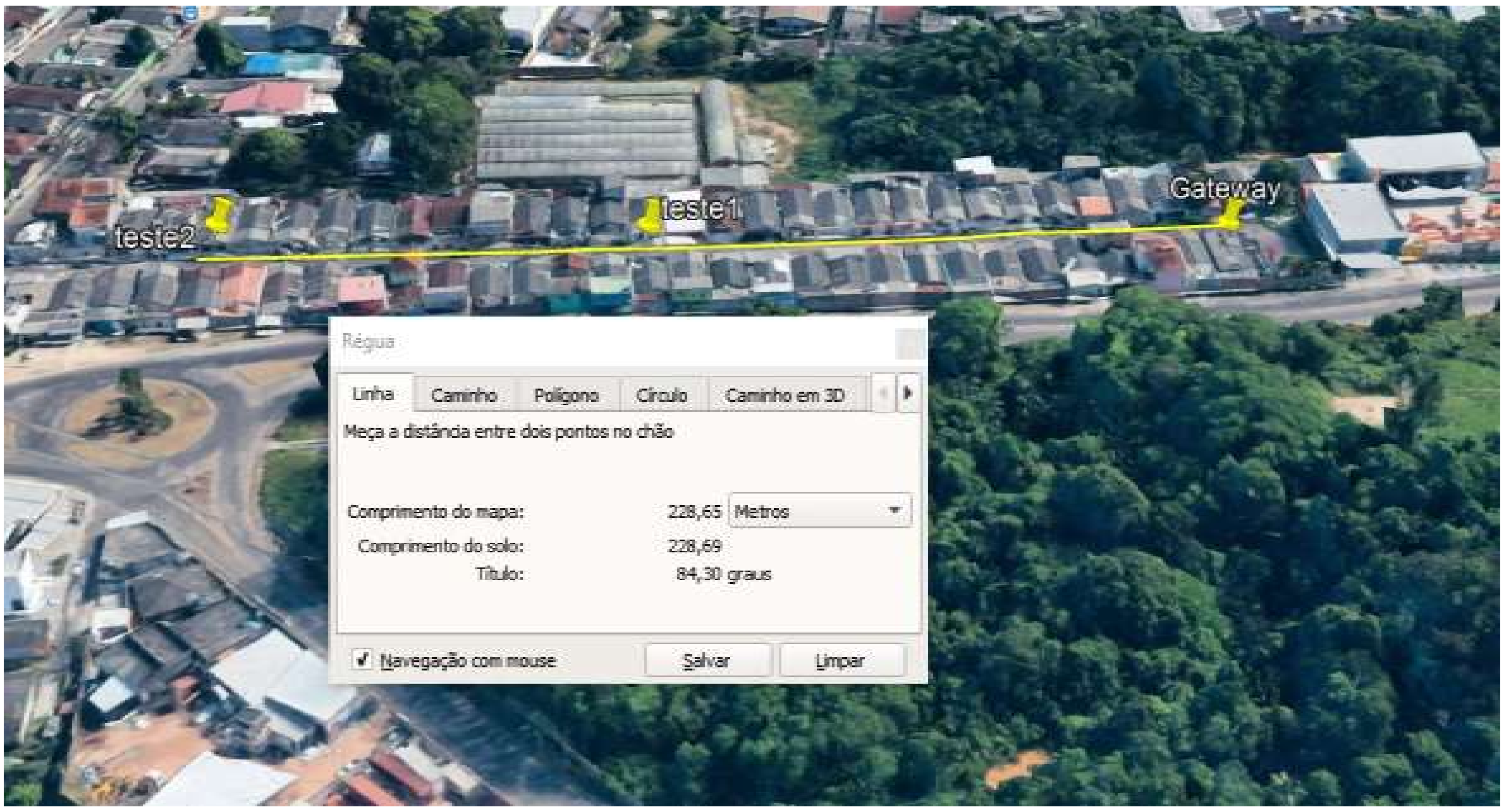

**Fonte:** Lucas Gonçalves, 2020

Figura 38 – Teste de alcance a 300 metros

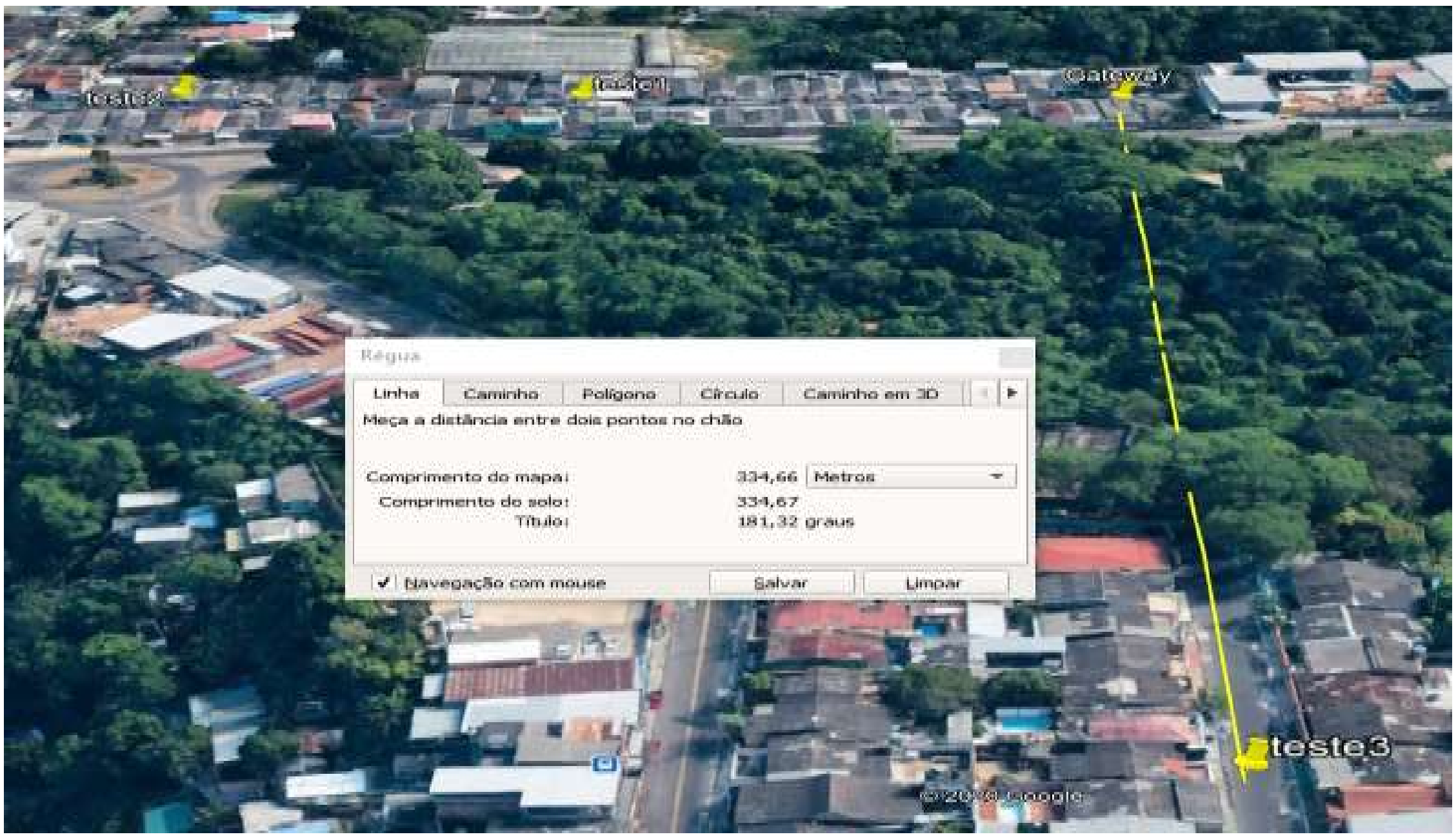

**Fonte:** Lucas Gonçalves, 2020

Figura 39 – Teste de alcance a 700 metros.

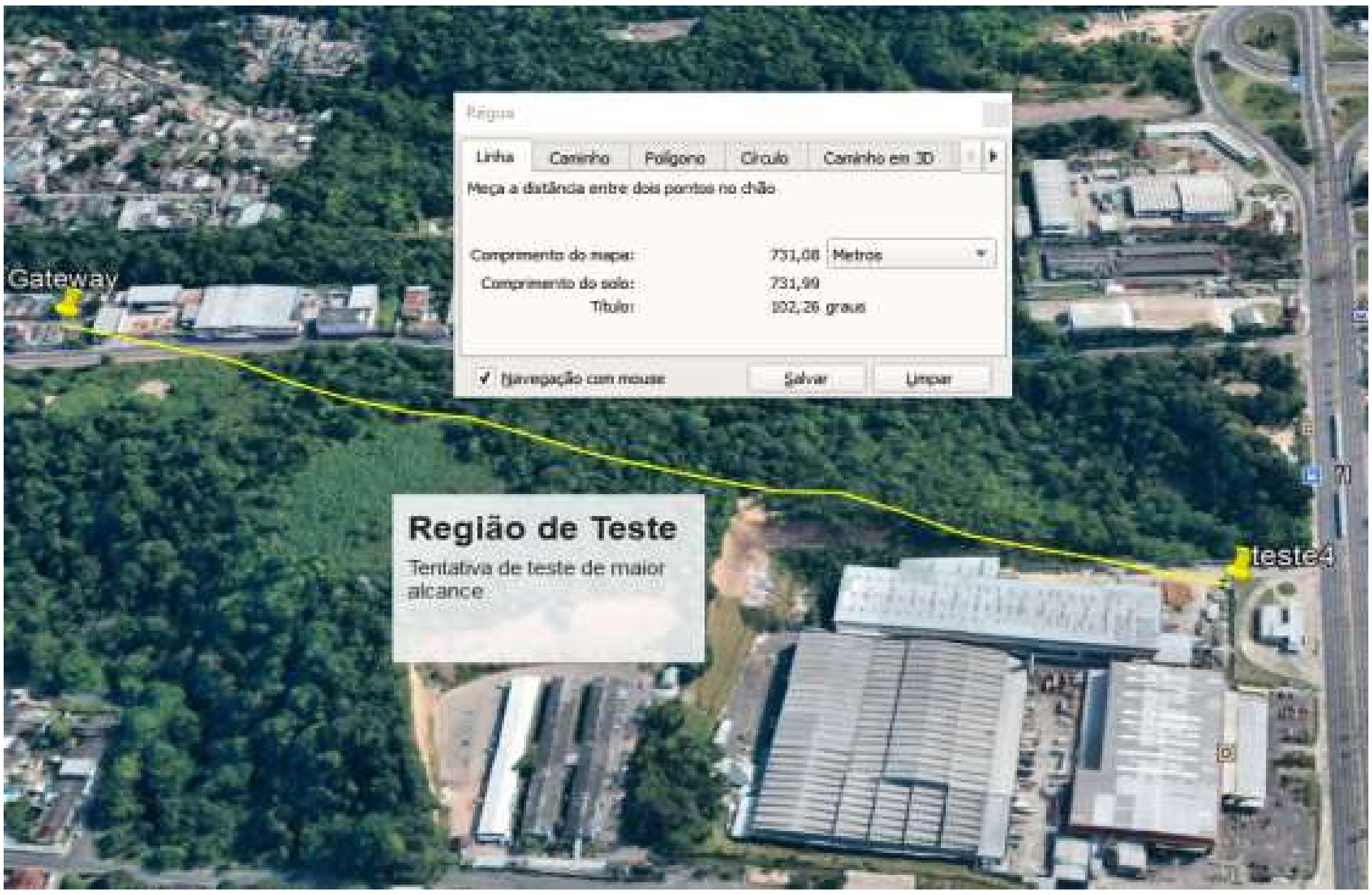

**Fonte:** Lucas Gonçalves, 2020

Observou-se que – nos pontos referentes aos testes 3 e 4, cruzando uma certa extensão florestal –, foram os mais prejudicados, enquanto os pontos com uma rota mais livre, conseguiram gerar resultados mais satisfatórios.

Tabela 8 – Valores coletados dentro dos maiores raios de alcance.

| **Testes** | **Distância (m)** | **Media_RSSI (dBm)** | **Tempo de Uplink(ms)** | **Tx recebidos (%)** |
|---|---|---|---|---|
| Teste 1 | 100 | -112 | 51,5 | 100 |
| Teste 2 | 200 | -112 | 102,9 | 95 |
| Teste 3 | 300 a 400 | -115 | 185,3 | 10 |
| Teste 4 | 700 | - | - | 0 |

A maior diferença observada foi no tempo de transmissão, como mensionado anteriormente, em que a distância e os obstáculos interferiram bastante na recepção dos pacotes nos testes 3 e 4. Ademais, a área coberta pelo sistema foi a seguinte:

Figura 40: Área urbana coberta pelo sistema

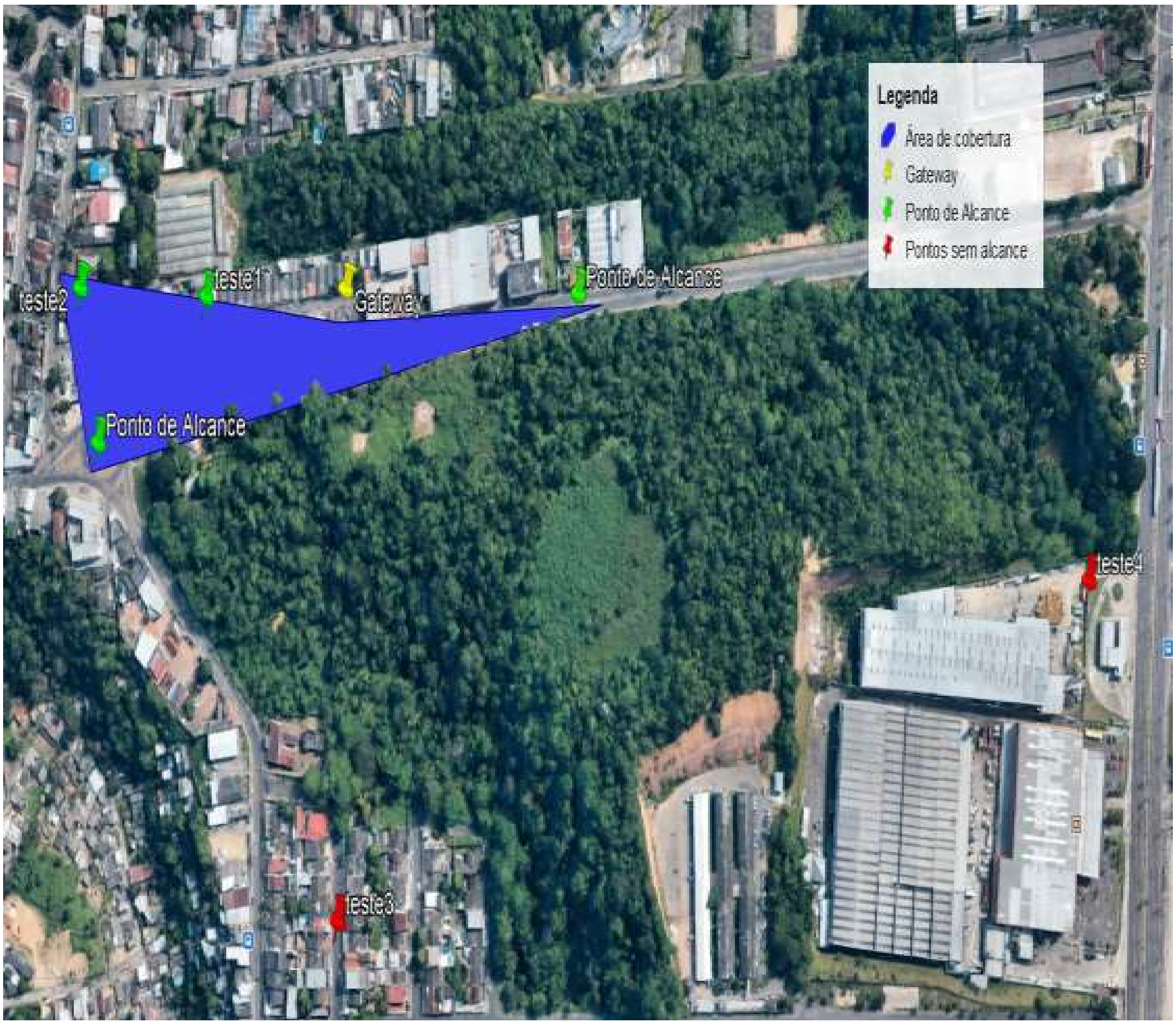

**Fonte:** Lucas Gonçalves, 2020

Em azul, encontra-se a área em que os dados são transmitos e entregues com alta eficiência. Em verdes, situa-se os limites em que são os extremos. A partir deles, observa-se uma queda no percentual de recebimento de pacotes, até ser igual a zero, como foi o caso nos pontos em vermelho.

### 3.7.4 Cenário 4 – Teste em Cenário Aberto

Este teste é para verificar se a sensibilidade dos sensores estava regulada de forma que trouxesse o máximo de precisão possível em ambientes mais próximos do real.

Figura 41 – Local de teste no cenário aberto – simulação 1

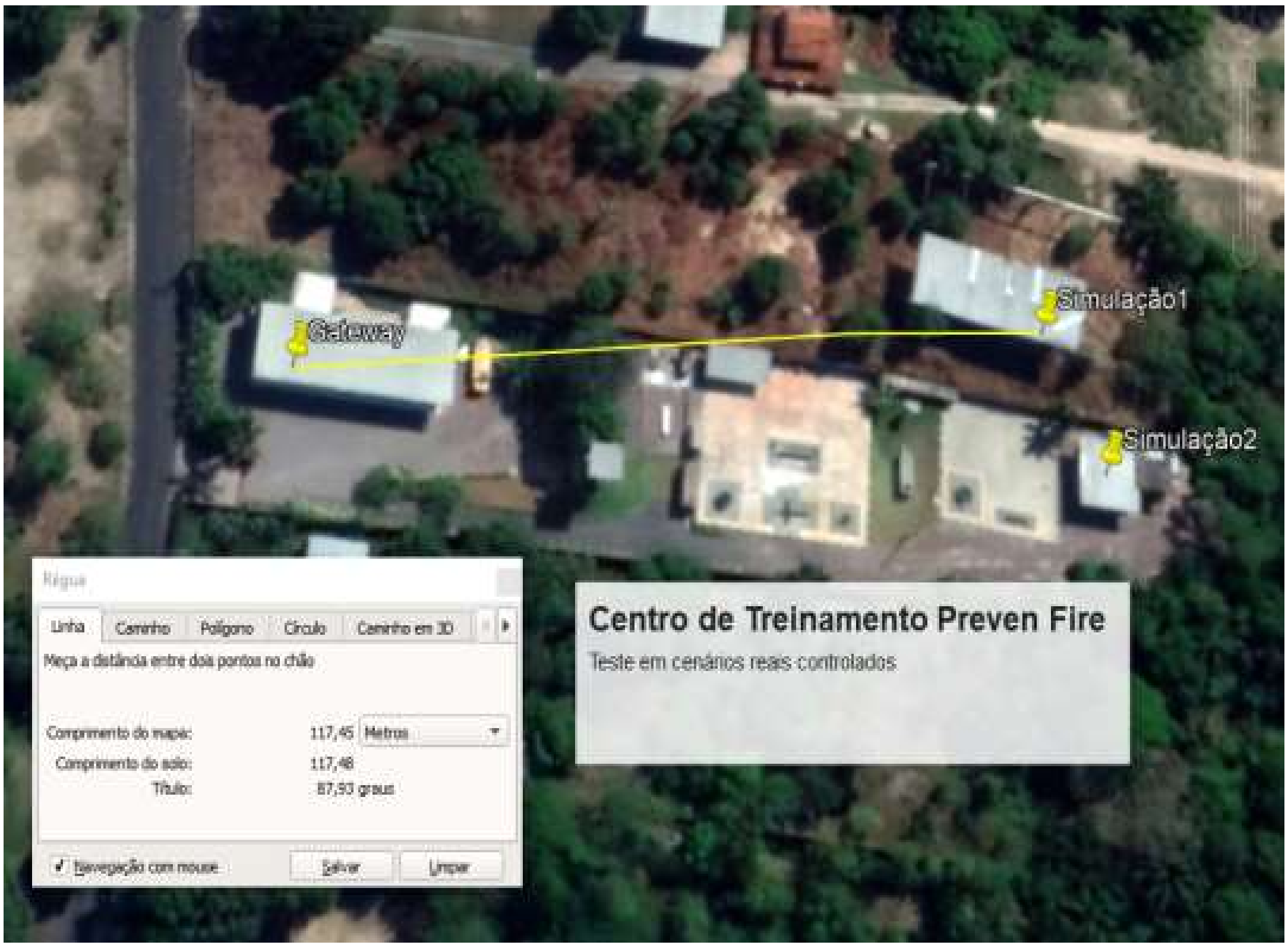

**Fonte:** Lucas Gonçalves, 2020

Foi realizado um pequeno incêndio de classe A, onde é feita uma fogueira com materiais florestais como galhos secos, capim, folhas secas, etc. Foi seguido o mesmo formato do teste anterior, distanciando o módulo sensor do *gateway*, porém em uma área aberta com um relevo mais natural, se aproximando mais de um incêndio florestal em sua fase inicial, porém em uma área com floresta menos densa.

#### 3.7.4.1 Verificação dos valores de risco outdoor

Tendo acesso a um cenário de incêndio real, com proporções controladas, foi possível obter as medidas dos sensores antes, durante e depois do incêndio, medindo em diferentes posições e em diferentes distâncias do fogo. Dessa forma os valores normais e valores que demonstram perigo, foram coletados para dar mais confiabilidade nas condições determinadas no algoritmo para tomada de decisão do envio ou não dos dados.

Figura 42 – Incêndio classe A.

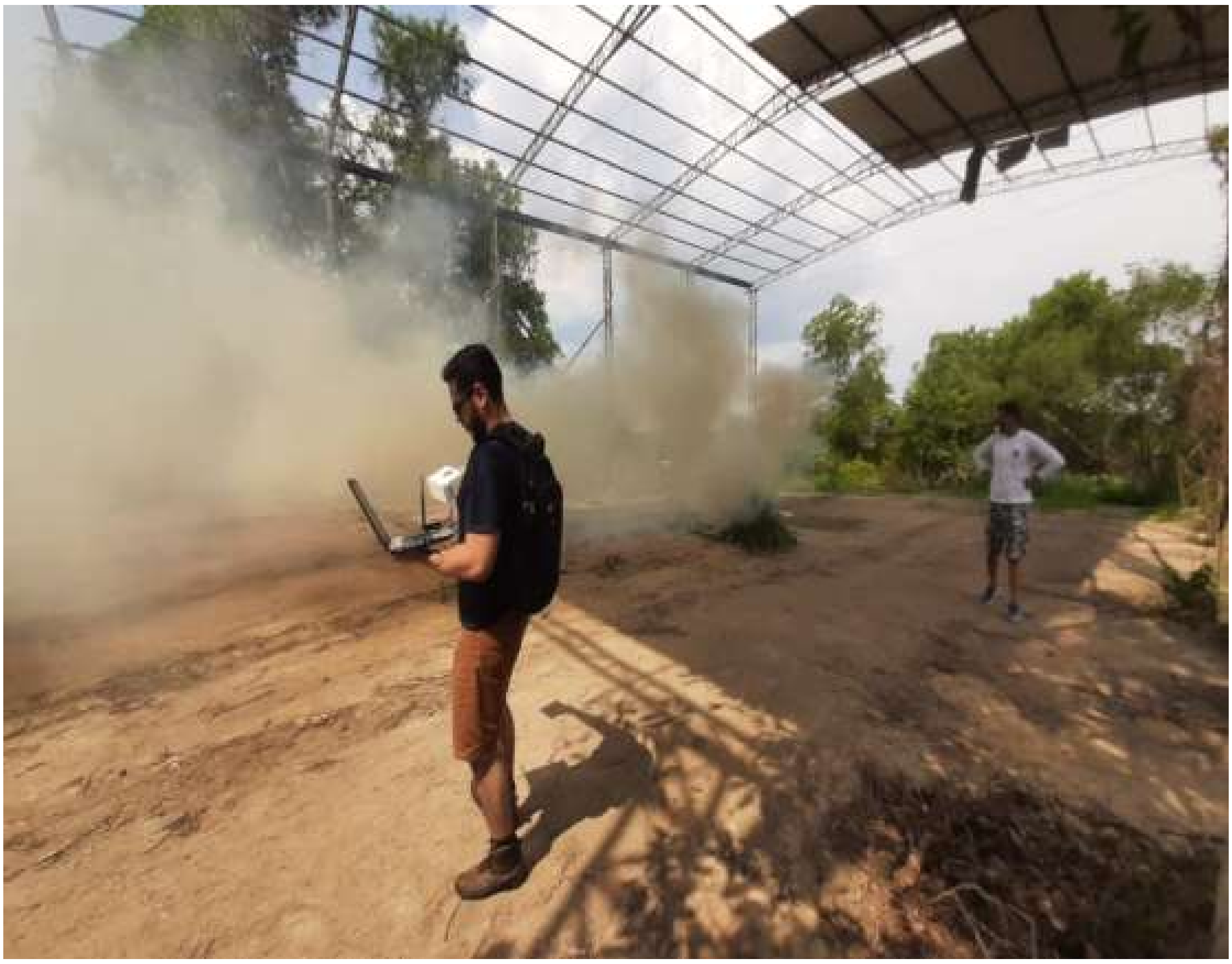

**Fonte:** Lucas Gonçalves, 2020

Tabela 9 – Valores de risco baseado nas medições em campo.

| Sensor | Valores sem risco | Valores de alerta | Valor de risco | Unidade |
|---|---|---|---|---|
| Gás | 10 a 20 | 100 a 400 | 600 a 1000 | ppm |
| Chamas | 760 a 800 | 900 | 1100 | nm |
| Temperatura | -40 a 10 | 30 a 50 | 60 a 80 | $^{o}C$ |

Essas medições foram observadas durante o experimento, e o único que conseguiu apresentar detecções a uma distância maior foi o sensor de fumaça, realizado medições até a 7 metros de distância do fogo. Utilizando o protótipo melhorado do *End-point LoRa.*

Figura 43 – Protótipo do End-Point

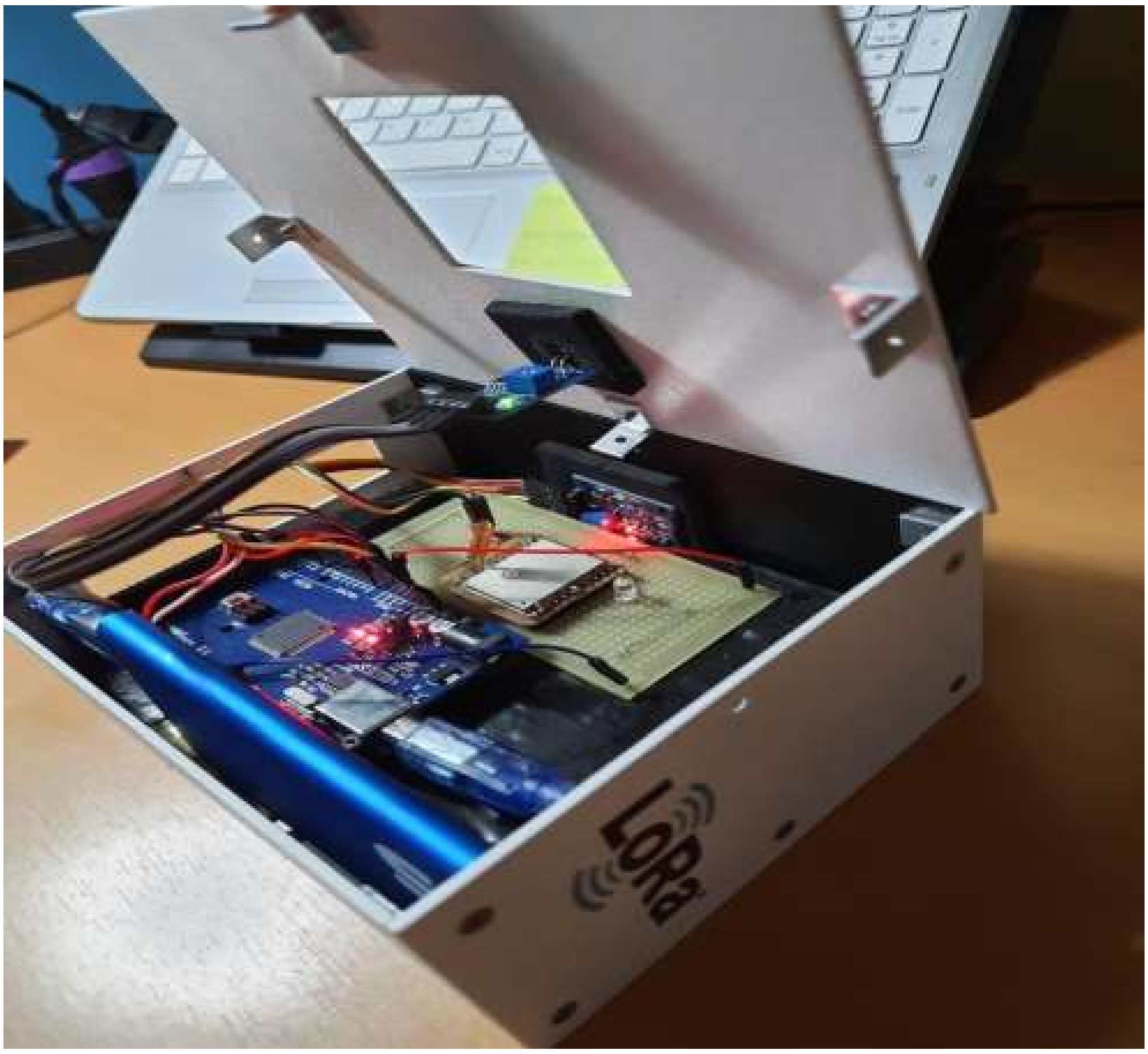

**Fonte:** Lucas Gonçalves, 2020.

# 4. Conclusão

A Internet of Things se refere a um novo nível de interação entre homem e máquina e entre as próprias máquinas também (Machine to Machine). Pois se baseia na comunicação entre sensores e atuadores com toda uma rede, possibilitando a integração de qualquer objeto com a internet. Dese modo, é evidente que a conexão de tais dispositivos será por meio de sistemas de comunicação wireless. Todavia, é necessário – com as inúmeras possibilidades de aplicação e serviços de IoT – um alcance maior, por exemplo, em ambientes *outdoor* bem distantes, variando de metros até quilômetros, onde uma taxa de transmissão de poucos bps (bits por segundo) já seria suficiente.

Ademais, tecnologias como ZigBee e Bluetooth não seriam capazes de cumprir com todos esses requisitos, uma vez que possuem alcance limitado. Já no caso da tecnologia das redes celulares, não haveria viabilidade financeira, tendo em vista a necessidade de uma grande área de cobertura e uma boa infraestrutura para uma aplicação com taxa de transmissão tão baixa, como seria no exemplo dado.

À luz dessas ideias – devido a faixa de frequência ser licenciada por empresas privadas –, a implementação de projetos IoT fica mais difícil, com uso de redes móveis. Pensando nisso, foram criadas normas para o uso da tecnologia de rede de área ampla e baixo consumo, definidas como padrões NB-IoT, objetivando a permissão da implementação no espectro destinado à tecnologia LTE (Long Term Evolution), dedicado para operações *open source*, assim como aos canais de frequência GSM (System for Mobile Communications).

Em suma, a disponibilidade de conectividade NB-IoT é muito escassa no Brasil, até porque as condições atuais para aplicações IoT de tecnologia LoRa são mais acessíveis e promissoras para os sistemas com baixa taxa de dados e a distâncias maiores. Então, neste trabalho, a tecnologia LoRa se mostrou muito viável para lidar com esse tipo de situação. Isso se deu por sua fácil integração com outros recursos de hardware e software, possibilitando uma exploração mais abrangente de suas funcionalidades e limites. Em suma, foi possível fazer levamentamentos de dados a boas distâncias, possibilitando a validação do sistema em ambientes de simulação bem próximos de um ambiente real, apresentando um bom desempenho em campo, levando em consideração o relevo das áreas que foi testado, e por ser uma adptação com Arduino e sensores compatíveis de grande acesso na indústria "*maker*", o que sugere a possibilidade de uma alta escalabilidade para projetos neste formato, voltados para aplicações em campo.

# Agradecimentos

Às instituições que ofereceram infraestrutura e apoio financeiro como: o Centro de PD em Eletrônica e Tecnologia da Informação da Universidade Federal do Amazonas (Ceteli-UFAM); Fundação de Amparo à Pesquisa do Estado do Amazonas (FAPEAM, programa PPP); o Fundo Setorial de Infraestrutura (CT-INFRA), MCT/CNPQ e Governo do Estado do Amazonas dando todo o suporte e acesso às condições necessárias para desenvolver este projeto.

# 6. Referências

# Apêndices

# APÊNDICE A – Código JavaScript para utilizar Api do Google Maps

```
// Initialize and add the map
function initMap() {
  // The location of Uluru
  const uluru = { lat: -25.344, lng: 131.036 };
  // The map, centered at Uluru
  const map = new google.maps.Map(document.getElementById("map"), {
    zoom: 4,
    center: uluru,
  });
  // The marker, positioned at Uluru
  const marker = new google.maps.Marker({
    position: uluru,
    map: map,
  });
}
```

# APÊNDICE B – Estrutura HTML para criar uma janela “Modal”

```
<div class="modal" tabindex="-1" role="dialog">
  <div class="modal-dialog" role="document">
    <div class="modal-content">
      <div class="modal-header">
        <h5 class="modal-title">Modal title</h5>
        <button type="button" class="close" data-dismiss="modal" aria-
label="Close">
          <span aria-hidden="true">&times;</span>
        </button>
      </div>
      <div class="modal-body">
        <p>Modal body text goes here.</p>
      </div>
      <div class="modal-footer">
        <button type="button" class="btn btn-primary">Save changes</button>
        <button type="button" class="btn btn-secondary" data-
dismiss="modal">Close</button>
      </div>
    </div>
  </div>
</div>
```

# APÊNDICE C – Integração do servidor web TTN com NodeJS

```
var ttn = require("ttn")

var appID = "foo"

var accessKey = "ttn-account-v2.eiPq8mEeYRL_PNBZsOpPy-O3ABJXYWulODmQGR5PZzg"

ttn.data(appID, accessKey)

then(function (client) {

client.on("uplink", function (devID, payload) {

console.log("Received uplink from ", devID)

console.log(payload)

})

})

.catch(function (error) {

console.error("Error", error)

process.exit(1)

})
```

# APÊNDICE D – Gauge para interface do dashboard

HTML:

```
<div class="gauge">
<div class="gauge__body">
<div class="gauge__fill"></div>
<div class="gauge__cover"></div>
</div>
</div>
```

CSS:

```
gauge {
width: 100%;
max-width: 250px;
font-family: "Roboto", sans-serif;
font-size: 32px;
color: #004033;
}

gauge__body {
width: 100%;
height: 0;
padding-bottom: 50%;
background: #b4c0be;
position: relative;
border-top-left-radius: 100% 200%;
border-top-right-radius: 100% 200%;
overflow: hidden;
}
```

```
gauge__fill {
position: absolute;
top: 100%;
left: 0;
width: inherit;
height: 100%;
background: #009578;
transform-origin: center top;
transform: rotate(0.25turn);
transition: transform 0.2s ease-out;
}

gauge__cover {
width: 75%;
height: 150%;
background: #ffffff;
border-radius: 50%;
position: absolute;
top: 25%;
left: 50%;
transform: translateX(-50%);
/* Text */
display: flex;
align-items: center;
justify-content: center;
padding-bottom: 25%;
box-sizing: border-box;
}
```

JavaScript:

```
const gaugeElement = document.querySelector(".gauge");
function setGaugeValue(gauge, value) {
if (value < 0 || value > 1) {
return;
}
gauge.querySelector(".gauge__fill").style.transform = `rotate(${
value / 2
}turn)`;
gauge.querySelector(".gauge__cover").textContent = `${Math.round(
value * 100
)}%`;
}
setGaugeValue(gaugeElement, 0.3);
```

# Anexos

# ANEXO A – Servidor local NodeJS

```
const express = require('express')//para API REST
const app = express(); //carregar dados
const http = require("http").createServer(app);
const io = require('socket.io')(http);
const indexRouter = require("./routes/index")(io);

app.engine(".html",require("ejs").renderFile);//reconhecer arquivos html
app.use("/js",express.static(__dirname + '/public/js/'));//mostra caminho
 do projeto
app.use("/css",express.static(__dirname + '/public/css/'));
app.use("/scripts",express.static(__dirname + '/node_modules/'));
app.use(indexRouter);

const port = 3000;
http.listen(port,()=>{
  console.log(`servidor ligado na porta: ${port}`);
})
```

# ANEXO B – HTML para pagina de interface com dashboard e mapa

```
<html lang="en">
    <head>
        <link rel="stylesheet" type="text/css" href="./scripts/bootstrap/
dist/css/bootstrap.min.css">
        <link rel="stylesheet" type="text/css" href="./css/styles.css">
        <link rel="stylesheet" type="text/css" href="./css/gauge.css">
        <meta charset="UTF-8">
        <meta name="viewport" content="width=device-width, initial-
scale=1.0">
        <title>Dashboard</title>
    </head>
        <body>
            <div id="map"></div>

            <div class="modal" id="modalDevice" data-
backdrop="static" data-keyboard="false" tabindex="-1" role="dialog" aria-
labelledby="staticBackdropLabel" aria-hidden="true">
                <div class="modal-dialog">
                    <div class="modal-content">

                        <div class="modal-header">
                            <h6 id="titleDevice"></h6>
                            <button type="button" class="close" data-
dismiss="modal" aria-label="Close">
                            <span aria-hidden="true">&times;</span>
                            </button>
                        </div>

                        <div class="modal-body">
                            <div class="container-fluid">
                                <div class="column">
                                    <div class="col-xs-12 col-md-6 m-
auto">
                                        <h6 class="text-
center">Gas <span class="gauge_gas_text"></span></h6>
                                        <div class="gauge gauge_gas">
                                            <div class="gauge__body">
                                                <div class="gauge__fill g
auge__fill--gas"></div>
                                                <div class="gauge__cover"
></div>
                                            </div>
                                        </div>
                                    </div><br>
```

```
                                    <div class="col-xs-12 col-md-6 m-
auto">
                                        <h6 class="text-
center">Fire <span class="gauge_fire_text"></span></h6>
                                        <div class="gauge gauge_fire">
                                            <div class="gauge__body">
                                                <div class="gauge__fill g
auge__fill--fire"></div>
                                                <div class="gauge__cover"
></div>
                                            </div>
                                        </div>
                                    </div><br>
                                    <div class="col-xs-12 col-md-6 m-
auto">
                                        <h6 class="text-
center">Temperature <span class="gauge_temp_text"></span></h6>
                                        <div class="gauge gauge_temp">
                                            <div class="gauge__body">
                                                <div class="gauge__fill g
auge__fill--temp"></div>
                                                <div class="gauge__cover"
></div>
                                            </div>
                                        </div>
                                    </div>
                                </div>
                            </div>
                        </div>
                    </div>
                </div>
            </div>

            <script src="./scripts/axios/dist/axios.min.js" type="text/ja
vascript"></script>
            <script src="./scripts/socket.io-
client/dist/socket.io.js" type="text/javascript"></script>
            <script src="./scripts/canvas-
gauges/gauge.min.js" type="text/javascript"></script>
            <script src="./scripts/jquery/dist/jquery.min.js" type="text/
javascript"></script>
            <script src="./scripts/bootstrap/dist/js/bootstrap.min.js" ty
pe="text/javascript"></script>


            <script type="text/javascript" src="./js/script.js"></script>
```

```
            <script type="text/javascript" src="./js/realtime.js"></scrip
t>

            <script async defer src="https://maps.googleapis.com/maps/api
/js?key=AIzaSyAn51vnthzG3sFMj3i-
            ozZ09jwwpYTOyAA&callback=initMap"></script>

        </body>
</html>
```

ANEXO C – Atualização/requisição de dados em tempo real

```
//direcionamento da msg do device para a pagina web

const socket = io('http://localhost:3000/');

const gaugeGas = document.querySelector(".gauge_gas");
const gaugeFire = document.querySelector(".gauge_fire");
const gaugeTemp = document.querySelector(".gauge_temp");
const gaugeGasText = document.querySelector(".gauge_gas_text");
const gaugeFireText = document.querySelector(".gauge_fire_text");
const gaugeTempText = document.querySelector(".gauge_temp_text");

function getDataFromDevice(deviceName){

    console.log('Listening data for device: ' + deviceName);

    socket.on("arrived", (msg)=>{
      let fire = msg.dados.payload_fields.payload_fire;
      let gas = msg.dados.payload_fields.payload_gas;
      let temp = msg.dados.payload_fields.payload_temp;

      gaugeGasText.innerHTML = "(" + gas + ")";
      gaugeFireText.innerHTML = "(" + fire + ")";
      gaugeTempText.innerHTML = "(" + temp + ")";

      setGaugeValue(gaugeGas, convertToMinimumDecimal(gas));
      setGaugeValue(gaugeFire, convertToMinimumDecimal(fire));
      setGaugeValue(gaugeTemp, convertToMinimumDecimal(temp));
    });
}

function toggleBounce(event) {

    getDataFromDevice(this.label);
    $('#modalDevice').modal();
    $('#titleDevice').html(this.label);
```

```
    //animação para node 1 interagir com o click
    if (this.getAnimation() !== null) {//condição seria payload diferente
 de vazio
        this.setAnimation(null);
    } else {
      this.setAnimation(google.maps.Animation.BOUNCE);
    }
}

function setGaugeValue(gauge, value) {
  if (value < 0 || value > 1) {
    return;
  }

  gauge.querySelector(".gauge__fill").style.transform = `rotate(${value /
 2}turn)`;

  if(gauge.classList.value === 'gauge gauge_fire'){
    gauge.querySelector(".gauge__fill").style.background = `#e44b25`;
  }
  if(gauge.classList.value === 'gauge gauge_temp'){
    gauge.querySelector(".gauge__fill").style.background = `#6a13dd`;
  }

  gauge.querySelector(".gauge__cover").textContent = `${Math.round(value
* 100)}%`;
}

function convertToMinimumDecimal(value){
  if(isNaN(value)) return 0;
  const val = parseInt(value, 10);

  if(val >= 1000){
    return val / 10000;
  }
  else if(val >= 100){
    return val / 1000;
  }
  else if(val >= 10){
    return val / 100;
  }
  else{
   return val / 10;
  }
}
```

```
setGaugeValue(gaugeGas, convertToMinimumDecimal(0));
setGaugeValue(gaugeFire, convertToMinimumDecimal(0));
setGaugeValue(gaugeTemp, convertToMinimumDecimal(0));
```

# ANEXO D – Firmware com algoritmo dos sensores e comandos AT

```
#include <dht.h>
dht&nbspDHT;
#define DHT22_PIN 5

#define GasPin&nbspA0
#define FirePin&nbspA1
#define DigitalGasPin 2
#define DigitalFirePin 3

void setup() {
  //&nbspput&nbspyour&nbspsetup&nbspcode&nbsphere,&nbspto&nbsprun&nbsponc
e:

 &nbspSerial.begin(9600);
  pinMode(GasPin,&nbspINPUT);
  pinMode(FirePin,&nbspINPUT);

 &nbspSerial.print("AT\r\n");
 //Serial.print("+AT:OK\r\n");
  delay(2000);
 &nbspSerial.print("AT+ID=DevAddr,\"2603172D\"\r\n");
  delay(2000);
 &nbspSerial.read();
 &nbspSerial.print("AT+ID=DevEui,\"00E0136E0847D7F8\"\r\n");
  delay(2000);
 &nbspSerial.print("AT+ID=AppEui,\"70B3D57ED0014F64\"\r\n");
  delay(2000);
 &nbspSerial.print("AT+KEY=NwkSKey,\"F6012FAD4F28BEA501A4E9841D8A0EBC\"\r
\n");
  delay(2000);
 &nbspSerial.print("AT+KEY=AppSKey,\"A484A36F909D5A74D7456BBB2C511058\"\r
\n");
  delay(2000);
 &nbspSerial.print("AT+ID=?\r\n");
  delay(1000);
 &nbspSerial.print("AT+DR=AU915\r\n");
  delay(1000);
 &nbspSerial.print("AT+CH=0,915.2,2,5\r\n");
  delay(1000);
 &nbspSerial.print("AT+CH=1,915.4,2,5\r\n");
  delay(1000);
 &nbspSerial.print("AT+CH=2,915.6,2,5\r\n");
  delay(1000);
 &nbspSerial.print("AT+CH=3,915.8,2,5\r\n");
```

```
 delay(1000);
&nbspSerial.print("AT+CH=4,916.0,2,5\r\n");
 delay(1000);
&nbspSerial.print("AT+CH=5,916.2,2,5\r\n");
 delay(1000);
&nbspSerial.print("AT+CH=6,916.4,2,5\r\n");
 delay(1000);
&nbspSerial.print("AT+CH=7,916.6,2,5\r\n");
 delay(1000);
&nbspSerial.print("AT+RXWIN2=923.3,DR8\r\n");
 delay(1000);
&nbspSerial.print("AT+RESET\r\n");
 delay(1000);
&nbspSerial.print("AT+ADR=ON\r\n");
 delay(1000);
 //Serial.print("AT+DR=DR2\r\n");
 //delay(20);
&nbspSerial.print("AT+MODE=LWABP\r\n");
 delay(1000);
&nbspSerial.print("AT+CLASS=A\r\n");
 delay(1000);
&nbspSerial.print("AT+CH=8,0,2,5\r\n");
 delay(1000);
&nbspSerial.print("AT+CH=9,0,2,5\r\n");
 delay(1000);
&nbspSerial.print("AT+CH=10,0,2,5\r\n");
 delay(1000);
&nbspSerial.print("AT+CH=11,0,2,5\r\n");
 delay(1000);
&nbspSerial.print("AT+CH=12,0,2,5\r\n");
 delay(1000);
&nbspSerial.print("AT+CH=13,0,2,5\r\n");
 delay(1000);
&nbspSerial.print("AT+CH=14,0,2,5\r\n");
 delay(1000);
&nbspSerial.print("AT+CH=15,0,2,5\r\n");
 delay(1000);
&nbspSerial.print("AT+CH=16,0,2,5\r\n");
 delay(1000);
&nbspSerial.print("AT+CH=17,0,2,5\r\n");
 delay(1000);
&nbspSerial.print("AT+CH=18,0,2,5\r\n");
 delay(1000);
&nbspSerial.print("AT+CH=19,0,2,5\r\n");
 delay(1000);
&nbspSerial.print("AT+CH=20,0,2,5\r\n");
 delay(1000);
```

```
&nbspSerial.print("AT+CH=21,0,2,5\r\n");
 delay(1000);
&nbspSerial.print("AT+CH=22,0,2,5\r\n");
 delay(1000);
&nbspSerial.print("AT+CH=23,0,2,5\r\n");
 delay(1000);
&nbspSerial.print("AT+CH=24,0,2,5\r\n");
 delay(1000);
&nbspSerial.print("AT+CH=25,0,2,5\r\n");
 delay(1000);
&nbspSerial.print("AT+CH=26,0,2,5\r\n");
 delay(1000);
&nbspSerial.print("AT+CH=27,0,2,5\r\n");
 delay(1000);
&nbspSerial.print("AT+CH=28,0,2,5\r\n");
 delay(1000);
&nbspSerial.print("AT+CH=29,0,2,5\r\n");
 delay(1000);
&nbspSerial.print("AT+CH=30,0,2,5\r\n");
 delay(1000);
&nbspSerial.print("AT+CH=31,0,2,5\r\n");
 delay(1000);
&nbspSerial.print("AT+CH=32,0,2,5\r\n");
 delay(1000);
&nbspSerial.print("AT+CH=33,0,2,5\r\n");
 delay(1000);
&nbspSerial.print("AT+CH=34,0,2,5\r\n");
 delay(1000);
&nbspSerial.print("AT+CH=35,0,2,5\r\n");
 delay(1000);
&nbspSerial.print("AT+CH=36,0,2,5\r\n");
 delay(1000);
&nbspSerial.print("AT+CH=37,0,2,5\r\n");
 delay(1000);
&nbspSerial.print("AT+CH=38,0,2,5\r\n");
 delay(1000);
&nbspSerial.print("AT+CH=39,0,2,5\r\n");
 delay(1000);
&nbspSerial.print("AT+CH=40,0,2,5\r\n");
 delay(1000);
&nbspSerial.print("AT+CH=41,0,2,5\r\n");
 delay(1000);
&nbspSerial.print("AT+CH=42,0,2,5\r\n");
 delay(1000);
&nbspSerial.print("AT+CH=43,0,2,5\r\n");
 delay(1000);
&nbspSerial.print("AT+CH=44,0,2,5\r\n");
```

```
 delay(1000);
&nbspSerial.print("AT+CH=45,0,2,5\r\n");
 delay(1000);
&nbspSerial.print("AT+CH=46,0,2,5\r\n");
 delay(1000);
&nbspSerial.print("AT+CH=47,0,2,5\r\n");
 delay(1000);
&nbspSerial.print("AT+CH=48,0,2,5\r\n");
 delay(1000);
&nbspSerial.print("AT+CH=49,0,2,5\r\n");
 delay(1000);
&nbspSerial.print("AT+CH=50,0,2,5\r\n");
 delay(1000);
&nbspSerial.print("AT+CH=51,0,2,5\r\n");
 delay(1000);
&nbspSerial.print("AT+CH=52,0,2,5\r\n");
 delay(1000);
&nbspSerial.print("AT+CH=53,0,2,5\r\n");
 delay(1000);
&nbspSerial.print("AT+CH=54,0,2,5\r\n");
 delay(1000);
&nbspSerial.print("AT+CH=55,0,2,5\r\n");
 delay(1000);
&nbspSerial.print("AT+CH=56,0,2,5\r\n");
 delay(1000);
&nbspSerial.print("AT+CH=57,0,2,5\r\n");
 delay(1000);
&nbspSerial.print("AT+CH=58,0,2,5\r\n");
 delay(1000);
&nbspSerial.print("AT+CH=59,0,2,5\r\n");
 delay(1000);
&nbspSerial.print("AT+CH=60,0,2,5\r\n");
 delay(1000);
&nbspSerial.print("AT+CH=61,0,2,5\r\n");
 delay(1000);
&nbspSerial.print("AT+CH=62,0,2,5\r\n");
 delay(1000);
&nbspSerial.print("AT+CH=63,0,2,5\r\n");
 delay(1000);
&nbspSerial.print("AT+CH=64,0,2,5\r\n");
 delay(1000);
&nbspSerial.print("AT+CH=65,0,2,5\r\n");
 delay(1000);
&nbspSerial.print("AT+CH=66,0,2,5\r\n");
 delay(1000);
&nbspSerial.print("AT+CH=67,0,2,5\r\n");
 delay(1000);
```

```
 &nbspSerial.print("AT+CH=68,0,2,5\r\n");
  delay(1000);
 &nbspSerial.print("AT+CH=69,0,2,5\r\n");
  delay(1000);
 &nbspSerial.print("AT+CH=70,0,2,5\r\n");
  delay(1000);
 &nbspSerial.print("AT+CH=71,0,2,5\r\n");
  delay(1000);
  //Serial.print("AT+DR=DR2\r\n");
 &nbspSerial.print("AT+DR=DR3\r\n");
  delay(1000);

}

void loop() {

  //&nbspput&nbspyour&nbspmain&nbspcode&nbsphere,&nbspto&nbsprun&nbsprepe
atedly:

 int&nbspgas_digital = digitalRead(DigitalGasPin);
 int&nbspfire_digital = digitalRead(DigitalFirePin);

 uint32_t&nbspgas_concentration = analogRead(GasPin);
 uint32_t&nbspfire_incidence = analogRead(FirePin);

//////////////////////////TESTAR&nbspSENSOR&nbspTEMP/////////////////////
////////////

 uint32_t&nbspstart = micros();
 int&nbspchk =&nbspDHT.read22(DHT22_PIN);
 uint32_t&nbspstop = micros();
 int&nbsptemp_value =&nbspDHT.temperature;

 //Temperatura&nbspem&nbspgraus&nbspcelsius&nbspem&nbspdecimal
Serial.print("Tempeture&nbspdecimal: ");
Serial.println(DHT.temperature);
delay(5000);


&nbspbyte&nbsppayload[6];
&nbsppayload[0] = highByte(gas_concentration);
&nbsppayload[1] = lowByte(gas_concentration);
&nbsppayload[2] = highByte(fire_incidence);
&nbsppayload[3] = lowByte(fire_incidence);
&nbsppayload[4] = highByte(temp_value);
&nbsppayload[5] = lowByte(temp_value);
```

```
////////////////////////////TESTAR&nbspSENSOR&nbspTEMP///////////////////
//////////////////////

  if(analogRead(GasPin)>255 and analogRead(FirePin)>255){
   &nbspSerial.println("MONITORING");

   &nbspSerial.print("fire_incidence&nbspDEC: ");
   &nbspSerial.println(analogRead(FirePin));
    delay(5000);
   &nbspSerial.print("fire_incidence&nbspHEX: ");
   &nbspSerial.println(fire_incidence,HEX);
    delay(5000);
   &nbspSerial.println("AT+MSGHEX=0"+String(payload[0],HEX)+String(payloa
d[1],HEX)+String(0)+String(payload[2],HEX)+String(payload[3],HEX)+String(
0)+String(payload[4],HEX)+String(payload[5],HEX));
    delay(5000);
   &nbspSerial.print("gas_concentration&nbspDEC: ");
   &nbspSerial.println(analogRead(GasPin));
    delay(5000);
   &nbspSerial.print("gas_concentration&nbspHEX: ");
   &nbspSerial.println(gas_concentration,HEX);
    delay(5000);

  }
  if(analogRead(GasPin)<16 and analogRead(FirePin)<40){
   &nbspSerial.println("TESTING");

   &nbspSerial.print("fire_incidence&nbspDEC: ");
   &nbspSerial.println(analogRead(FirePin));
    delay(5000);
   &nbspSerial.print("fire_incidence&nbspHEX: ");
   &nbspSerial.println(fire_incidence,HEX);
    delay(5000);
   &nbspSerial.println("AT+MSGHEX=0"+ String(0)+String(payload[0],HEX)+St
ring(payload[1],HEX)+String(0)+String(payload[2],HEX)+String(payload[3],H
EX)+String(0)+String(payload[4],HEX)+String(payload[5],HEX));
    delay(5000);
   &nbspSerial.print("gas_concentration&nbspDEC: ");
   &nbspSerial.println(analogRead(GasPin));
    delay(5000);
   &nbspSerial.print("gas_concentration&nbspHEX: ");
   &nbspSerial.println(gas_concentration,HEX);
    delay(5000);
  }
  while(analogRead(GasPin)>300 and analogRead(FirePin)<400){
    if(gas_digital==0 and&nbspfire_digital==0){
      &nbspSerial.println("FIRE!");
```

```
      &nbspSerial.println("AT+MSGHEX=0"+String(payload[0],HEX)+String(pay
load[1],HEX)+String(0)+String(0)+String(payload[2],HEX)+String(payload[3]
,HEX)+String(0)+String(payload[4],HEX)+String(payload[5],HEX));
       delay(5000);

      }
      if(gas_digital==1 and&nbspfire_digital==0){
      &nbspSerial.println("High&nbspincidence&nbsphere!");
      &nbspSerial.println("AT+MSGHEX=0"+String(payload[0],HEX)+String(pay
load[1],HEX)+String(0)+String(payload[2],HEX)+String(payload[3],HEX)+Stri
ng(0)+String(payload[4],HEX)+String(payload[5],HEX));
       delay(5000);

      }
      if(gas_digital==0 and&nbspfire_digital==1){
      &nbspSerial.println("High&nbspconcentration&nbsphere");
      &nbspSerial.println("AT+MSGHEX=0"+String(payload[0],HEX)+String(pay
load[1],HEX)+String(0)+String(payload[2],HEX)+String(payload[3],HEX)+Stri
ng(0)+String(payload[4],HEX)+String(payload[5],HEX));
       delay(5000);

      }
   &nbspSerial.println("DANGER!!!");

   &nbspSerial.print("fire_incidence&nbspDEC: ");
   &nbspSerial.println(analogRead(FirePin));
    delay(5000);
   &nbspSerial.print("fire_incidence&nbspHEX: ");
   &nbspSerial.println(fire_incidence,HEX);
    delay(5000);
   &nbspSerial.println("AT+MSGHEX=0"+String(0)+String(payload[0],HEX)+Str
ing(payload[1],HEX)+String(0)+String(payload[2],HEX)+String(payload[3],HE
X)+String(0)+String(payload[4],HEX)+String(payload[5],HEX));
    delay(5000);
   &nbspSerial.print("gas_concentration&nbspDEC: ");
   &nbspSerial.println(analogRead(GasPin));
    delay(5000);
   &nbspSerial.print("gas_concentration&nbspHEX: ");
   &nbspSerial.println(gas_concentration,HEX);
    delay(5000);
  }

 }
```

# Formulário de Identificação

Exemplo de Formulário de Identificação, compatível com o Anexo A (informativo) da ABNT NBR 10719:2011. Este formulário não é um anexo. Conforme definido na norma, ele é o último elemento pós-textual e opcional do relatório.

<table>
<tr><th colspan="4">Dados do Relatório Técnico e/ou científico</th></tr>
<tr><td colspan="3" rowspan="2">Título e subtítulo</td><td>Classificação de segurança</td></tr>
<tr><td>No.</td></tr>
<tr><td colspan="3">Tipo de relatório</td><td>Data</td></tr>
<tr><td colspan="3">Título do projeto/programa/plano</td><td>No.</td></tr>
<tr><td colspan="4">Autor(es)</td></tr>
<tr><td colspan="4">Instituição executora e endereço completo</td></tr>
<tr><td colspan="4">Instituição patrocinadora e endereço completo</td></tr>
<tr><td colspan="4">Resumo</td></tr>
<tr><td colspan="4">Palavras-chave/descritores</td></tr>
<tr><td>Edição</td><td>No. de páginas</td><td>No. do volume</td><td>N<sup>o</sup> de classificação</td></tr>
<tr><td>ISSN</td><td colspan="2">Tiragem</td><td>Preço</td></tr>
<tr><td colspan="4">Distribuidor</td></tr>
<tr><td colspan="4">Observações/notas</td></tr>
</table>